\documentclass[%
preprint,
 amsmath,amssymb,
 aps,
longbibliography
]{revtex4-1}

\usepackage{graphicx}
\usepackage{dcolumn}
\usepackage{bm}
\usepackage[colorlinks,linkcolor=blue, urlcolor=blue,anchorcolor=blue, citecolor=blue]{hyperref}
\usepackage{lineno}
\usepackage{multirow}

\begin{document}

\preprint{}

\title{A wall model based on neural networks for LES of turbulent flows over periodic hills}

\author%
{Zhideng Zhou$^{1,2}$, 
Guowei He$^{1,2,}$ 
and Xiaolei Yang$^{1,2,}$
}
\email[Corresponding author: ]{xyang@imech.ac.cn}

\affiliation{$^1$The State Key Laboratory of Nonlinear Mechanics, Institute of Mechanics, Chinese Academy of Sciences, Beijing 100190, China\\
$^2$School of Engineering Sciences, University of Chinese Academy of Sciences, Beijing 100049, China}


\begin{abstract}
In this work, a data-driven wall model for turbulent flows over periodic hills is developed using the feedforward neural network (FNN) and wall-resolved LES (WRLES) data. To develop a wall model applicable to different flow regimes, the flow data in the near wall region at all streamwise locations are grouped together as the training dataset. In the developed FNN wall models, we employ the wall-normal distance, near-wall velocities and pressure gradients as input features and the wall shear stresses as output labels, respectively. The prediction accuracy and generalization capacity of the trained FNN wall model are examined by comparing the predicted wall shear stresses with the WRLES data. For the instantaneous wall shear stress, the FNN predictions show an overall good agreement with the WRLES data with some discrepancies observed at locations near the crest of the hill. The correlation coefficients between the FNN predictions and WRLES predictions are larger than 0.7 at most streamwise locations. For the mean wall shear stress, the FNN predictions agree very well with WRLES data. More importantly, overall good performance of the FNN wall model is observed for different Reynolds numbers, demonstrating its good generalization capacity.
\end{abstract}

\maketitle


\section{\label{sec:Introduction}Introduction}
Separation and reattachment in turbulent flows over curved surfaces exist in numerous environmental and industrial processes, e.g. underwater vehicle, fuselages at high incidence, curved ducts, and stalled wings and turbine blades. Such flows are difficult to predict accurately using Reynolds-Averaged Navier-Stokes (RANS) method due to non-equilibrium spatial and temporal fluctuations, although it is widely used in engineering applications. On the other hand, large-eddy simulation, which directly solve energetic turbulence scales and model the subgrid scales, being significantly less computational expensive than direct numerical simulation (DNS) \cite{Moin_Mahesh_ARFM_1998, Choi_Moin_PoF_2012, He_Jin_Yang_ARFM_2017}, provides a feasible way for simulating complex turbulent flows with separation and reattachment at a reasonable cost. However, it is still not applicable to employ wall-resolved large-eddy simulation (WRLES) in the design and optimization of high Reynolds number turbulent flow problems in real life because of the extremely high resolution needed to resolve the viscous scale near the wall \cite{Piomelli_Balaras_ARFM_2002, Choi_Moin_PoF_2012}. To reduce the computational cost of WRLES, wall models are employed in the literature \cite{Cabot_Moin_FTC_2000, Davidson_IJHF_2009} to avoid the need to resolve the small scale turbulence in the near wall region, providing a feasible way for LES of wall-bounded flows at high Reynolds number. However, most existing wall models \cite{Piomelli_PAS_2008, Bose_Park_ARFM_2018} based on equilibrium hypothesis are incapable of predicting flow separations and reattachments. The development of machine learning methods \cite{Duraisamy_etal_ARFM_2019, Brunton_etal_ARFM_2019} and the availability of high-resolution data from experiments and high-fidelity simulations provide another possible approach for developing advanced wall models for complex turbulent flows. As the first step, in this work we develop the wall models based on neural networks for turbulent flows over periodic hills.

We first briefly review different wall models developed in the literature. In wall-modeled LES, the turbulent flow near the wall is described by a wall-layer model with its influence on the outer flow represented by appropriate boundary conditions. The wall-stress model is the most widely used one in the literature, in which the wall shear stress is computed and provided as boundary conditions for outer flow simulations. Different models have been developed in the literature for computing wall shear stress, which include the equilibrium-stress model and zonal model (also dubbed as two-layer model) \cite{Bose_Park_ARFM_2018}. The algebraic equilibrium-stress models assume a constant-stress layer near the wall \cite{Pope_Cambridge_2000} and calculate the wall shear stress using the law of the wall of deterministic form \cite{Piomelli_PAS_2008}. The algebraic model has the advantage of low computational cost, but it cannot accurately predict the wall shear stress in non-equilibrium flows, for which the equilibrium-stress hypothesis is no longer valid. The zonal model, on the other hand, solves the thin-boundary-layer equation (TBLE) on an embedded grid between the first grid point and the wall. Wang and Moin \cite{Wang_Moin_PoF_2002} systematically studied the efficacy of zonal models and found that the instantaneous wall shear stress cannot be accurately predicted when the non-equilibrium terms are ignored or the pressure gradient term is only considered. Later, the dynamic zonal models were proposed, which adjust mixing-length eddy viscosity in TBLE, and shown being able to predict low-order turbulence statistics \cite{Kawai_Larsson_PoF_2013, Park_Moin_PoF_2014}. The integral wall model was also developed in the literature \cite{Yang_etal_PoF_2015}, which introduces an additional linear term into the equilibrium logarithmic velocity profile and accounts for near-wall non-equilibrium effects by solving the vertically integrated momentum equation. However, this model has only been tested in applications in which the non-equilibrium effects is insignificant. Besides the wall-stress type models, the virtual-wall model was also developed by aligning the slip velocity in the integrated TBLE on the virtual wall \cite{Chung_Pullin_JFM_2009, Cheng_etal_JFM_2015}. It has been demonstrated being capable of capturing the quantitative features of a separation-reattachment turbulent boundary-layer flow at low to moderately large Reynolds numbers. However, the identification of virtual wall in a virtual-wall model is challenging for flows with complex geometries. Recently, the dynamic slip wall model was developed \cite{Bose_Moin_PoF_2014, Bae_etal_PRF_2018} to model the wall shear stress from the derivation of the LES equations using a differential filter, but its accuracy is sensitive to the subgrid-scale (SGS) models and numerical methods. The conventional wall models have been applied to different kinds of flows \cite{Larsson_etal_MER_2016, Frere_etal_FTC_2018, Bose_Park_ARFM_2018, Bae_etal_JFM_2019}, but they still cannot accurately predict the flow separation and reattachment. Advanced wall model accounting for such non-equilibrium effects is still yet to develop.

Thanks to the exponetial growth in computing power, the increasing amount of high-fidelity data provides a possibility to develop data-driven wall models to resolve the above issues. The data-based approches, particularly the machine learning (ML) method, have been applied to various turbulence problems, e.g. the development of turbulence models \cite{Ling_etal_JFM_2016, Zhou_etal_CAF_2019}, temporal prediction of turbulence \cite{King_etal_arXiv_2018, Lee_You_JFM_2019}, and reconstruction of the turbulent flow fields \cite{Fukami_Fukagata_Taira_JFM_2019, Liu_etal_PoF_2020}. For the applications of the ML method in developing RANS models, Ling et al. \cite{Ling_etal_JFM_2016} presented a deep neural network for RANS turbulence modeling on an invariant tensor basis \cite{Ling_etal_JCP_2016}. Xiao and co-workers \cite{Wang_Wu_Xiao_PRF_2017, Wu_Xiao_Paterson_2018} developed a physics-informed ML framework to learn Reynolds stress discrepancies between RANS and DNS. Duraisamy et al. \cite{Duraisamy_etal_ARFM_2019} reviewed in detail the recent developments of RANS turbulence models based on ML. As for LES models, the ML has been applied to model the SGS stresses in different flows including the turbulent channel flows \cite{Gamahara_Hattori_PRF_2017}, two-dimensional decaying turbulence \cite{Maulik_etal_PoF_2018} and isotropic turbulence \cite{Zhou_etal_CAF_2019}, and to model the SGS scalar flux \cite{Vollant_etal_JOT_2017}. For the wall-bounded turbulent flows, the ML was also employed to develop wall models \cite{Milano_Koumoutsakos_JCP_2002}, which is the major concern of this work. Recently, Yang et al. \cite{Yang_etal_PRF_2019} developed a wall-stress model for LES of turbulent channel flow using DNS data and physics-informed neural networks. They found that the trained wall model outperforms the conventional equilibrium wall model in simulating the three-dimensional boundary-layer flow. A simliar neural network was then applied to spanwise rotating turbulent channel flows with a discussion on the performance of physics-based and data-based approaches \cite{Huang_etal_PoF_2019}. However, to the best of our knowledge, the non-equilibrium effects, e.g. pressure gradients, curvature and separation, which are important for complex turbulent flows in engineering applications, have not been taken into account in the existing data-driven wall model and need to be systematically investigated.

Characteristics of complex wall-bounded turbulent flows depends on the geometry of the boundary and the corresponding boundary conditions. Development of a wall model applicable to different turbulent flows does not seem feasible or at least beyond the scope of this work. The flow over periodic hills, in which the flow is featured by separation from a curved surface, recirculation, reattachment and strong pressure gradient, is an ideal generic test case for developing statistical closures for separated flow \cite{Frohlich_etal_JFM_2005}. Different wall models have been applied to simulate the flow over periodic hills. For instance, Temmerman et al. \cite{Temmerman_etal_IJHF_2003} applied the equilibrium wall models to simulate the flow over periodic hills, and found that the sensitivity of the solution to the SGS model is lower than that to grid resolution and wall model. Furthermore, they demonstrated that the WMLES cannot accurately predict the flow separation, reattachment and related statistics. To simulate the flow over periodic hills, Manhart et al. \cite{Manhart_etal_TCFD_2008} proposed an extended inner scaling for the wall layer of wall-bounded flows under the influence of both wall shear stress and adverse pressure gradient. Duprat et al. \cite{Duprat_etal_PoF_2011} constructed a new wall model based on the simplified TBLE, which takes into account both the streamwise pressure gradient and the Reynolds stresses effects, and applied it to simulate the flow over periodic hills. It was shown that their proposed model yields good results for predictions of first order statistics and reproduction of flow separation. To investigate in detail the separation and reattachment process, Breuer et al. \cite{Breuer_etal_CAF_2009} carried out numerical and wind tunnel experiment of the flow over periodic hills at various Reynolds numbers in the range of 100 to 10595. Rapp and Manhart \cite{Rapp_Manhart_EF_2011} experimentally investigated the flow over periodic hills at four Reynolds numbers ranging from 5600 to 37000. Krank et al. \cite{Krank_etal_FTC_2018} carried out DNS of the flow over periodic hills at Reynolds number 10595, which is the highest fidelity to date. Moreover, Xiao et al. \cite{Xiao_etal_CAF_2020} constructed benchmark datasets for the flow over periodic hills by performing DNS with varing flow configurations to alleviate the lack of data for training and testing data-driven models.

The objective of this work is to develop a data-driven wall model for the flow over periodic hills using the feedforward neural network (FNN) and WRLES data. The datasets employed for training the model consist of flow field data in the near-wall flow region at all streamwise locations with different flow features. To train the FNN wall model, we employ the wall-normal distance, near-wall velocities and pressure gradients as input features and the wall shear stresses as output labels, respectively. The trained wall model is evaluated using the data from different snapshots and spanwise slices for both training and testing datasets.

The rest of the paper is organized as follows: In Section \ref{sec:Numerical_data}, WRLES of the flow over periodic hills is briefly described, which is followed by the procedure for preparing datasets for training and testing FNN models. The feedforward neural network is introduced and trained in Section \ref{sec:FNN}. Then, the evaluation of the trained wall model is presented in Section \ref{sec:Results_and_discussion}. At last, conclusions from this work are drawn in Section \ref{sec:Conclusion}.

\section{\label{sec:Numerical_data}Data generation and preparation}

\subsection{\label{sec:2.1}Data generation using wall-resolved large-eddy simulation}
\begin{figure}[!ht]
\centering{\includegraphics[width=0.6\textwidth]{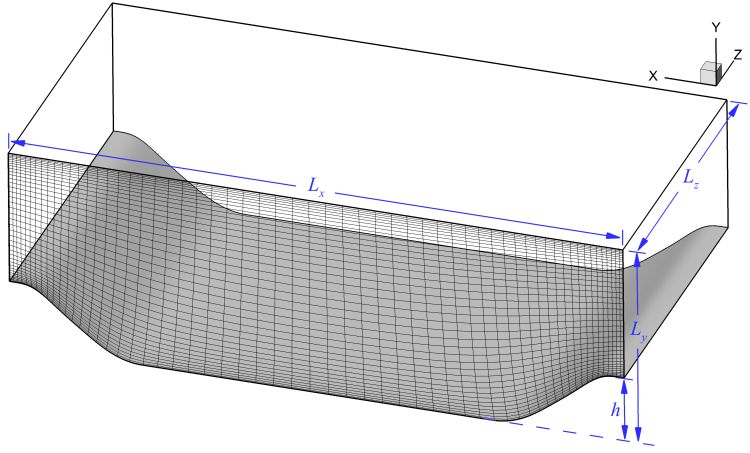}}
  \caption{Schematic of the geometry for the periodic hills, computational domain ($L_x=9.0h$, $L_y=3.036h$, $L_z=4.5h$), and the employed curvilinear mesh on a $x-y$ plane (on which every fifth grid line is displayed).}
\label{fig:fig1}
\end{figure}

In this section, we describe the numerical method and the case setup for generating the data employed for developing a data-driven wall model, which can take into account the non-equilibrium effects, e.g. flow separation and reattachment, for turbulent flows over periodic hills.

We employ the Virtual Flow Simulator (VFS-Wind) \cite{Yang_etal_WE_2015, Yang_Sotiropoulos_WE_2018} code for WRLES of turbulent flows over periodic hills. The VFS code has been successfully applied to industrial and environmental turbulent flows \cite{Kang_Sotiropoulos_AWR_2012, Kang_etal_JFM_2014, Khosronejad_Sotiropoulos_JFM_2014, Yang_etal_RE_2017, Yang_Sotiropoulos_BE_2019, Yang_Sotiropoulos_PRF_2019, Foti_etal_JFM_2019, Khosronejad_etal_JHE_2020}. In VFS-Wind code, the governing equations are the three-dimensional unsteady spatially filtered incompressible Navier-Stokes equations in non-orthogonal, generalized curvilinear coordinates shown as follows:
\begin{equation}
  \begin{aligned}
    J \frac{ \partial U^{j} }{\partial \xi^{j}} &= 0, \\
    \frac{1}{J} \frac{\partial U^{i}}{\partial t} &= \frac{\xi^{i}_{l}}{J} \left( -\frac{\partial}{\partial \xi^{j}} (U^{j}u_{l}) - \frac{1}{\rho} \frac{\partial}{\partial \xi^{j}} ( \frac{\xi^{j}_{l} p}{J} ) + \frac{\mu}{\rho} \frac{\partial}{\partial \xi^{j}}( \frac{g^{jk}}{J} \frac{\partial u_{l}}{\partial \xi^{k}} ) - \frac{1}{\rho} \frac{\partial \tau_{lj}}{\partial \xi^{j}} + f_{l} \right),
  \end{aligned}
\label{eq_1}
\end{equation}
where $x_{i}$ and $\xi^{i}$ are the Cartesian and curvilinear coordinates, respectively, $\xi^{i}_{l} = \partial \xi^{i} / \partial x_{l}$ are the transformation metrics, $J$ is the Jacobian of the geometric transformation, $u_{i}$ is the $i$-th component of the velocity vector in Cartesian coordinates, $U^{i} = (\xi^{i}_{m} / J) u_{m}$ is the contravariant volume flux, $g^{jk} = \xi^{j}_{l} \xi^{k}_{l}$ are the components of the contravariant metric tensor, $\rho$ is the fluid density, $\mu$ is the dynamic viscosity, and $p$ is the pressure. In the momentum equation, $\tau_{ij}$ represents the anisotropic part of the subgrid-scale stress tensor, which is modeled by the dynamic eddy viscosity subgrid-scale model,
\begin{equation}
\tau_{ij} - \frac{1}{3} \tau_{kk} \delta_{ij} = - 2 \nu_{t} \overline{S}_{ij},
\label{eq_2}
\end{equation}
where $\overline{S}_{ij}$ is the filtered strain-rate tensor and $\nu_t$ is the eddy viscosity calculated by
\begin{equation}
\nu_{t} = C \Delta^{2} |\overline{S}|,
\label{eq_3}
\end{equation}
where $C$ is the model coefficient calculated dynamically using the procedure of Germano et al. \cite{Germano_etal_PoF_1991}, $|\overline{S}| = \sqrt{2\overline{S}_{ij}\overline{S}_{ij}}$ and $\Delta = J^{-1/3}$ is the filter size, where $J^{-1}$ is the cell volume.

The governing equations are spatially discretized using a second-order accurate central differencing scheme, and integrated in time using the fractional step method. An algebraic multigrid acceleration along with generalized minimal residual method (GMRES) solver is used to solve the pressure Poisson equation. A matrix-free Newton-Krylov method is used for solving the discretized momentum equation. More details about the flow solver can be found in the references\cite{Ge_Sotiropoulos_JCP_2007, Kang_etal_AWR_2011, Yang_etal_WE_2015}.

The geometry of the periodic hills is shown in Figure \ref{fig:fig1}, which has been extensively employed in experiments \cite{Breuer_etal_CAF_2009, Rapp_Manhart_EF_2011} and numerical simulations \cite{Temmerman_etal_IJHF_2003, Frohlich_etal_JFM_2005, Krank_etal_FTC_2018}. As seen, the height of the hill is $h$, with a flat wall placed $2.036h$ above the crest of the hill and the distance between the crests of two hills is $L_x = 9h$. In the spanwise direction, the size of the computational domain is $L_z=4.5h$. The Reynolds number based on the bulk velocity $U_b$, which is defined as $U_b = Q/(\rho L_z \times (L_y-h))$ (where $Q$ is the mass flux), and the height of the hill is $Re_h = \rho U_b h/\mu$. No slip boundary condition is applied at the top wall and the surface of the hills. In the streamwise and spanwise directions, periodic boundary condition is applied. The flow is driven by a pressure gradient uniformly applied to whole domain to maintain a constant mass flux.

\newcommand{\tabincell}[2]{\begin{tabular}{@{}#1@{}}#2\end{tabular}}
\begin{table}[!ht]
\centering
\caption{\label{tab:table1}Parameters of the WRLES of flow over periodic hills.}
\begin{tabular}{p{1.6cm}<{\centering}p{1.6cm}<{\centering}p{4.5cm}<{\centering}p{2.2cm}<{\centering}p{3.2cm}<{\centering}}
  \hline
  Case   &  $Re_h$    &  Mesh ($N_x \times N_y \times N_z$)  &  $\Delta t$ ($\times 10^{-2}$)  &  $\Delta y_{cc}/h$ ($\times 10^{-3}$) \\
  \hline
  1      &  5600      &  $297 \times 193 \times 187$         &  1.0                            &  1.5                                  \\
  2      &  10595     &  $297 \times 193 \times 187$         &  1.0                            &  1.5                                  \\
  3      &  19000     &  $460 \times 300 \times 290$         &  0.5                            &  0.75                                 \\
  \hline
\end{tabular}
\end{table}

In this work, the WRLES of flow over periodic hills at three Reynolds numbers are carried out for the training and testing of data-driven wall model, as shown in Table \ref{tab:table1}. The computational domain is discretized using a body-fitted curvilinear grid (as shown in Figure \ref{fig:fig1}). The height of the first off-wall grid nodes in wall units, $\Delta y^{+} = \Delta y_{cc} u_{\tau}/\nu$, is in the range of 0.056 to 3.95 at $Re_h = 10595$ as shown in Figure \ref{fig:fig_A1} in Appendix \ref{Appendix_A}. Here, $\Delta y_{cc}$ is half height of the first off-wall grid, $u_{\tau} = \sqrt{\tau_w/\rho}$ denotes the friction velocity and $\nu = \mu/\rho$.

The size of time step is $\Delta t = 0.01 h/U_b$. The simulation is first carried out for about $22T$ (flow-through time $T = L_x/U_b$) for the flow to achieve a fully developed state. Then the flow is further simulated for about $50T$ for time-averaged quantities and flowfield data on slices for training the data-driven wall model.

To validate the employed numerical method and case setup, we compare the profiles of the mean velocity, Reynolds shear stress, turbulence kinetic energy, and the skin friction and pressure coefficients computed in this work with the results from measurements \cite{Temmerman_etal_IJHF_2003} and DNS by Krank et al. \cite{Krank_etal_FTC_2018} and demonstrate an overall good agreement as shown in Appendix \ref{Appendix_A} for validating the employed numerical method and case setup.

\subsection{\label{sec:2.2}Data preparation}
The WRLES data are further processed to prepare the data for training the data-driven wall model. In WMLES, the wall shear stress is often computed using the velocity at the first off-wall grid node. If the data-driven model is developed using the velocity at a specific location, it may only be applicable to grids of grid spacings. Moreover, if the wall model is developed using the data at a certain streamwise location, e.g. the location where the flow is attached or the location where flow separation occurs, it may only be valid for a certain flow condition. To avoid these two issues, we are devoted to develop a data-driven wall model applicable to different spatial resolutions and not limited to certain flow conditions using the data in the near-wall region of the periodic hills at all streamwise locations. Specifically in this work, the flow data in the near-wall region with wall-normal distance in the range of $0.006 \le \eta/h \le 0.1$ are employed, where $\eta$ denotes the wall-normal coordinate. The top boundary of the region at $\eta/h = 0.1$ is determined considering that the flow field above is less correlated with the wall shear stress and is usually well-resolved by WMLES. It is noticed that the region with $\eta/h<0.1$ is defined as the inner layer for a turbulent channel flow, where $h$ is the half width of the channel. The bottom boundary at $\eta/h = 0.006$ is defined to preclude the effects of viscous sublayer and with the consideration that no wall model is needed if the viscous scale is resolved.

\begin{figure}[!ht]
\centering{\includegraphics[width=1.0\textwidth]{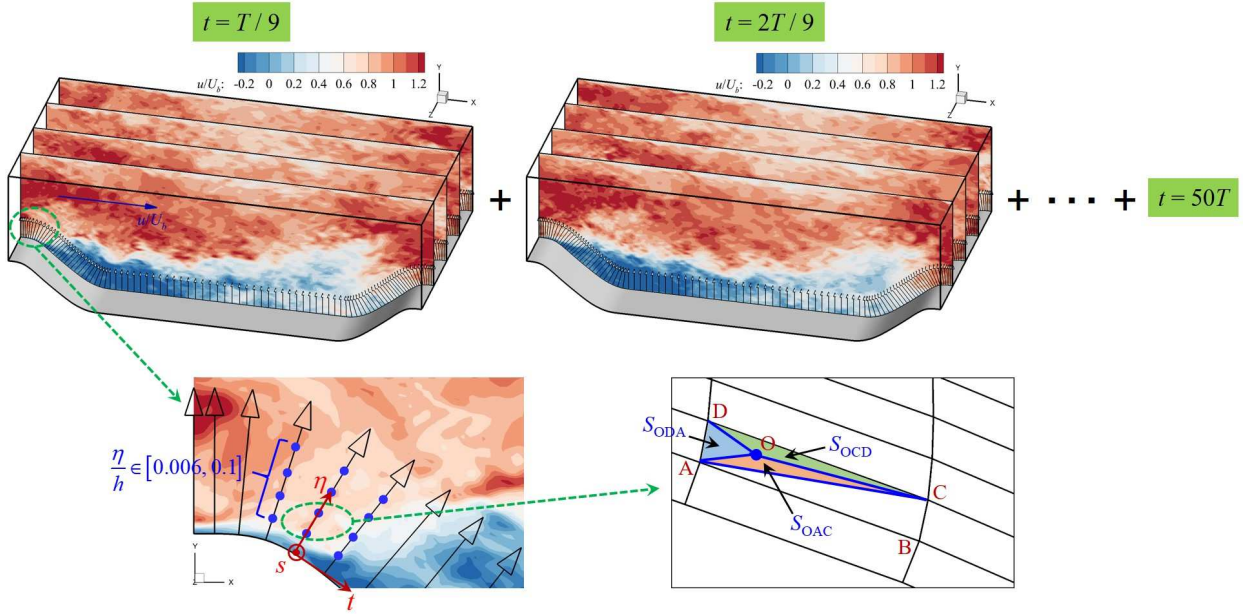}}
  \caption{Schematic diagram of the data preparation from the WRLES data for training the data-driven wall model. To prepare the data, 450 snapshots covering $50$ flow-through times are extracted on four $x-y$ slices located at $z/h=0.0$, $1.125$, $2.25$ and $3.375$. At each snapshot, the flow field data at 95 locations along the wall-normal direction in the region $0.006 \leq \eta/h \leq 0.1$ are extracted at ${N_x}-1$ streamwise locations using the triangulation with linear interpolation approach.}
\label{fig:fig2}
\end{figure}

A step-by-step diagram for preparing the training data is shown in Figure \ref{fig:fig2}. Saving the three-dimensional flow fields at every time step, which requires a significant amount of disk space, is not feasible. Instead, we save the WRLES data at four spanwise ($x-y$) slices located at $z/h=0.0$, $1.125$, $2.25$ and $3.375$, respectively. To make the most of the WRLES data and meanwhile keep the cost for training the model at a reasonable level (in other words, avoid using the flowfields close in time, which can be very similar), 9 snapshots of the instantaneous flowfields are extracted on each slice for one flow-through time. And in total we obatain 450 snapshots for the whole $50$ flow-through times at each Reynolds number. For each snapshot, the flow field data at different wall-normal locations are further extracted at ${N_x}-1$ streamwise locations due to periodic boundary conditon to include different flow features along the lower wall of periodic hills. At each streamwise location of the lower wall, the flowfield data at 95 nodes, which are uniformly distributed in $\eta/h \in [0.006, 0.1]$, are then extracted to form 95 pairs of input-output data along the wall-normal direction. Finally, the flowfield data along the wall-normal direction at different streamwise locations for all considered spanwise slices and snapshots form the complete training and testing datasets, which contain approximately $1.8 \times 10^8$ input-output pairs. It is noticed that the grid nodes in the wall-normal ($\eta$) direction in general do not coincide with the curvilinear grid nodes employed for solving the flow. The linear interpolation approach based on triangulation is employed to obtain the flowfield data at the 95 points along the wall-normal coordinate.

Input features and output labels are critical for the success of data-driven models. The wall shear stress including the streamwise and spanwise stresses $\tau_{w,t}$, $\tau_{w,s}$, which are often applied as boundary conditions for outer flow simulations in WMLES, are employed as the output labels. To prepare the dataset for model training, the wall shear stresses are directly calculated using the WRLES data. As for the input features, the wall-normal distance $\eta$, the three velocity components $u_{w,t}$, $u_{w,n}$ and $u_s$ in the wall-tangential, -normal and spanwise directions and the pressure gradients $\frac{\partial p}{\partial w_t}$, $\frac{\partial p}{\partial w_n}$ in the wall-tangential and -normal directions are employed. It has been shown by Duprat et al. \cite{Duprat_etal_PoF_2011} that using a near-wall scaling defined with the classic friction velocity and the streamwise pressure gradient can improve the performance of wall models for separated flows. To take into account such knowledge when constructing the neural networks for data-driven wall models, the wall-normal distance normalized using the near-wall scale and written in the logarithmic form, i.e. $\ln \left( \eta/y^{*} \right)$, where $y^{*} = \nu/u_{\tau p}$, $u_{\tau p} = \sqrt{u_{\tau}^2+u_p^2}$, $u_{\tau} = \sqrt{\left| \tau_w \right|/\rho}$, $u_p = \left| (\mu/\rho^2)(\partial P/\partial x) \right|^{1/3}$, is employed as the input feature for training the data-driven wall model. Here the wall shear stress is computed as $\tau_w = \mu u_t/\eta$, which implicitly accounts for no-slip boundary conditions at the wall and approximates the wall shear stress resolved in WMLES. To further improve the generality of the trained model, the pressure gradients are multiplied by $\frac{\eta}{h}$ before taken as input features as suggested by Yang, Bose and Moin \cite{Yang_etal_CTR_2017}.

\section{\label{sec:FNN}Construction of data-driven wall models}

\subsection{\label{sec:3.1}Feedforward neural network}
\begin{figure}[!ht]
\centering{\includegraphics[width=0.5\textwidth]{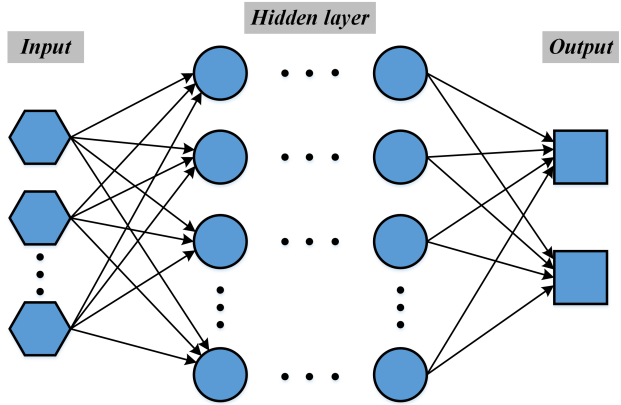}}
  \caption{Schematic diagram of the feedforward neural network (FNN) with multi hidden layers.}
\label{fig:fig3}
\end{figure}

We use a multi-hidden-layer feedforward nueral network (FNN) to establish the relation between the near-wall flow and the wall shear stress on the surface of periodic hills. As shown in Figure \ref{fig:fig3}, the employed FNN consists of an input layer, multi hidden layers and an output layer. Each layer has a number of neurons, which are computational units that propagate the weighted sums of the inputs to an activation function and calculate the output. The detailed procedures for calculating the output based on the input in the FNN is shown in Appendix \ref{Appendix_B}, which includes the linear matrix manipulation of the weight and bias coefficients and the nonlinear mapping using the activation function.

The activation function used in this paper is the rectified linear unit (ReLU) \citep{Goodfellow_etal_2016},
\begin{equation}
f\left( x \right)=\left\{ \begin{matrix}
   0, &  \quad\quad  if\ \ x<0,  \\
   x, &  \quad\quad  if\ \ x\ge 0.  \\
\end{matrix} \right.
\label{eq_4}
\end{equation}

The prepared input and output data are normalized using the Min-Max scaling,
\begin{equation}
x^{\text{*}} = \frac{x - x_{\min}}{x_{\max} - x_{\min}}.
\label{eq_5}
\end{equation}

The loss function is defined as
\begin{equation}
{{E}_{\text{WM}}}=\frac{1}{N_s}\sum\limits_{i=1}^{N_s}{{{\left( {{\mathbf{Y}}_i}-{{\mathbf{Y}}_i}^{\text{*}} \right)}^{2}}}+\frac{\lambda_0}{2{{N}_{s}}}\left\| w \right\|_{2}^{2},
\label{eq_6}
\end{equation}
where $N_s$ is the number of training samples, $w$ is the weight coefficient, and $\lambda_0$ is the regularization rate, which is set to 0.001. The first term in Eq. (\ref{eq_6}) denotes the mean square error (MSE) between the FNN output ${\mathbf{Y}}^{\text{*}}$ and the labeled output $\mathbf{Y}$ from the WRLES. The second term is an L2 regularization term included to avoid overfitting.

We use the error backpropagation (BP) scheme [56] implemented with TensorFlow [57] to train the FNN by optimizing the weight and bias coefficients to minimize the loss function. The key steps for FNN training are as follows:

(1)	Provide training data to the input layer, propagate data signal forward layer by layer, and compute the result in the output layer. Details on this step can be found in Appendix \ref{Appendix_B}.

(2)	Compute the loss function according to Eq. (\ref{eq_6}) using FNN output and the labeled output.

(3)	Adjust the weight and bias coefficients using the gradient descent algorithm,
\begin{equation}
v = v+\Delta v, \quad  \Delta v = -\eta \frac{\partial {{E}_{\text{WM}}}}{\partial v},
\label{eq_7}
\end{equation}
where $v$ denotes the weight and bias coefficients in the FNN, $\eta \in \left( 0,1 \right)$ denotes the learning rate, which is dynamically adjusted using the Adam optimizer \cite{Kingma_Ba_arXiv_2014}.

(4)	Repeat the above steps until the stop criterion is met.

\subsection{\label{sec:3.2}Training of the data-driven wall model}
In this section, three sets of cases, i.e. one with different number of input features, the other one with different number of neurons in each hidden layer, and the third one with different number of output labels, are carried out to evaluate the performance of different setups for training wall model using FNN and the flow data at $Re_h = 5600$ and $10595$. The key requirement for a wall model is to accurately predict the wall shear stress. In conventional wall models, the wall shear stress is determined by an empirical relation (e.g. the power law or the logarithmic law) or simplified Navier-Stokes equations (e.g. the thin boundary layer equation) using the velocity at one off-wall grid node (usually the first or the second off-wall node). To compensate the lack of governing equations in data-driven wall models, flowfield data at more than one off-wall grid nodes are probably needed. In the first set of cases, we test the effects of input features obtained using different numbers of wall-normal points (five inputs per point) with the distance between two adjacent points $0.03h$. How well a data-driven model represents near-wall turbulence with flow separations and reattachments depends on the complexity of the employed neural network, i.e. the number of hidden layers and the number of neurons in each layer. In the second set of cases, we examine the effects of the number of neurons ranging from 3 to 100 on the training performance with fixed number of hidden layers and input features. Details on these two sets of cases can be found in Table \ref{tab:table2}.

\begin{table}[!ht]
\centering
\caption{\label{tab:table2}Two sets of cases for examining the effects of different numbers of input features (case set 1) and different number of neurons (case set 2) on training performance. In this table, ``N\#'' denotes the number of neurons and ``H\#'' denotes the number of hidden layers. For both sets of cases, we only consider one output, i.e. the wall-tangential shear stress.}
\begin{tabular}{p{2.2cm}<{\centering}p{7.0cm}<{\centering}p{6.4cm}<{\centering}}
  \hline
              &  \# of inputs (\# per point $\times$ \# of points)  &  \# of neurons (\# of hidden layer)  \\
  \hline
  Case set 1  &  $5 \times$ (1, 2, 3, 4, 5, 6)                      &  N20 (H6)                            \\
  Case set 2  &  $5 \times 3$                                       &  N3, N5, N10, N20, N50, N100 (H6)    \\
  \hline
\end{tabular}
\end{table}

\begin{figure}[!ht]
\centering{\includegraphics[width=0.48\textwidth]{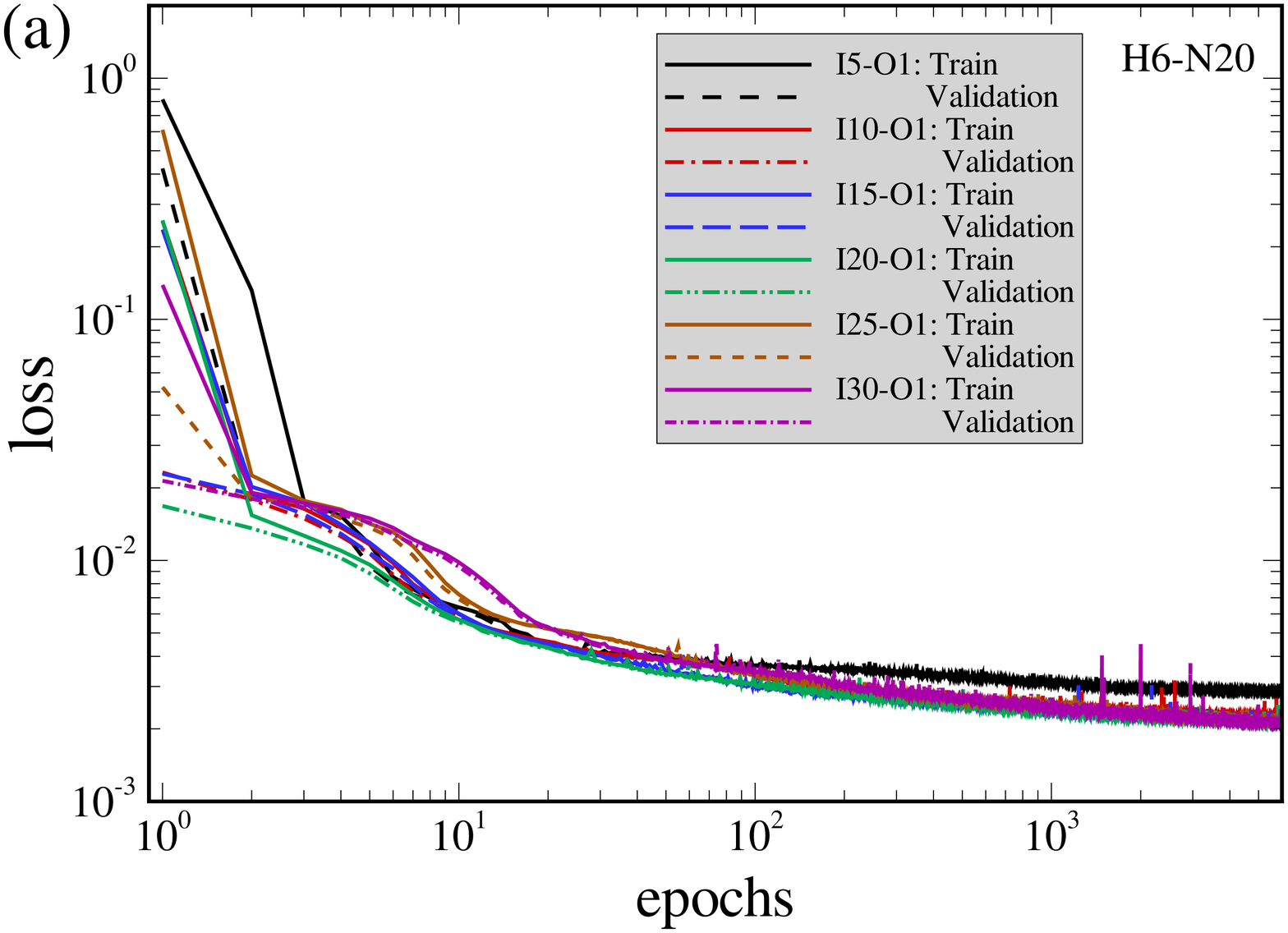}} \quad
\centering{\includegraphics[width=0.48\textwidth]{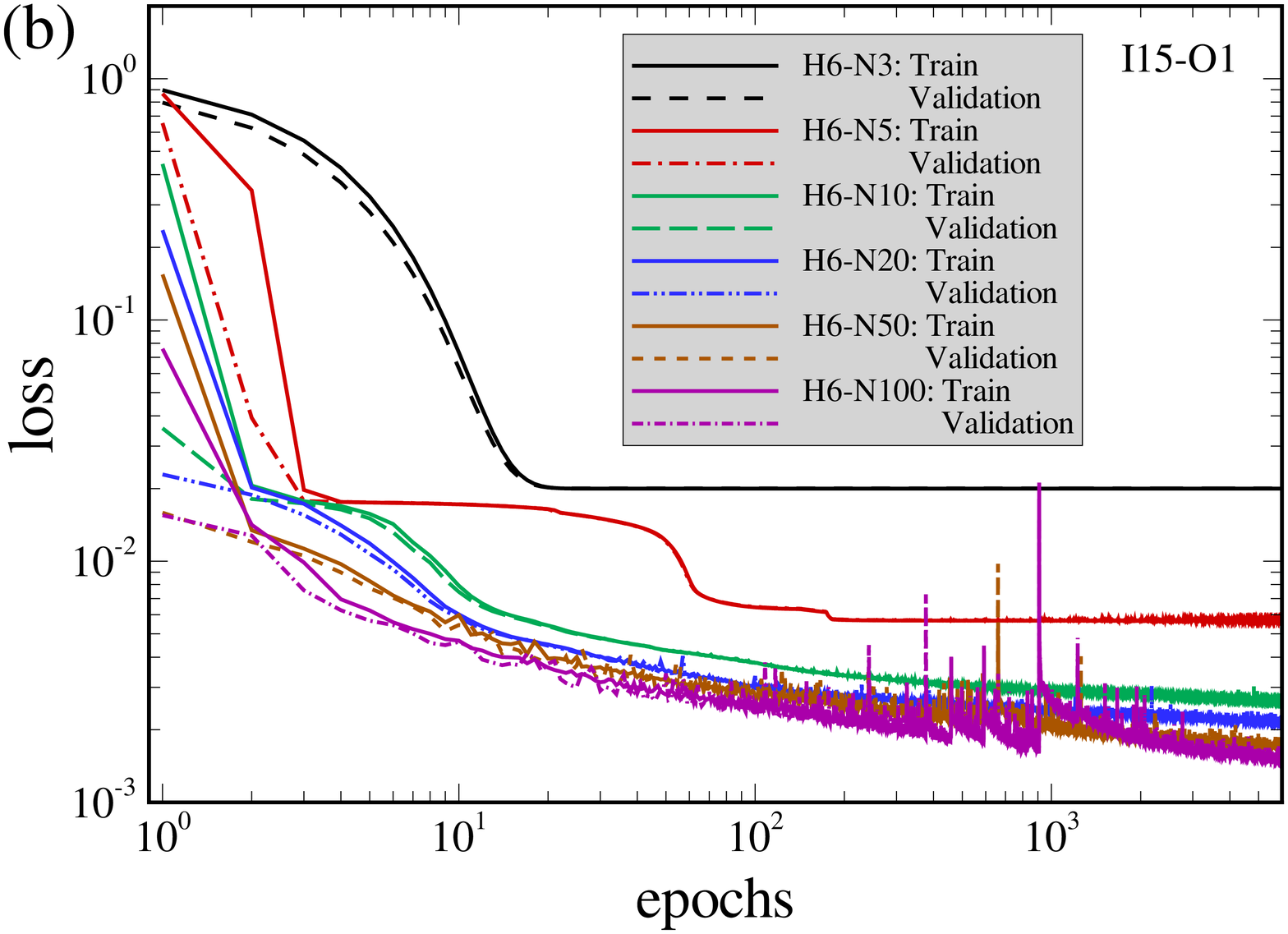}}
  \caption{The variations of loss with training epochs for both the training and validation datasets for (a) results from case set 1 with different numbers of input features and (b) results from case set 2 with different numbers of neurons. In this figure, the number after letter ``I'' , ``O'', ``H'' and ``N'' denotes the number of input features, output labels, hidden layers and neurons, respectively.}
\label{fig:fig4}
\end{figure}

Figure \ref{fig:fig4} plots the variations of loss with the training epochs for both training and validation datasets for the two sets of cases. The number of training samples is $1.24 \times 10^7$, of which $90\%$ are used as the training dataset and the rest $10\%$ are used as validation dataset, and the batch size is $2 \times 10^5$. Intially, the loss is large because the weight coefficients are randomly set and the bias coefficients are set to zero. Then, the weight and bias coefficients are adjusted and the loss rapidly decreases during the first few epochs. After the initial stage, the loss tends to approch a steady small value after approximately 1000 training epochs.

In Figure \ref{fig:fig4}(a), the loss in the FNN model I5-O1 (only using input features at one wall-normal point) is significantly worse than the other models, that the loss at 6000 epochs is about 1.5 times larger than others, while the values of loss from the models using input features at more than 1 points are similar with each other. This indicates that only using the input features at one point is not surfficient to accurately model the complex near-wall turbulence encountered in this periodic hill case, while adding just one point can significantly improve the training performance. To ensure the training performance without increasing the computational cost in the meantime, we choose 15 input features from 3 wall-normal points for case set 2 and other cases in this work. Figure \ref{fig:fig4}(b) compares the loss of FNN models with different number of neurons. If no overfitting occurs, the more neurons are employed, the smaller the loss can be obtained. In this work we use 20 neurons for the proposed FNN wall model.

\begin{table}[!ht]
\centering
\caption{\label{tab:table3}The training of two FNN models with different inputs and outputs.}
\begin{tabular}{p{2.0cm}<{\centering}p{8.8cm}<{\centering}p{2.6cm}<{\centering}p{1.8cm}<{\centering}}
  \hline
  FNN     &  Input    &   Output   &   HL size     \\
  \hline
  FNN-1   &  ($\ln \left( \eta/y^{*} \right)$, $\frac{u_{w,t}}{h_{wm}}$, $\frac{u_{w,n}}{h_{wm}}$, $\frac{\partial p}{\partial w_t} \cdot \frac{h_{wm}}{h}$, $\frac{\partial p}{\partial w_n} \cdot \frac{h_{wm}}{h}$) $\times$ 3                    &  $\tau_{w, t}$                   &  \multirow{2}*{H6-N20}  \\
  FNN-2   &  ($\ln \left( \eta/y^{*} \right)$, $\frac{u_{w,t}}{h_{wm}}$, $\frac{u_{w,n}}{h_{wm}}$, $\frac{u_s}{h_{wm}}$, $\frac{\partial p}{\partial w_t} \cdot \frac{h_{wm}}{h}$, $\frac{\partial p}{\partial w_n} \cdot \frac{h_{wm}}{h}$) $\times$ 3  &  $\tau_{w, t}$, $\tau_{w, s}$    &  ~                      \\
  \hline
\end{tabular}
\end{table}

\begin{figure}[!ht]
\centering{\includegraphics[width=0.48\textwidth]{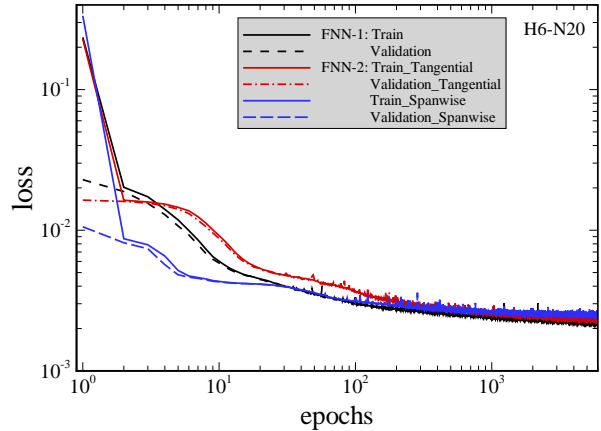}}
  \caption{The variations of loss with training epochs for both the training and validation datasets for FNN-1 and FNN-2.}
\label{fig:fig5}
\end{figure}

To consider the influence of output labels on the FNN training, the model `FNN-1' with only one output label ($\tau_{w, t}$) and `FNN-2' with two output labels ($\tau_{w, t}$, $\tau_{w, s}$) are trained, validated and tested, respectively. The settings of the two FNN wall models are shown in Table \ref{tab:table3}. Figure \ref{fig:fig5} shows the variations of loss with training epochs for FNN-1 and FNN-2. It can be observed that including the spanwise wall shear stress as the output label has some effects on the model training process for epochs less than 100 but little influence on the final loss of tangential wall shear stress.

\section{\label{sec:Results_and_discussion}Evaluation of data-driven wall models}
\subsection{\label{sec:4.1}Accuracy test}

To evaluate the prediction accuracy and generalization capacity of the trained FNN wall model, we apply it to predict the wall shear stress for different snapshots and spanwise slices for both training datasets and testing datasets obtained from the cases with $Re_h = 5600$ and $10595$.

\begin{figure}[!ht]
\centering{\includegraphics[width=0.7\textwidth]{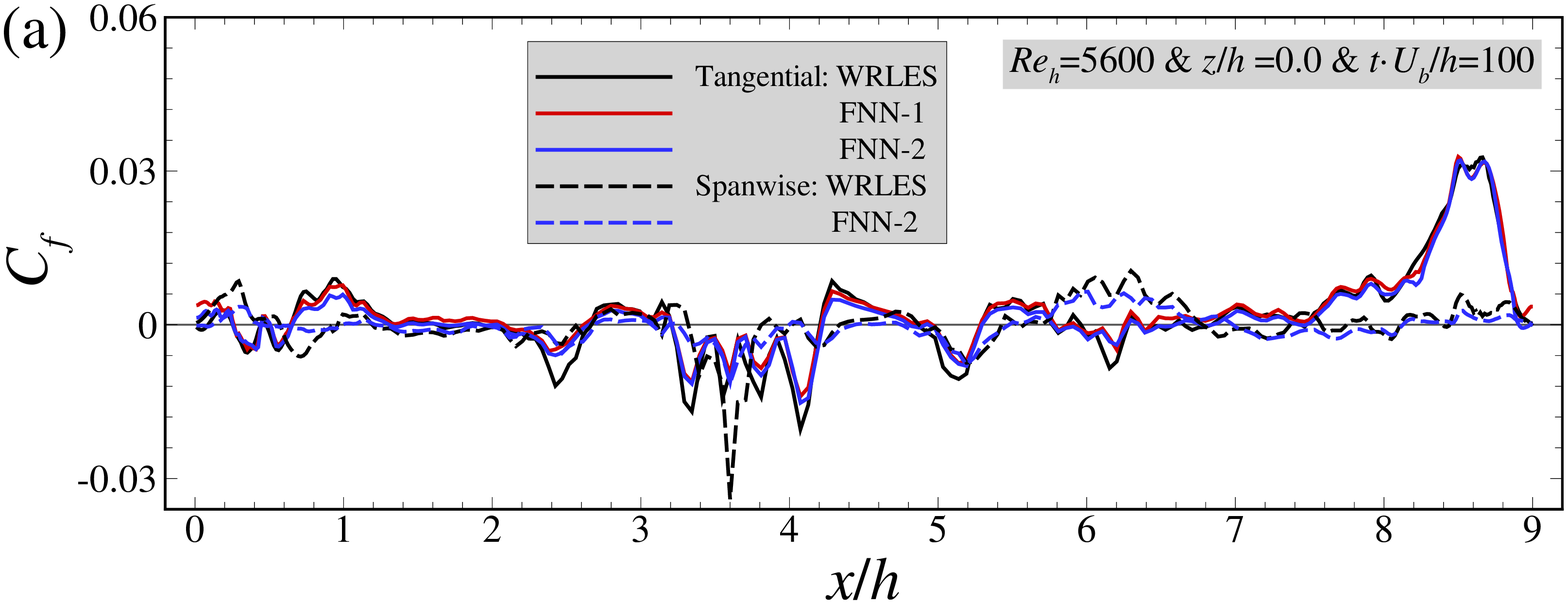}}
\centering{\includegraphics[width=0.7\textwidth]{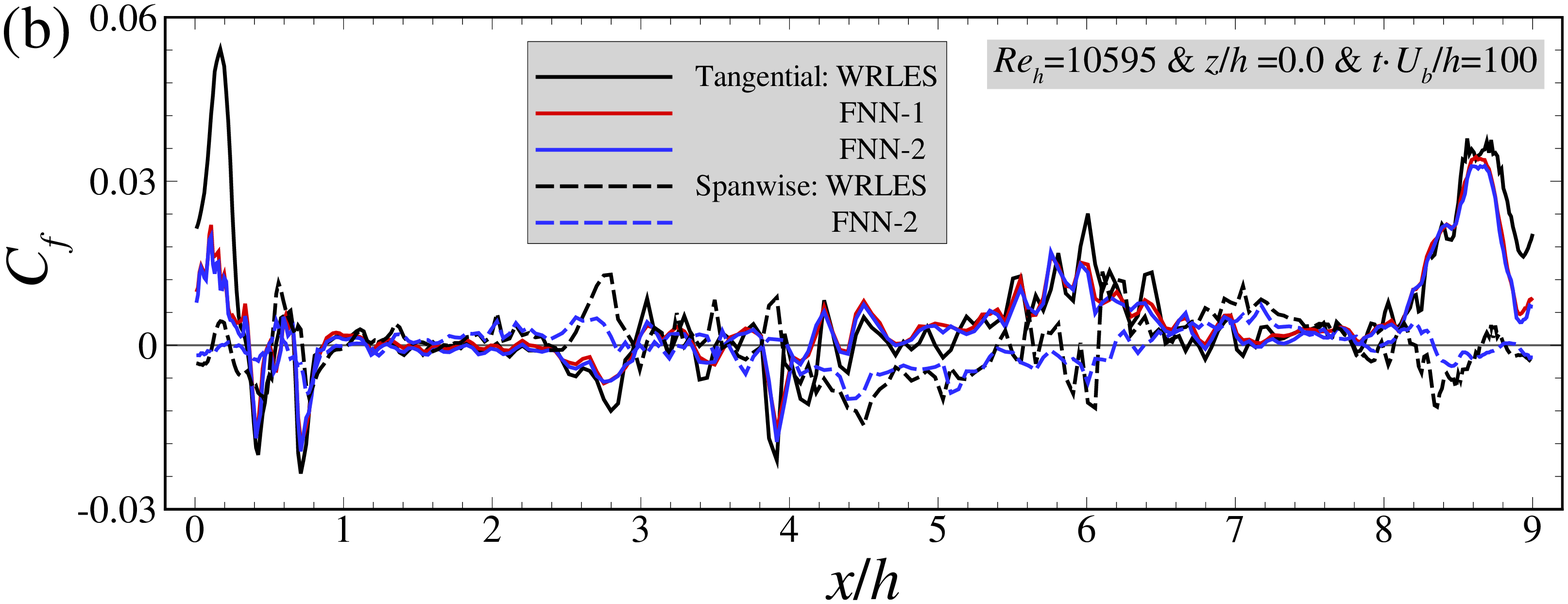}}
\centering{\includegraphics[width=0.7\textwidth]{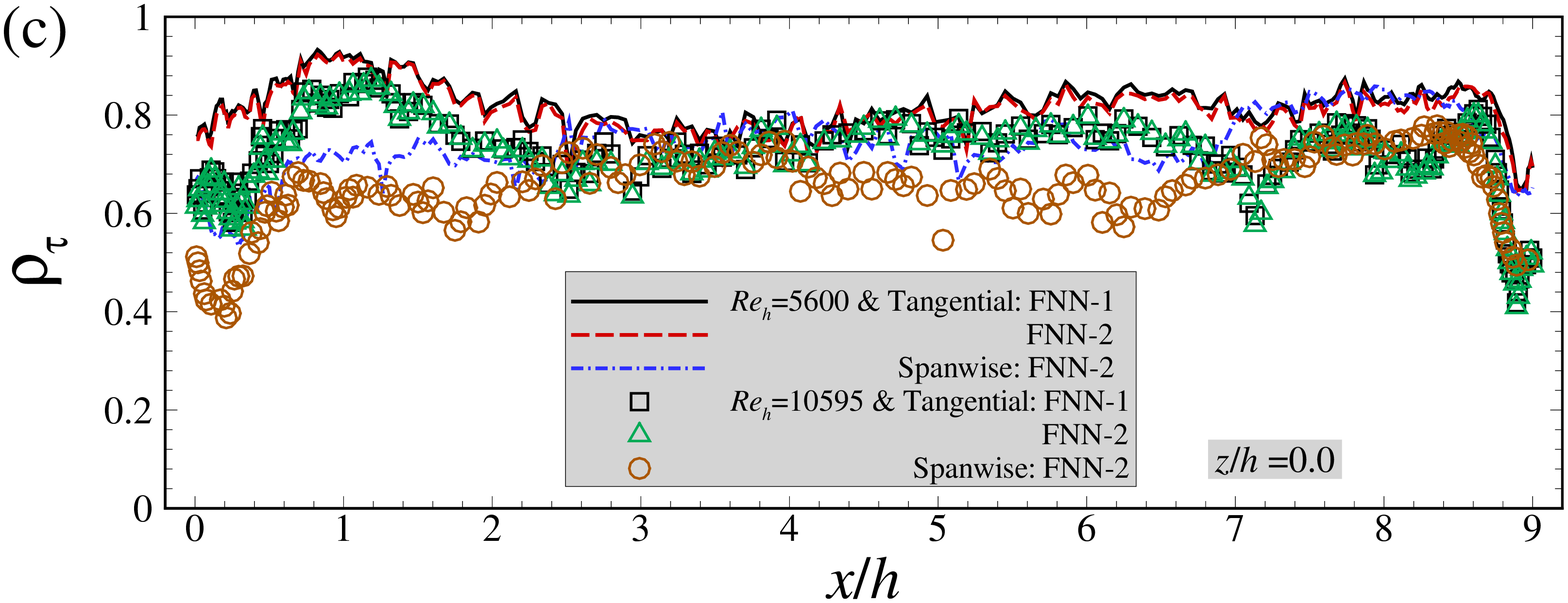}}
\centering{\includegraphics[width=0.7\textwidth]{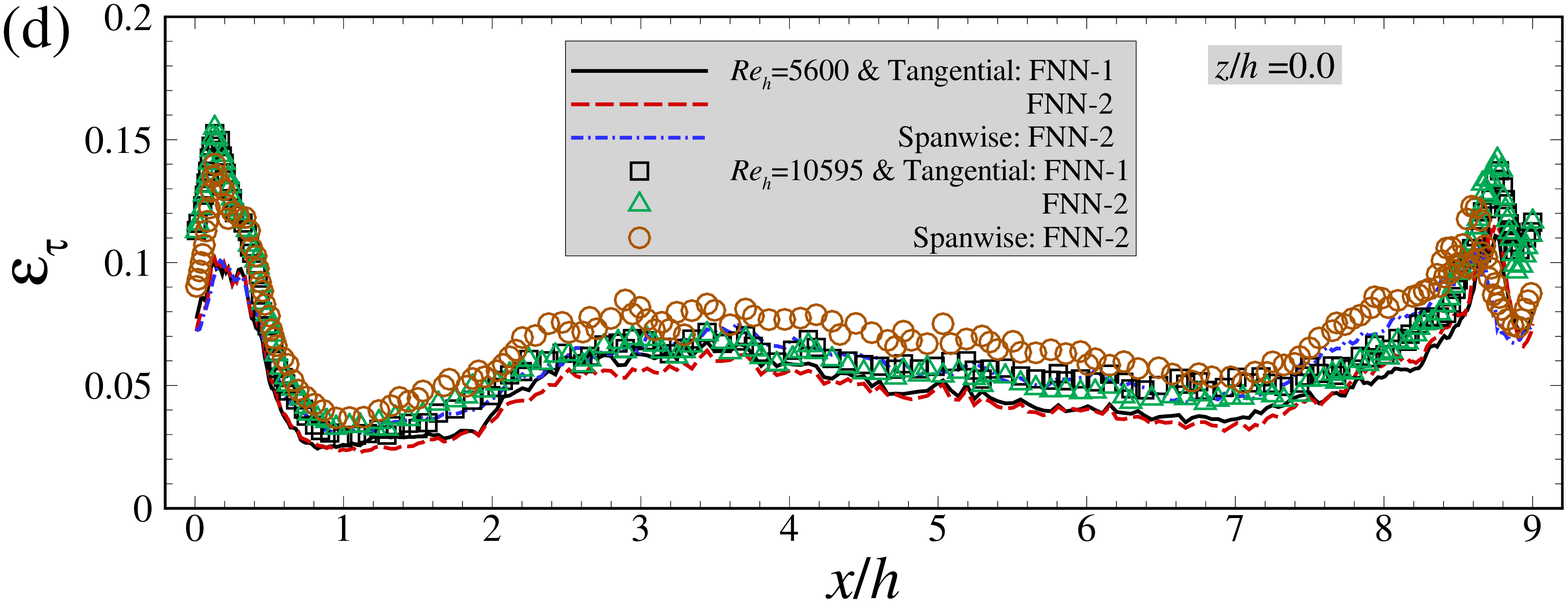}}
  \caption{Evaluation of the FNN wall model using the training dataset for: (a)(b) Comparison of instantaneous skin friction coefficients computed by different FNN wall models with that from WRLES; (c) Correlation coefficients (Eq. (\ref{eq_8})) of instantaneous wall shear stresses between the predictions from the FNN wall models and the WRLES; (d) Relative error (Eq. (\ref{eq_9})) for different FNN wall models for instantaneous snapshots on the $x-y$ slice located at $z/h=0.0$.}
\label{fig:fig6}
\end{figure}

We first evaluate the FNN wall models for predicting wall shear stresses using the training dataset. In Figure \ref{fig:fig6} we show the comparison of the instantaneous friction coefficient (which is defined as $C_f = \tau_w / \frac{1}{2} \rho U_b^2$) at an instant, the correlation coefficient $\rho_{\text{FNN-LES}}$ between the instantaneous wall shear stress predicted by the FNN model and the WRLES, and the error $\varepsilon_\tau$ of the instantaneous wall shear stress predicted by the FNN model relative to the WRLES predictions. Here, the correlation coefficient $\rho_\tau$ is defined as
\begin{equation}
\rho_{\text{FNN-LES}} = \frac{\left\langle \left( \tau_{w}^{\text{FNN}} - \left\langle \tau_{w}^{\text{FNN}} \right\rangle \right) \cdot \left( \tau_{w}^{\text{LES}} - \left\langle \tau_{w}^{\text{LES}} \right\rangle \right) \right\rangle }{ {\left\langle {{ \left( \tau_{w}^{\text{FNN}} - \left\langle \tau_{w}^{\text{FNN}} \right\rangle \right) }^2} \right\rangle }^{1/2} {\left\langle {{ \left( \tau_{w}^{\text{LES}} - \left\langle \tau_{w}^{\text{LES}} \right\rangle \right) }^2} \right\rangle}^{1/2} },
\label{eq_8}
\end{equation}
where $\left\langle {\;} \right\rangle $ denotes the average over snapshots.

\begin{figure}[!ht]
\centering{\includegraphics[width=0.7\textwidth]{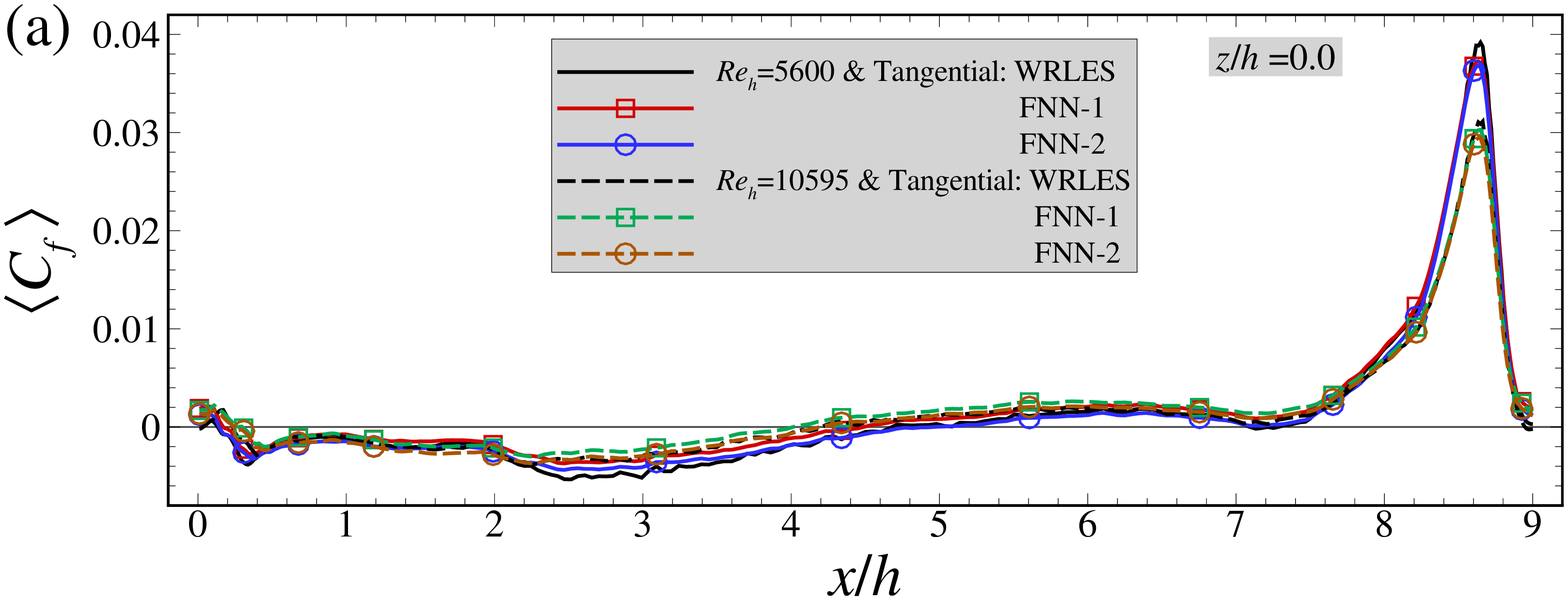}}
\centering{\includegraphics[width=0.7\textwidth]{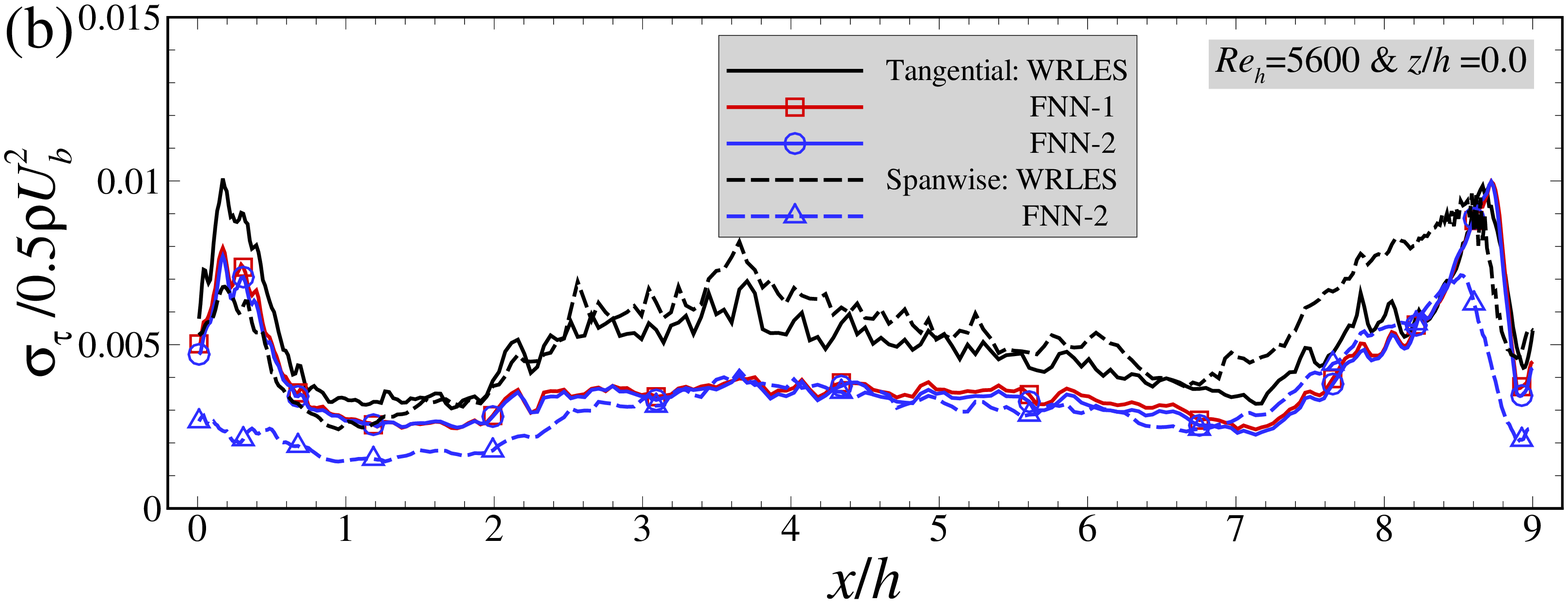}}
\centering{\includegraphics[width=0.7\textwidth]{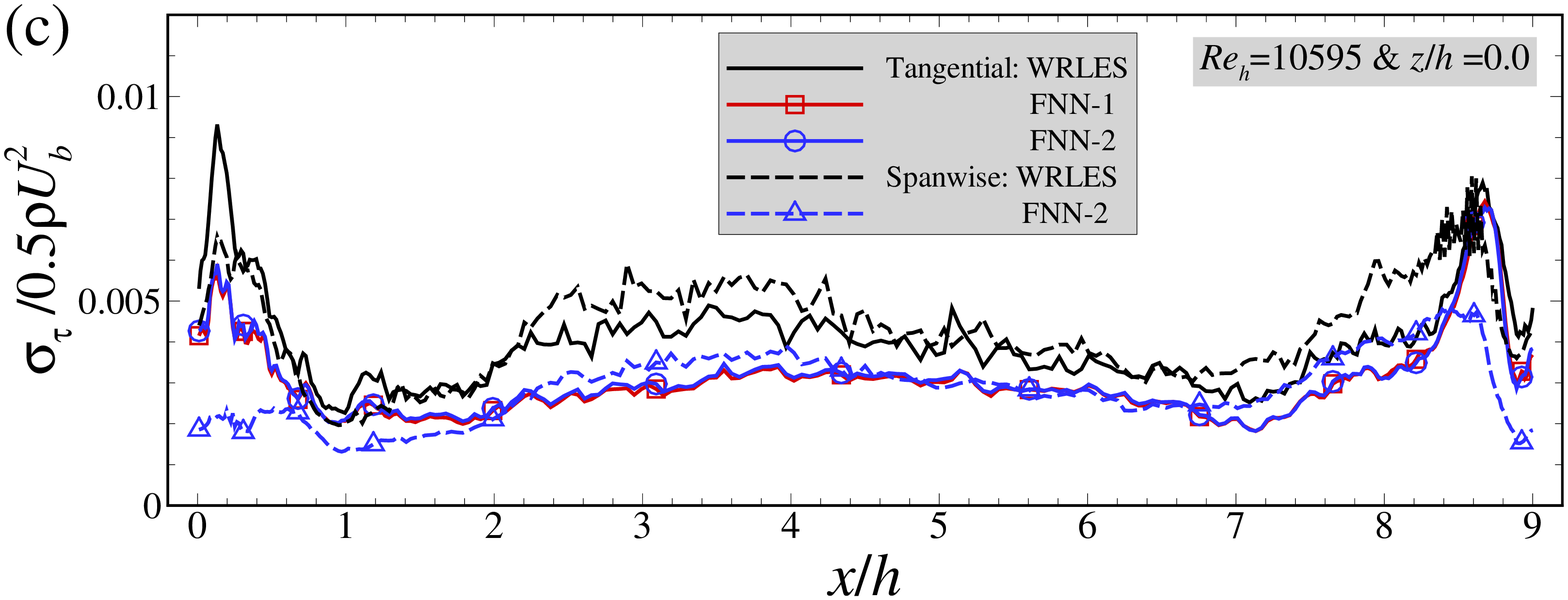}}
\centering{\includegraphics[width=0.7\textwidth]{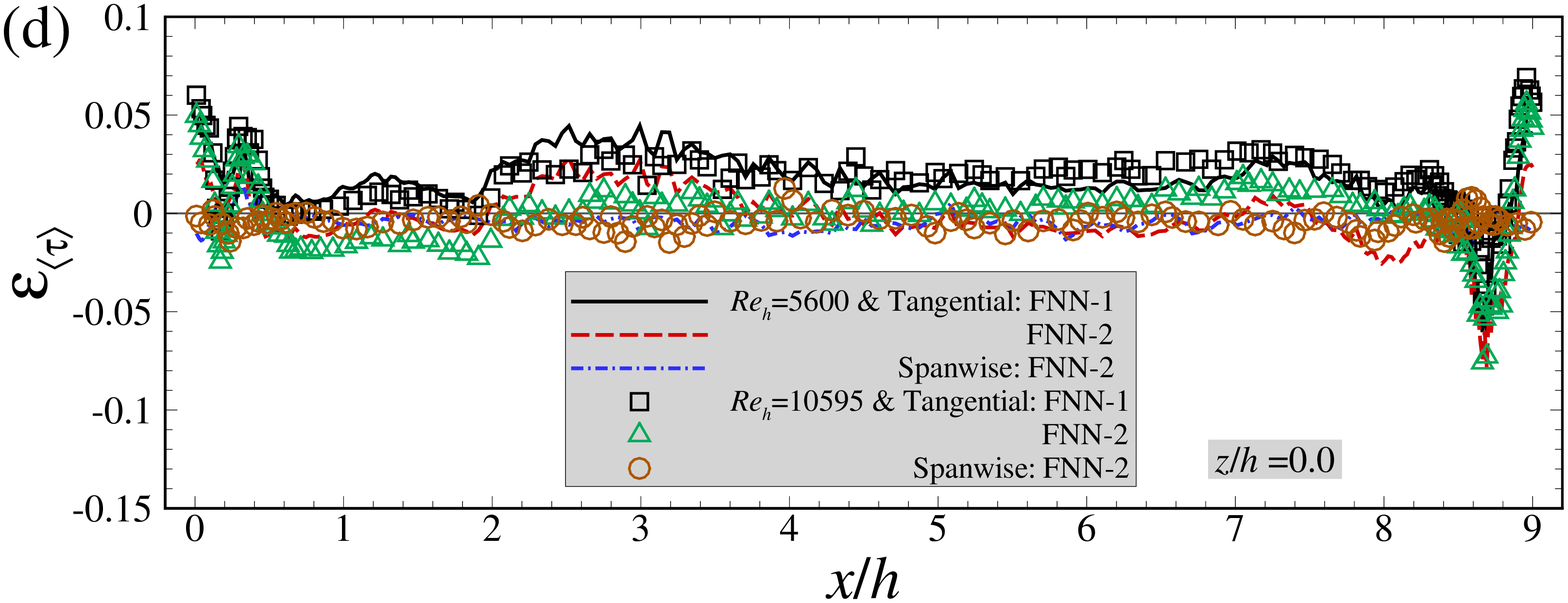}}
  \caption{Evaluation of the FNN wall model using the training dataset for: (a) Comparison of the time-averaged skin friction coefficients computed by different FNN wall models with that from WRLES; (b)(c) Normalized standard deviations of the wall shear stresses computed by the FNN models and the WRLES; (d) Relative error (Eq. (\ref{eq_10})) based on the time-averaged wall shear stress for different FNN wall models on the $x-y$ slice located at $z/h=0.0$.}
\label{fig:fig7}
\end{figure}

To further assess the prediction accuracy of FNN model on the fluctuations of wall shear stress over time, we define the instantaneous relative error $\varepsilon_{\tau}$ as follows:
\begin{equation}
\varepsilon_{\tau} = \left\langle \frac{ \left| \tau_{w}^{\text{FNN}} - \tau_{w}^{\text{LES}} \right| }{ \left| \left\langle \tau_{w}^{\text{LES}} \right\rangle \right|_{\max} } \right\rangle,
\label{eq_9}
\end{equation}
where $\left| \left\langle\tau_w^{\text{LES}} \right\rangle \right|_{\max}$ denotes the peak value of averaged wall shear stress among all the streamwise locations. The relative error of the time-averaged wall shear stress, which will be shown in Figure \ref{fig:fig7}(d), is defined as follows:
\begin{equation}
\varepsilon_{\left\langle \tau \right\rangle} = \frac{ \left\langle \tau_{w}^{\text{FNN}} \right\rangle - \left\langle \tau_{w}^{\text{LES}} \right\rangle }{ \left| \left\langle \tau_{w}^{\text{LES}} \right\rangle \right|_{\max} }.
\label{eq_10}
\end{equation}


As seen in Figure \ref{fig:fig6}(a)$\sim$(b), the instantaneous skin friction coefficient $C_f$ predicted by the FNN model in general agrees with that from WRLES at most streamwise locations. Many sharp peaks are observed in streamwise variation of the instantaneous wall shear stress. The trained FNN wall model is observed being able to predict these abrupt variations at most streamwise locations, although the peak amplitude is underpredicted at some locations, e.g. at $x/h \approx 0.18$ for this instant for the $Re_h = 10595$ case.. It is also noticed that the wall tangential component of wall shear stress predicted by the FNN-1 and FNN-2 almost collapse with each other. Moreover, the spanwise component of the wall stress predicted by the FNN-2 model is also in good agreement with the WRLES predictions. In Figure \ref{fig:fig6}(c)$\sim$(d), it is observed that the correlation coefficients are larger than 0.7 and the instantaneous relative errors are smaller than 0.1 at almost all streamwise locations except near the crest of the hill at $x/h=0.2$ and $x/h=8.5$ (where the correlation coefficient is around 0.6, the instantaneous relative error is around 0.15), indicating that the large temporal fluctuations there are not well captured by the FNN model. It is noticed that correlation coefficient for the spanwise wall shear stress is similar with that of the tangential wall shear stress although the magnitude of the instantaneous spanwise wall shear stress is one order of magnitude smaller than the tangential component, which makes it difficult to train the corresponding FNN model for both components.

In Figure \ref{fig:fig7}, we evaluate the capability the FNN model in predicting the mean wall shear stress and the standard deviation of wall shear stress. As seen in Figure \ref{fig:fig7}(a), the mean skin friction coefficients at both $Re_h = 5600$ and $10595$ predicted by the FNN models (i.e. FNN-1 and FNN-2) are in perfect agreement with WRLES results at all streamwise locations. The mean spanwise wall shear stress component, on the other hand, is close to zero for both FNN-2 and WRLES predictions (not shown in the figure). In Figure \ref{fig:fig7}(b)$\sim$(c), we compare the normalized standard deviations of the fluctuations of wall shear stresses predicted by the FNN models with those from WRLES. It is observed that the FNN predictions are smaller than the WRLES predictions for both tangential and spanwise components. Interestingly, it is observed that the standard deviation of the spanwise shear stress is similar with the tangential component although its instantaneous value is one order of magnitude smaller than the tangential component. Figure \ref{fig:fig7}(d) shows the error of time-averaged wall shear stress predicted by the FNN model relative to that from WRLES. It is seen that the absolute values of the errors are smaller than 0.05 at most streamwise locations, except at locations close to the crest of the hill.

\begin{figure}[!ht]
\centering{\includegraphics[width=0.7\textwidth]{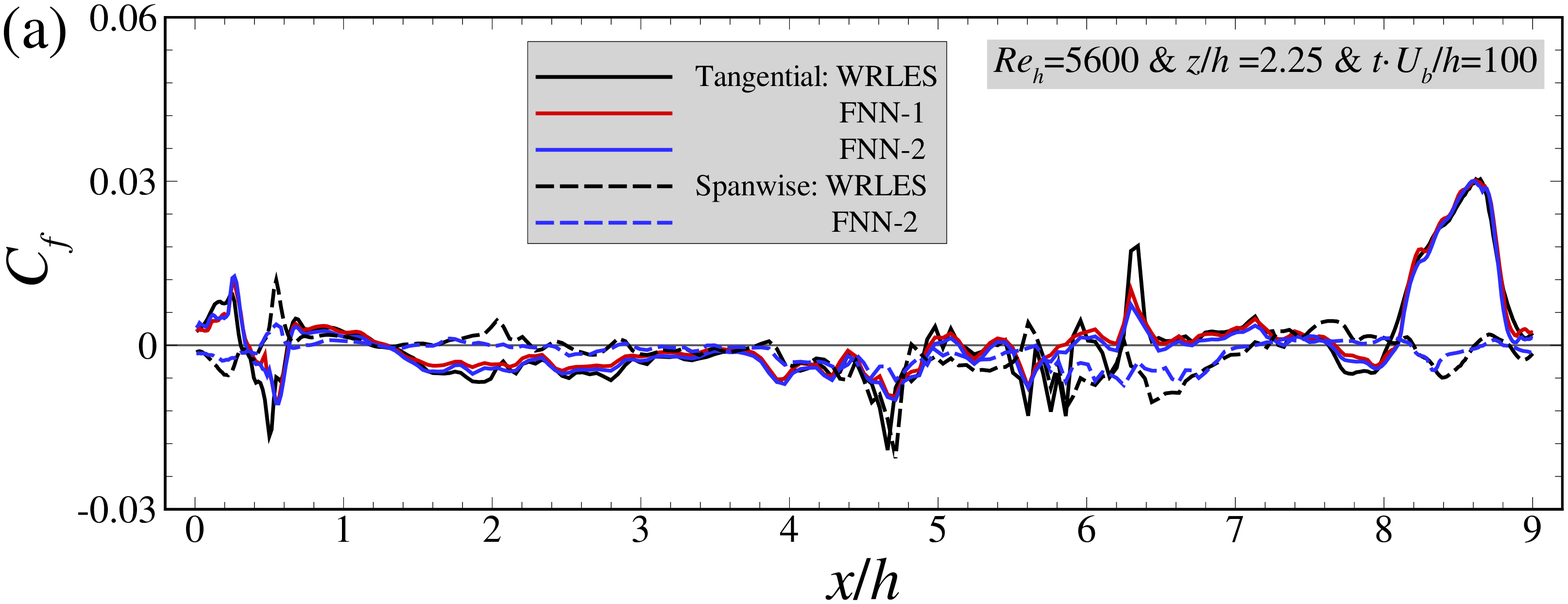}}
\centering{\includegraphics[width=0.7\textwidth]{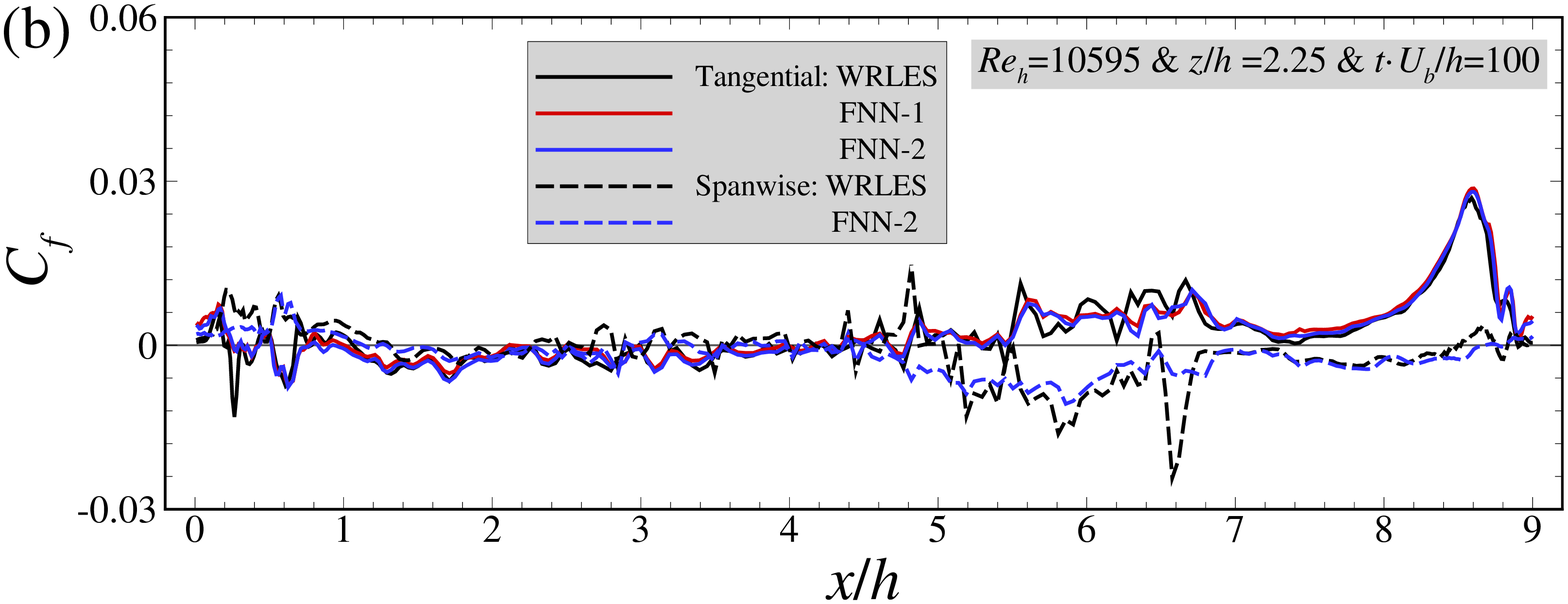}}
\centering{\includegraphics[width=0.7\textwidth]{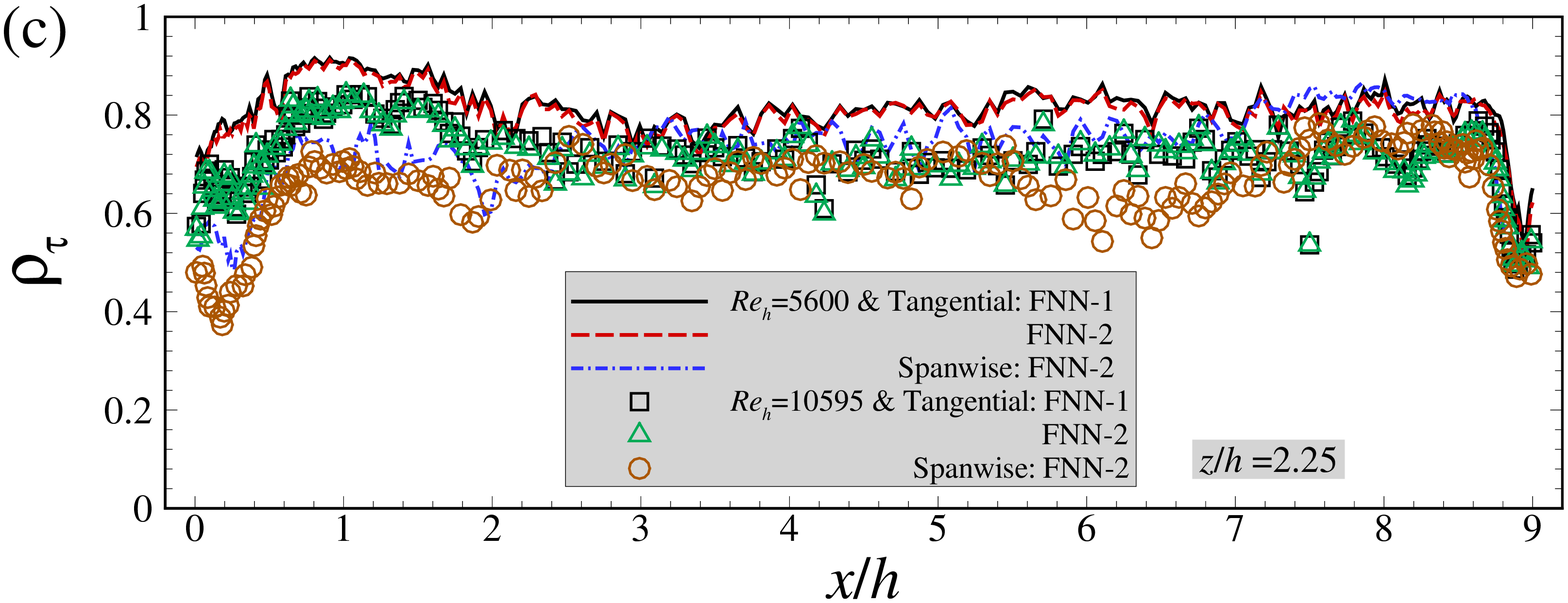}}
\centering{\includegraphics[width=0.7\textwidth]{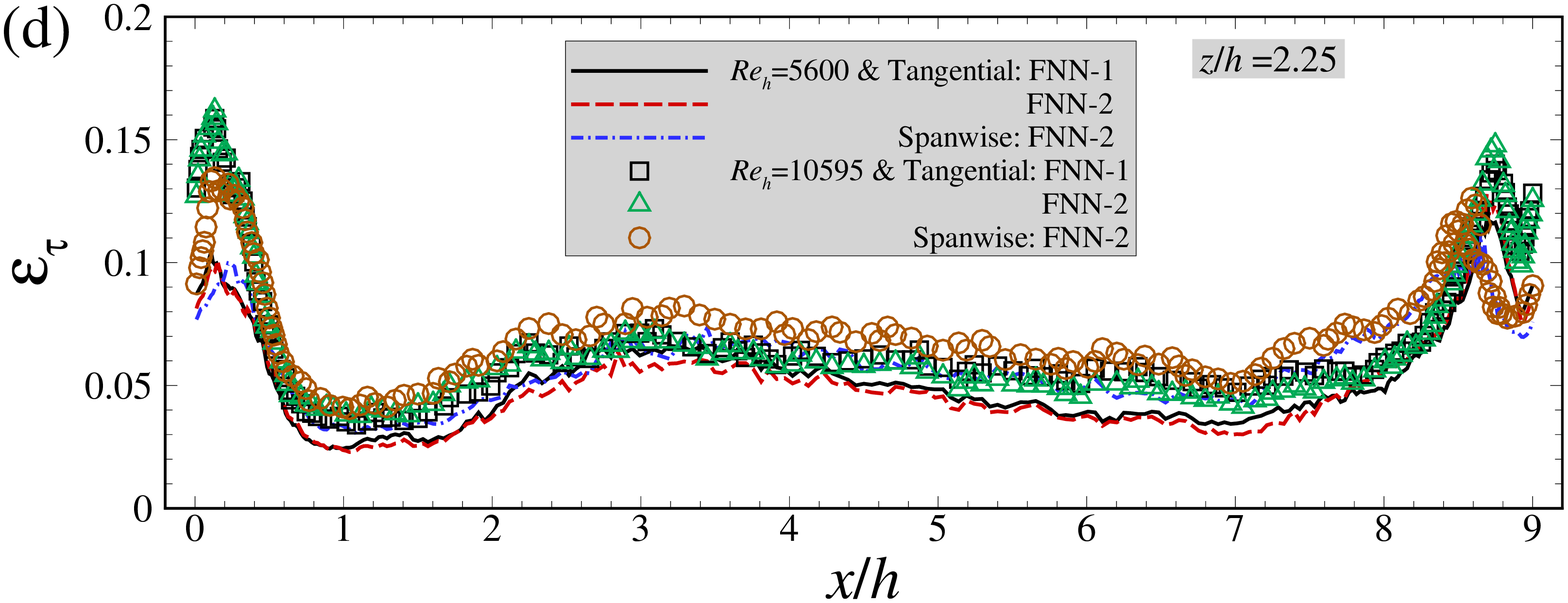}}
  \caption{Evaluation of the FNN wall model using the testing datasets at $Re_h = 5600$ and $10595$ for: (a)(b) Comparison of instantaneous skin friction coefficients computed by different FNN wall models with that from WRLES; (b) Correlation coefficients of instantaneous wall shear stresses between the predictions from the FNN wall models and the WRLES predictions; (c) Relative error for different FNN wall models for instantaneous snapshots on the $x-y$ slice located at $z/h=2.25$.}
\label{fig:fig8}
\end{figure}
\begin{figure}[!ht]
\centering{\includegraphics[width=0.7\textwidth]{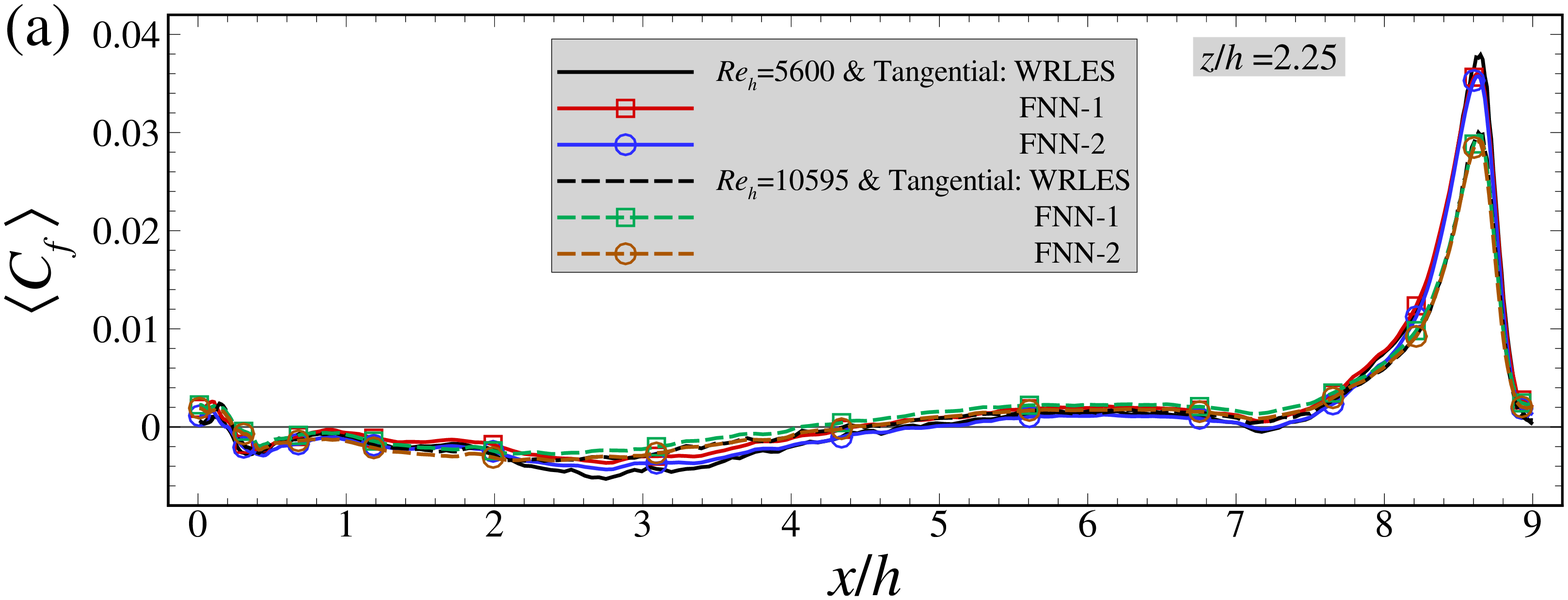}}
\centering{\includegraphics[width=0.7\textwidth]{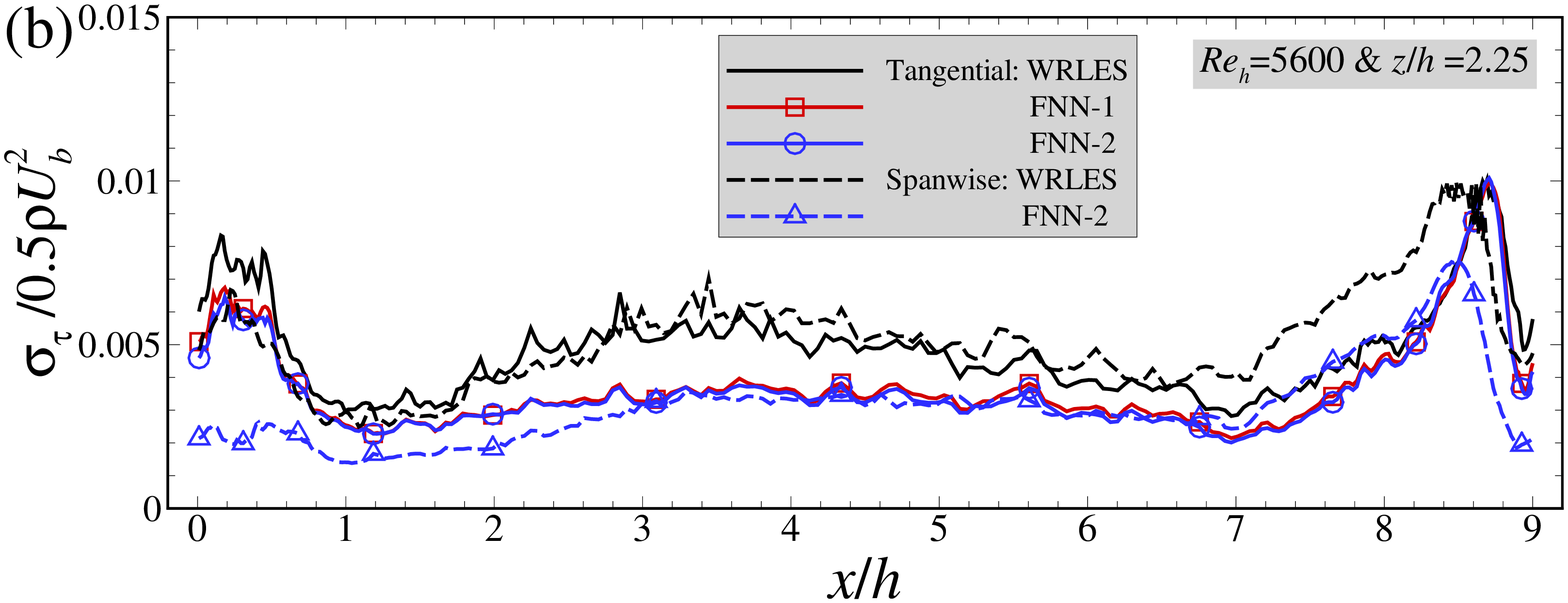}}
\centering{\includegraphics[width=0.7\textwidth]{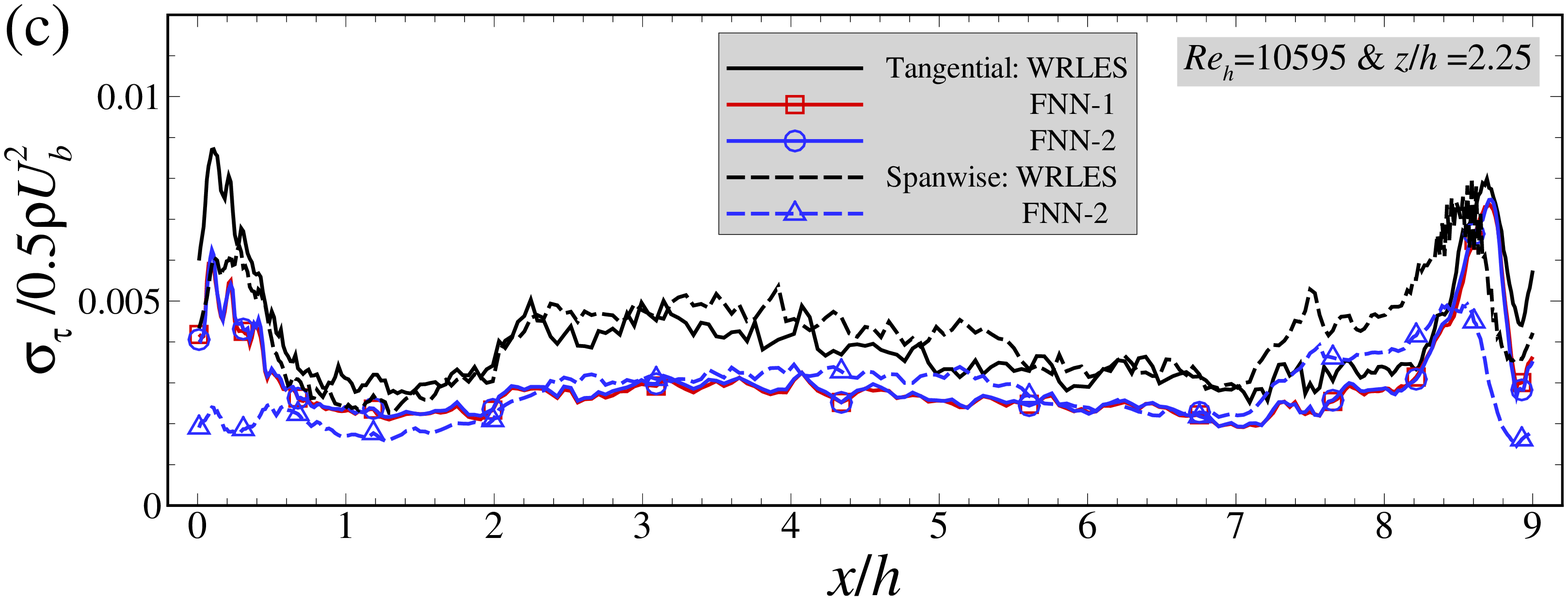}}
\centering{\includegraphics[width=0.7\textwidth]{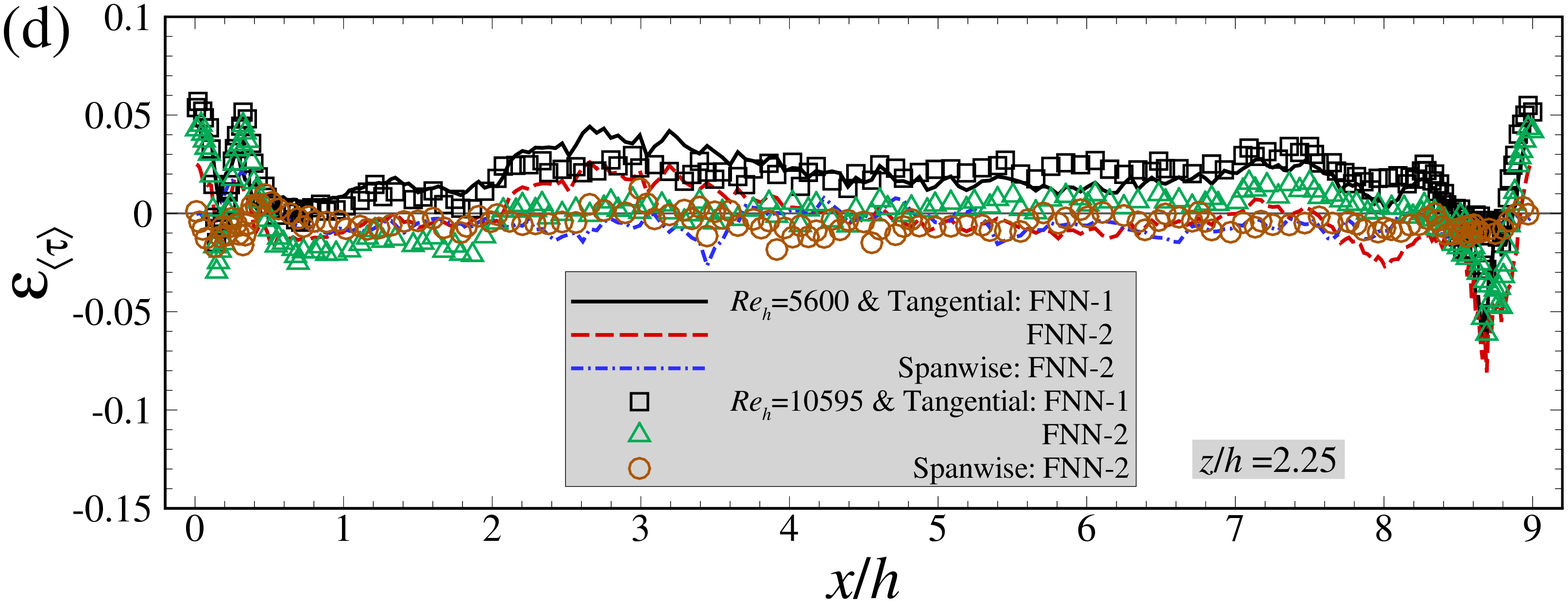}}
  \caption{Evaluation of the FNN wall model using the testing datasets at $Re_h = 5600$ and $10595$ for: (a) Comparison of the time-averaged skin friction coefficients computed by different FNN wall models with that from WRLES; (b)(c) Normalized standard deviations of the wall shear stresses computed by the FNN models and the WRLES; (d) Relative error based on the time-averaged wall shear stress for different FNN wall models on the $x-y$ slice located at $z/h=2.25$.}
\label{fig:fig9}
\end{figure}

Overall, we have shown that both FNN-1 and FNN-2 wall models can accurately predict the instantaneous and time-averaged wall shear stresses for the training dataset. Next, we will evaluate the performance of the FNN wall models using the testing dataset, which is obtained from a different $x-y$ slice (located at $z/h=2.25$) from the cases with $Re_h = 5600$ and $10595$ totally different from the training dataset.

In Figure \ref{fig:fig8}, we first evaluate the capability of the FNN models in predicting the instantaneous wall shear stress using the testing dataset. As seen in Figure \ref{fig:fig8}(a)$\sim$(b), both tangential and spanwise instantaneous skin friction coefficients predicted by the FNN models are in an overall good agreement with the WRLES predictions except for some sharp peaks. Figure \ref{fig:fig8}(c)$\sim$(d) show the correlation coefficient and relative error between the FNN and LES predictions. As seen in the range of $x/h = 1$ to $8.5$, the correlation coeffcients are in general larger than 0.7 and the relative errors are smaller than 0.1 for both tangential and spanwise components. At locations close to the crest of the hill, lower coefficients and larger errors are observed especially close the separation point for the spanwise component.

In Figure \ref{fig:fig9}, we compare the mean skin friction coefficients (averaged over snapshots), the standard deviations of the fluctuations and the time-averaged relative error of wall shear stresses predicted by different FNN wall models. As seen in Figure \ref{fig:fig9}(a), good agreements between the FNN and WRLES predictions at $Re_h = 5600$ and $10595$ are obtained for the mean friction coefficients for both FNN models. For the standard deviations of the wall shear stresses, the predictions by the FNN models are significantly smaller than those from WRLES for both tangential and spanwise components. In Figure \ref{fig:fig9}(d), the relative errors are smaller than 0.05 at most streamwise locations, which are close to the results in the training dataset.

\subsection{\label{sec:4.2}Application to different Reynolds numbers}

To further evaluate the generalization capacity of the FNN wall model, we apply the trained FNN wall models to the testing dataset at $Re_h = 19000$ in this section.

\begin{figure}[!ht]
\centering{\includegraphics[width=0.7\textwidth]{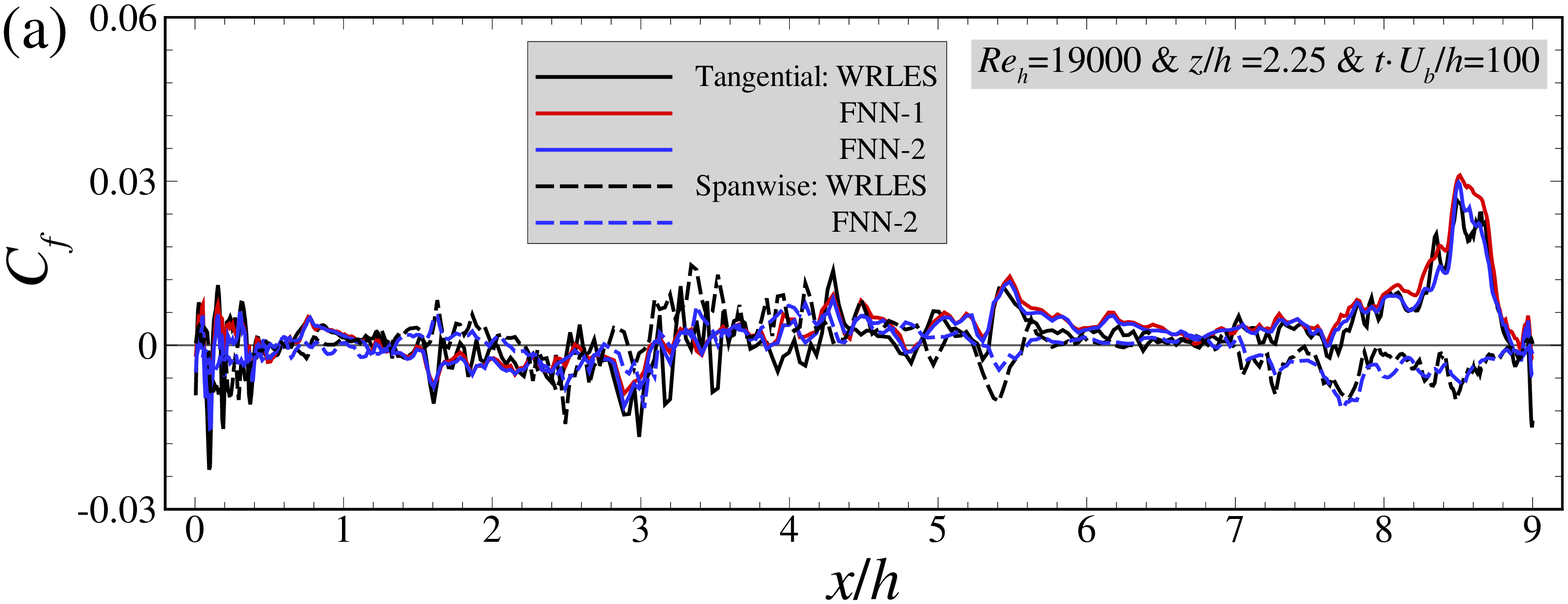}}
\centering{\includegraphics[width=0.7\textwidth]{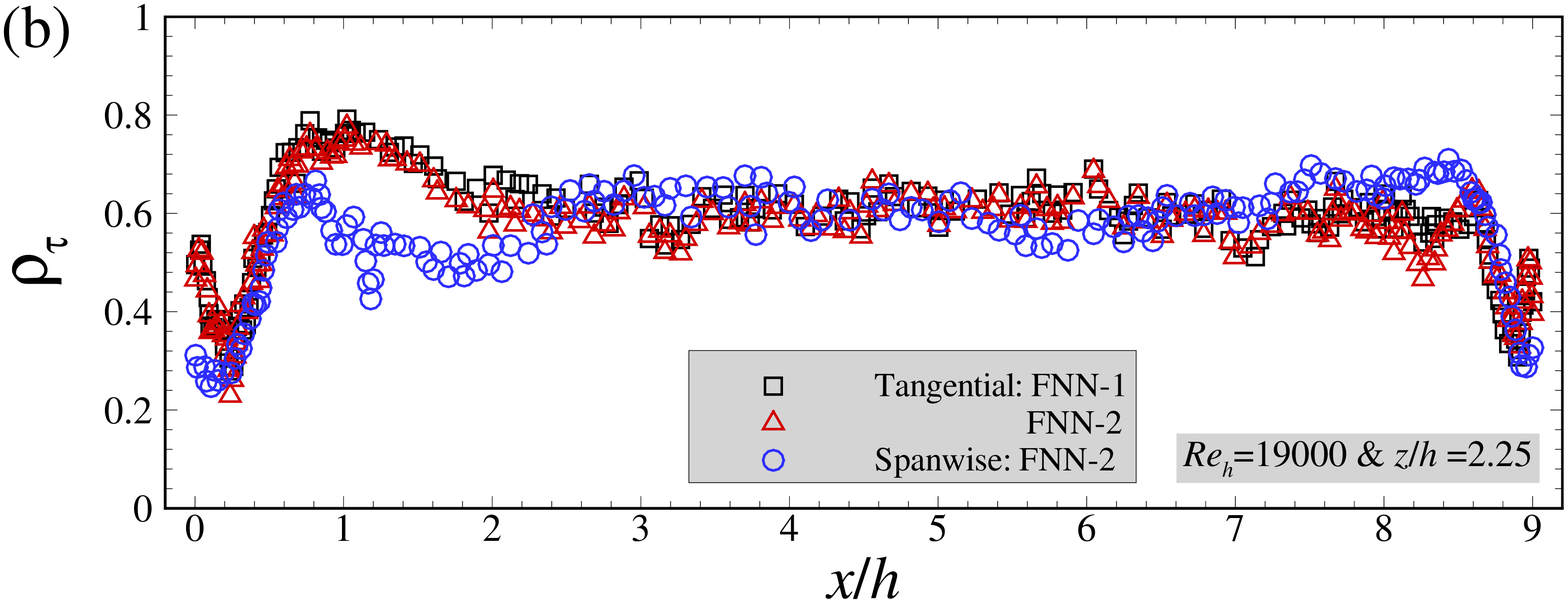}}
\centering{\includegraphics[width=0.7\textwidth]{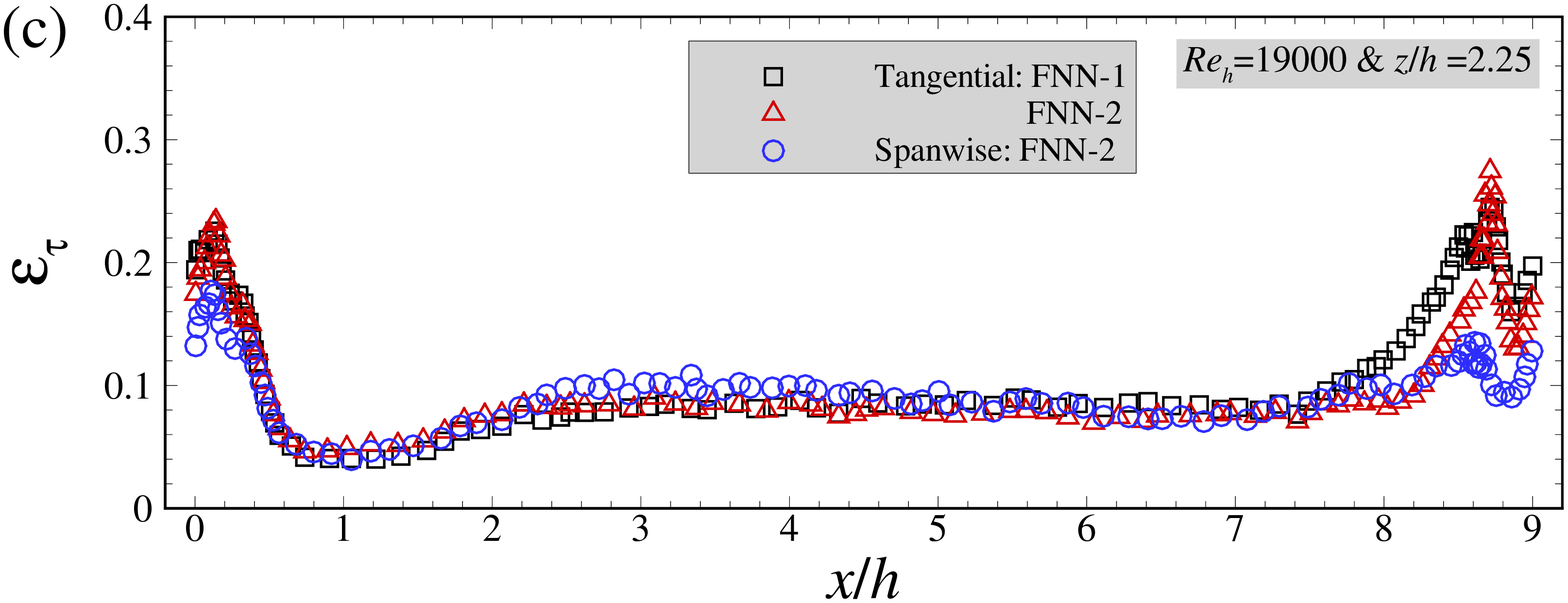}}
  \caption{Evaluation of the FNN wall model using the testing dataset at $Re_h = 19000$ for: (a) Comparison of instantaneous skin friction coefficients computed by different FNN wall models with that from WRLES; (b) Correlation coefficients of instantaneous wall shear stresses between the predictions from the FNN wall models and the WRLES predictions; (c) Relative error for different FNN wall models for instantaneous snapshots on the $x-y$ slice located at $z/h=2.25$.}
\label{fig:fig10}
\end{figure}
\begin{figure}[!ht]
\centering{\includegraphics[width=0.7\textwidth]{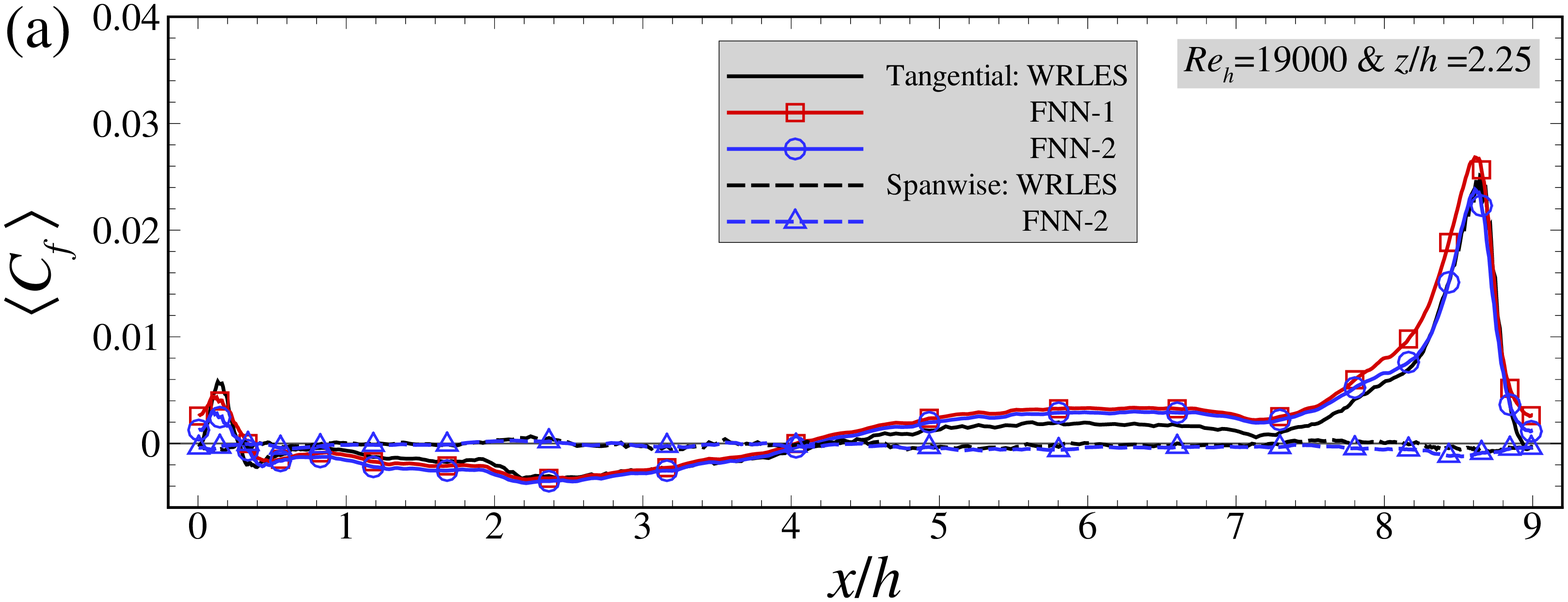}}
\centering{\includegraphics[width=0.7\textwidth]{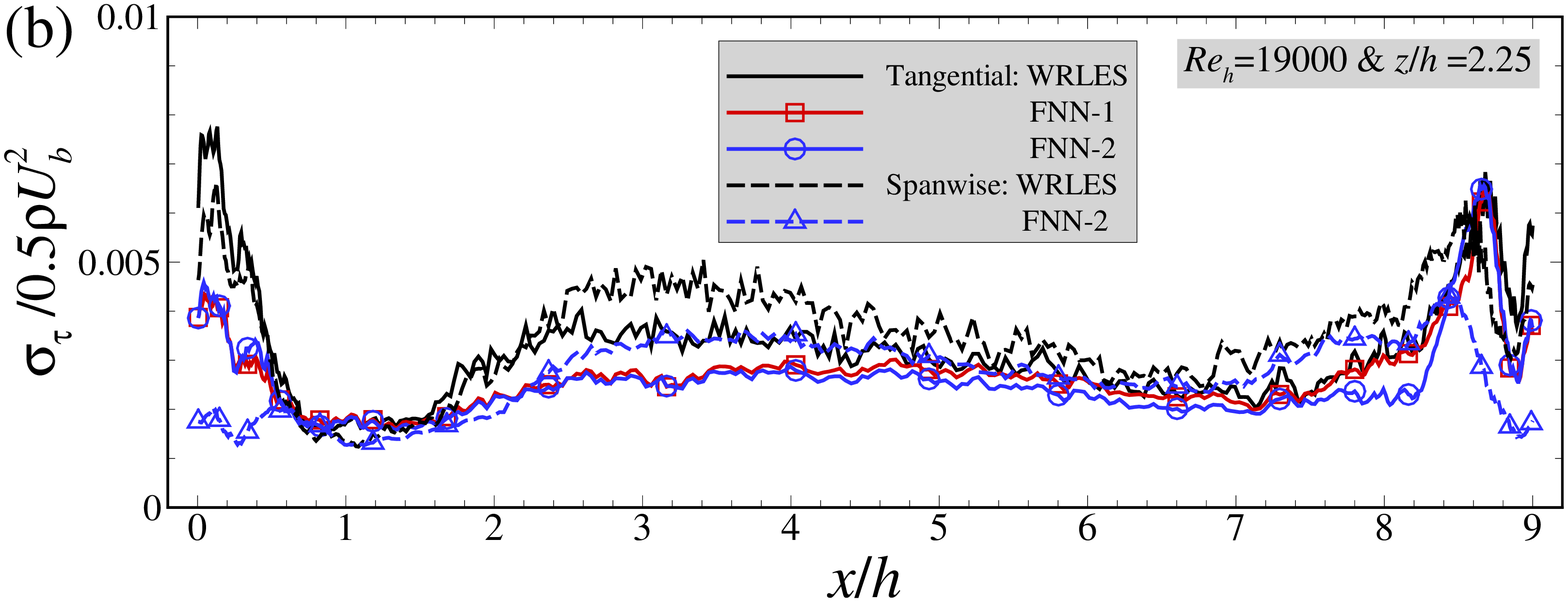}}
\centering{\includegraphics[width=0.7\textwidth]{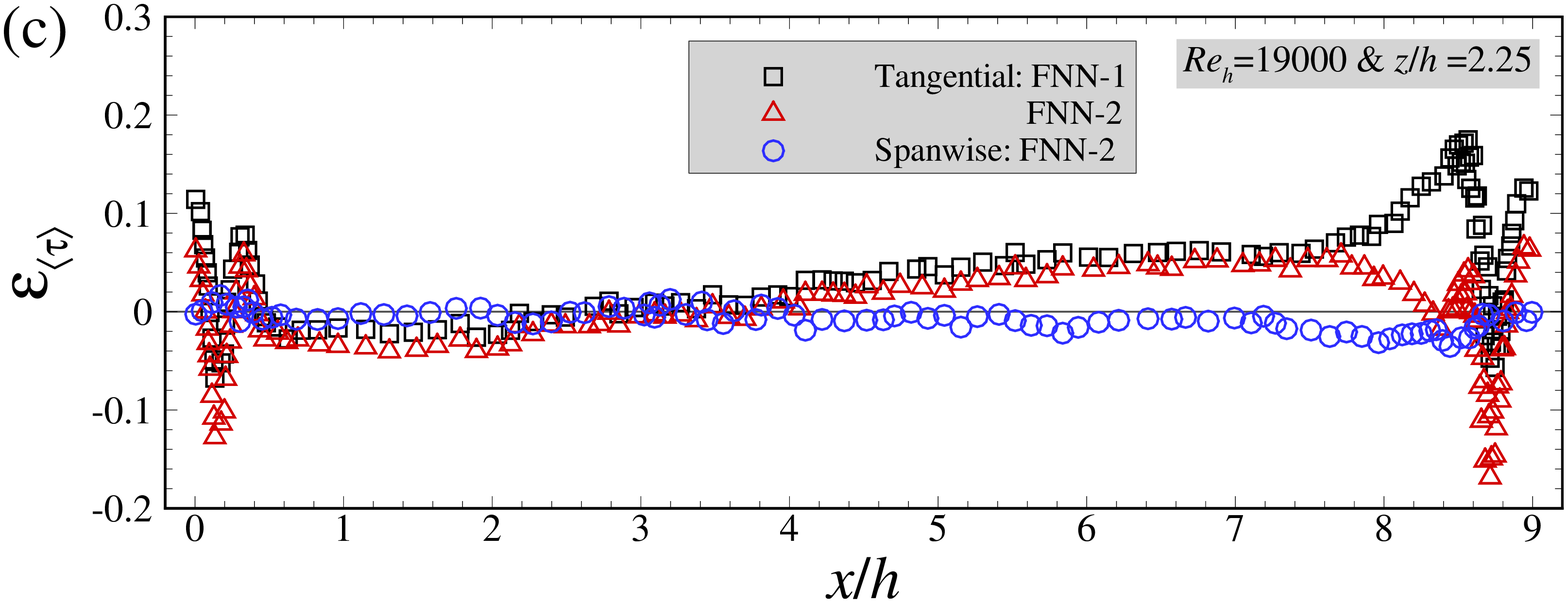}}
  \caption{Evaluation of the FNN wall model using the testing dataset at $Re_h = 19000$ for: (a) Comparison of the time-averaged skin friction coefficients computed by different FNN wall models with that from WRLES; (b) Normalized standard deviations of the wall shear stresses computed by the FNN models and the WRLES; (c) Relative error based on the time-averaged wall shear stress for different FNN wall models on the $x-y$ slice located at $z/h=2.25$.}
\label{fig:fig11}
\end{figure}

Figures \ref{fig:fig10} and \ref{fig:fig11} evaluate the predictive capability of the FNN models on the instantaneous and mean wall shear stresses, respectively. We first examine the predictons of the instantaneous skin friction coefficient. As shown in Figure \ref{fig:fig10}(a), the FNN predictions are consistent with those from WRLES at most streamwise locations. The correlation coeffcients (Figure \ref{fig:fig10}(b)) of instantaneous wall shear stresses between the FNN and LES predictions are also observed in general larger than 0.6 in the range of $x/h = 1$ to $8.5$, although they are somewhat smaller than those computed using the training dataset. The instantaneous relative errors are then examined in Figure \ref{fig:fig10}(d). It is observed that the magnitude of the errors are slightly larger than those in training dataset, but still smaller than 0.1 at most streamwise locations.

Here we examine the performance of the FNN models for predicting the mean skin friction coefficient.  As seen in Figure \ref{fig:fig11}(a), the mean skin friction coefficients at $Re_h = 19000$ predicted by the FNN-2 model are in good agreement with WRLES results at all streamwise locations. Some discrepancies are observed between the FNN-1 predictions and the WRLES predictions highlighting the importance of having the spanwise wall shear stress as one of the output labels. As expected, the mean spanwise wall shear stress component is close to zero for both FNN model predictions and WRLES predictions. Similar with those observed in the cases with $Re_h = 5600$ and $10595$, the normalized standard deviations of wall shear stress fluctuations predicted by the FNN models are smaller than those from WRLES, as shown in Figure \ref{fig:fig11}(b). In Figure \ref{fig:fig11}(d), the relative errors of the time-averaged wall shear stress are smaller than 0.1 at most streamwise locations, which are slightly larger than those computed using the training dataset. It is also noticed that the magnitude of the relative error for the predictions from the FNN-2 model is smaller than that from the FNN-1 model especially at locations around $x/h = 8.4$.

Although further improvement is still needed for the FNN wall model to accurately predict the standard deviations of wall shear stress fluctuations, the evaluations against the testing datasets at different Reynolds numbers demonstrate the excellent generalization capacity of the trained FNN wall models.

\section{\label{sec:Conclusion}Conclusion}
As a first step towards developing a general wall model for complex turbulent flows, in this work we developed a data-driven wall model for LES of flow over periodic hills using the physics-informed feedforward neural networks and WRLES data.

Data preparation is critical for the success of training data-driven wall models. As the objective of this work is to develop a wall model that is applicable to different streamwise locations (of different flow regime, i.e. attached wall turbulence, flow separation and reattachment) of the periodic hill, the flow data near the surface of the hill at all streamwise locations are grouped together as the training data. The wall shear stresses are taken as the ouput labels. The input features include wall-normal distance, different components of velocity and pressure gradient at different wall-normal locations.

Effects of number of input features and number of neurons in the hidden layers on training performance were tested. It was found that using the flow data at more than two off-wall locations (in addition to the velocity at the boundary, which is implicitly taken into account) are adequate for training the data-driven wall model. Further increasing the number of input features does not improve the convergence rate when training the model. Employing more than 20 neurons in each hidden layer is found enough for this case. In the data-driven model developed in this work, flow data at three off-wall locations are employed as input features with 20 neurons for each hidden layer. Two different wall models, i.e. one using only the tangential wall shear stress as the output label (FNN-1), and the other one using both wall shear stress components as the output labels (FNN-2), are tested.

The prediction accuracy and generalization capacity of the trained FNN wall model were examined by comparing the predicted wall shear stresses with the WRLES data. The instantaneous wall shear stresses predicted by the FNN wall model show an overall good agreement with the WRLES data with some discrepancies observed at locations near the crest of the hill. For the mean wall shear stress, the predictions from the FNN wall models agree very well with WRLES data. However, the standard deviations of the fluctuations of the wall shear stress are underpredicted by the FNN wall model. Furthermore, it is noticed that the predictions from the two models, i.e. FNN-1 and FNN-2, are very similar with each other for the $Re_h=5600$ and $10595$ cases, which are employed for training the model. For the $Re_h=19000$ case, for which the flow data are not employed for training the model, the FNN-2 model is observed performing better than the FNN-1 model. In summary, good performance and generalization capacity are observed for the developed FNN wall models. Implementation of the developed FNN wall model in WMLES and evaluation of its performance using \emph{a posteriori} LES will be carried out in our future work.

\begin{acknowledgments}
This work is partly supported by NSFC Basic Science Center Program for ``Multiscale Problems in Nonlinear Mechanics'' (No. 11988102) and National Natural Science Foundation of China (No. 12002345).
\end{acknowledgments}

\appendix
\section{Details on the employed grid and validation of the present WRLES case}\label{Appendix_A}
\begin{figure}[!ht]
\centering{\includegraphics[width=0.72\textwidth]{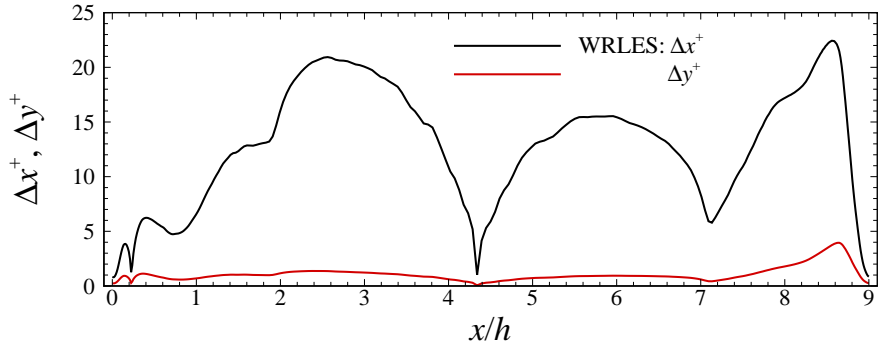}}
  \caption{Distribution of the grid spacings in wall units for $x-$ and $y-$ direction along the lower wall.}
\label{fig:fig_A1}
\end{figure}
\begin{figure}[!ht]
\centering{\includegraphics[width=0.72\textwidth]{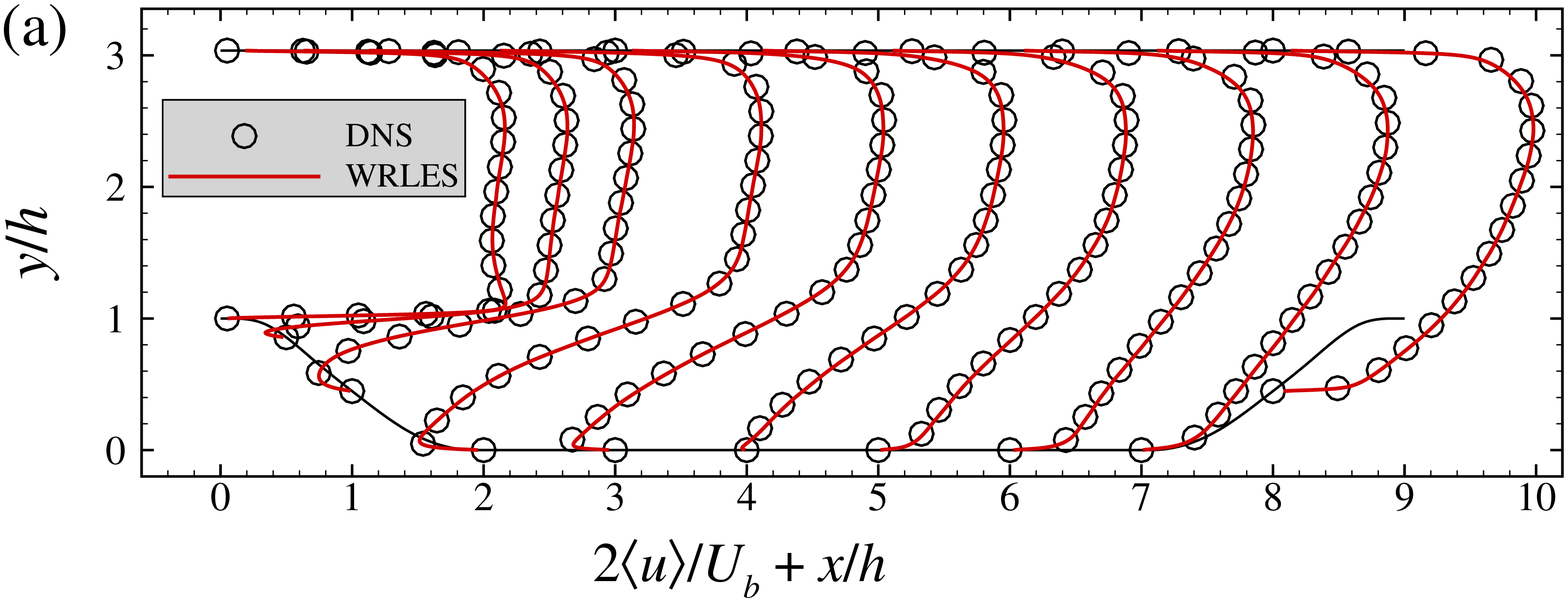}}
\centering{\includegraphics[width=0.72\textwidth]{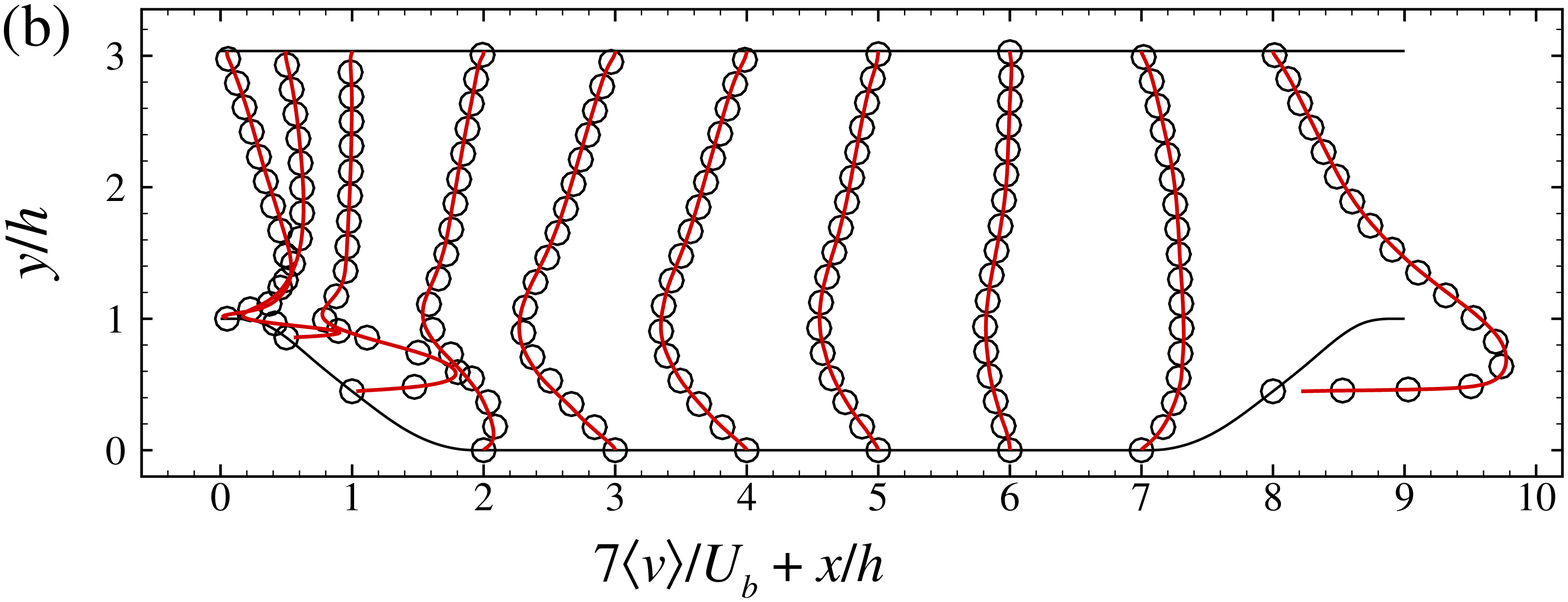}}
\centering{\includegraphics[width=0.72\textwidth]{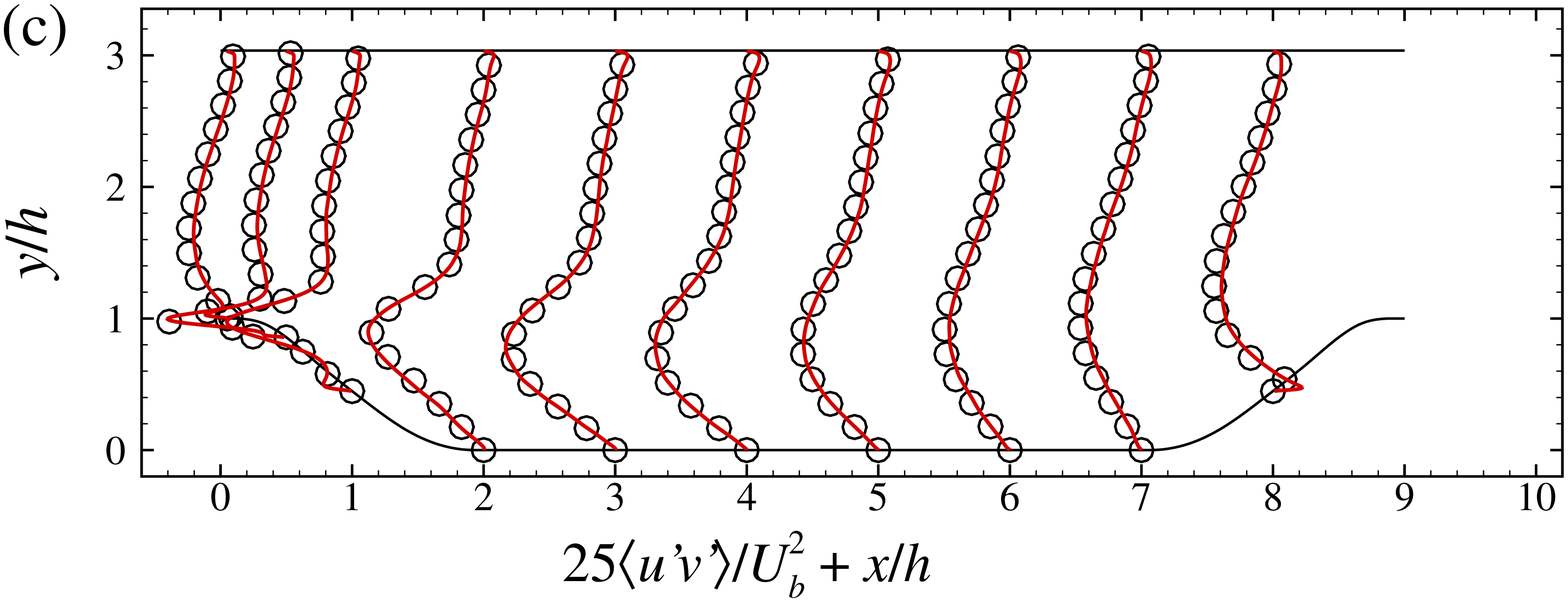}}
\centering{\includegraphics[width=0.72\textwidth]{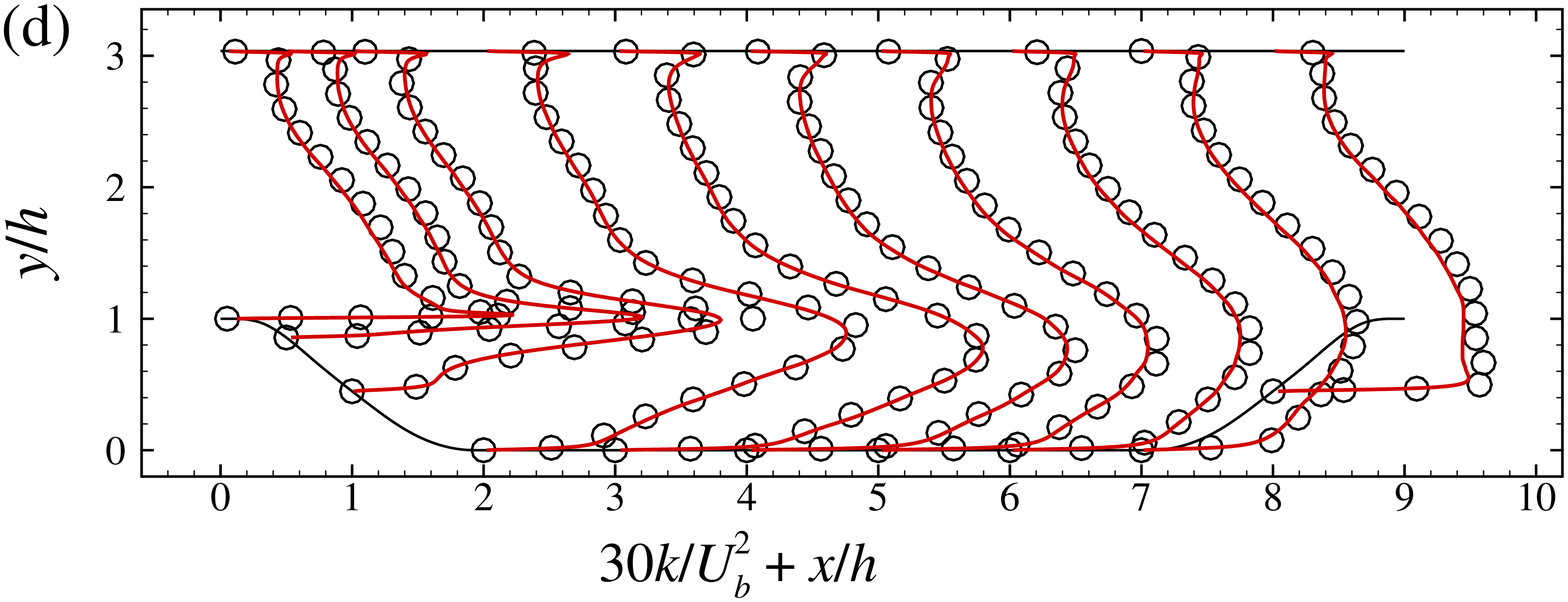}}
  \caption{Comparison of vertical profiles from the present WRLES with DNS data from Krank et al. [5] for (a) time-averaged streamwise velocity $\left\langle u \right\rangle$, (b) time-averaged vertical velocity $\left\langle v \right\rangle$, (c) primary Reynolds shear stress $\left\langle u'v' \right\rangle$, and (d) turbulence kinetic energy $k = \frac{1}{2}\left\langle u'u'+v'v'+w'w' \right\rangle$.}
\label{fig:fig_A2}
\end{figure}
\begin{figure}[!ht]
\centering{\includegraphics[width=0.72\textwidth]{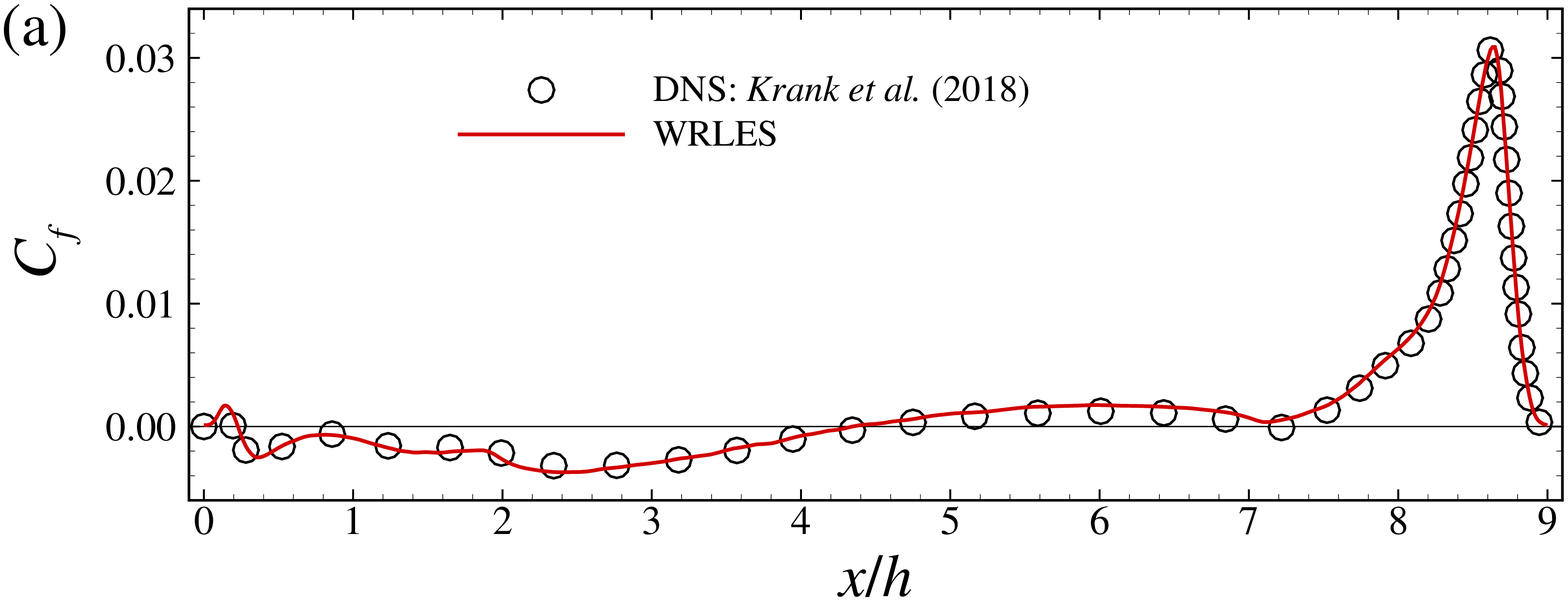}}
\centering{\includegraphics[width=0.72\textwidth]{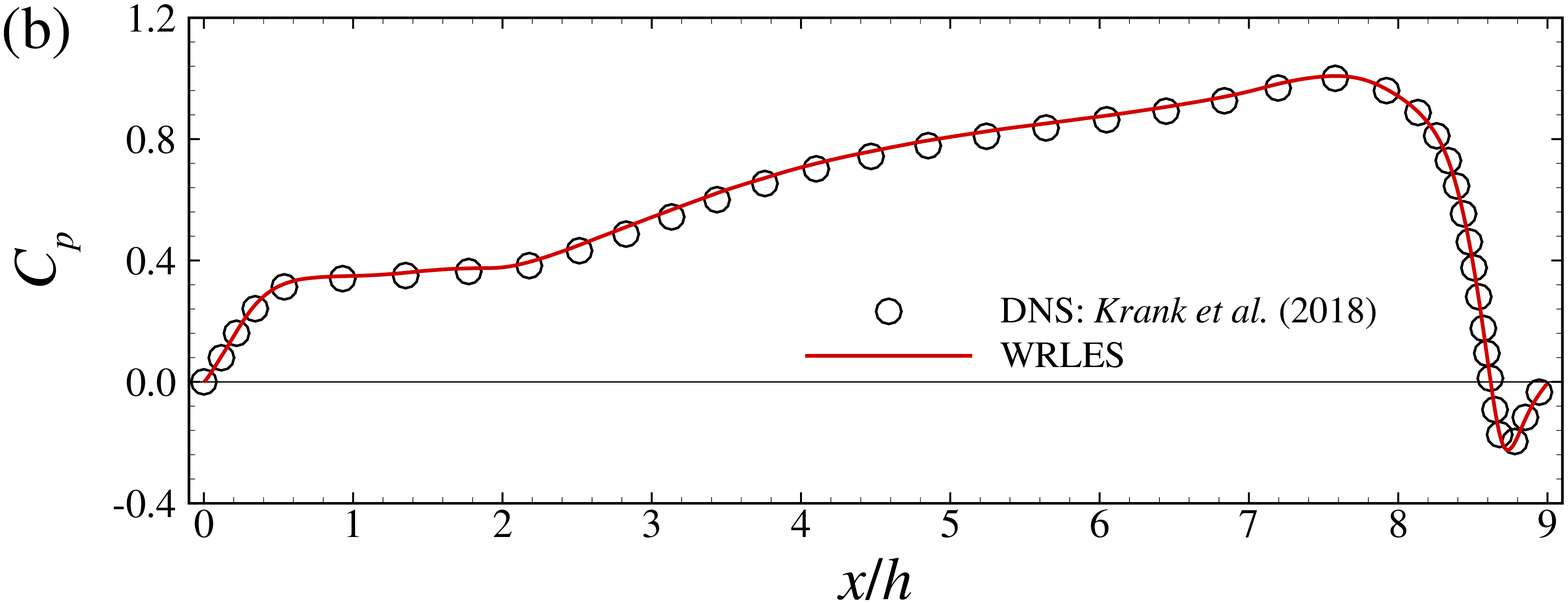}}
  \caption{Comparison of (a) skin friction coefficient and (b) pressure coefficient at the lower wall between the present WRLES and DNS by Krank et al. \cite{Krank_etal_FTC_2018}}
\label{fig:fig_A3}
\end{figure}

In this appendix, we show some details on the employed grid and validate the employed VFS-Wind code and the case setup for simulating turbulent flows over periodic hills at $Re_h = 10595$ by comparing our WRLES results with the DNS results by Krank et al. \cite{Krank_etal_FTC_2018} ($896 \times 448 \times 448$ grid points). Figure \ref{fig:fig_A1} shows the distribution of the grid spacings in wall units for $x-$ and $y-$ direction along the lower wall, where $\Delta x^{+} = \Delta x u_{\tau}/\nu$ denotes the streamwise grid spacing in wall unit. The height of the first off-wall grid nodes in wall units, $\Delta y^{+}$, is in the range of 0.056 to 3.95.

In Figure \ref{fig:fig_A2}, we plot the vertical profiles of the time-averaged streamwise velocity $\left\langle u \right\rangle$ and vertical velocity $\left\langle v \right\rangle$, primary Reynolds shear stress $\left\langle u^{\prime} v^{\prime} \right\rangle$, turbulence kinetic energy $k$. As seen, the WRLES results are in good agreement with the DNS results \cite{Krank_etal_FTC_2018} except for some minor differences observed in the Reynolds shear stress and the turbulence kinetic energy (with the relative error less than $12\%$). Figure \ref{fig:fig_A3} shows the comparison of the skin friction coefficient $C_f$ and the pressure coefficient $C_p$. Again, the $C_f$ and $C_p$ from the present WRLES agree very well with the DNS predictions.

\section{Feedforward neural network}\label{Appendix_B}
The detailed procedures for calculating the output based on the input in the FNN are described in this appendix.

The input layer is
\begin{equation}
\mathbf{X} = {\left[ x_1,x_2,\ldots,x_{n_I} \right]}^{\text{T}},
\label{eq_B1}
\end{equation}
where $x_i$ denotes the $i^{\text{th}}$ input feature, $n_I$ is the number of neurons in the input layer. The matrixes of weight and bias coefficient connecting the input layer and the hidden layer are
\begin{equation}
{{\mathbf{W}}^{1}} = \left[ \begin{matrix}
   w_{1,1}^{1} & w_{1,2}^{1} & \cdots  & w_{1,{n_I}}^{1}  \\
   w_{2,1}^{1} & w_{2,2}^{1} & \cdots  & w_{2,{n_I}}^{1}  \\
   \vdots  & \vdots  & \ddots  & \vdots   \\
   w_{{n_H},1}^{1} & w_{{n_H},2}^{1} & \cdots  & w_{{n_H},{n_I}}^{1}  \\
\end{matrix} \right], \quad  {{\mathbf{B}}^{1}} = \left[ \begin{matrix}
   b_{1}^{1}  \\
   b_{2}^{1}  \\
   \vdots   \\
   b_{{{n}_{H}}}^{1}  \\
\end{matrix} \right],
\label{eq_B2}
\end{equation}
where $w_{i,j}^{1}\left( i=1,2,\ldots ,{{n}_{H}}; j=1,2,\ldots ,{{n}_{I}} \right)$ denotes the weight coefficient connecting the $i^{\text{th}}$ neuron in the hidden layer and the $j^{\text{th}}$ neuron in the input layer, $b_{i}^{1}$ denotes the bias coefficient for the $i^{\text{th}}$ neuron in the hidden layer, $n_H$ is the number of neurons in the hidden layer. Initially, the weight coefficients are set to be random numbers from truncated normal distribution (0.0 mean and 0.1 standard deviation) and the bias coefficients are set to zero.

The output of the hidden layer is
\begin{equation}
{{\mathbf{H}}^{\text{T}}} = f\left( {{\mathbf{W}}^{1}}\mathbf{X}+{{\mathbf{B}}^{1}} \right) = {{\left[ {{h}_{1}},{{h}_{2}},\ldots ,{{h}_{{{n}_{H}}}} \right]}^{\text{T}}}, \quad  {{h}_{i}} = f\left( \sum\limits_{j=1}^{n_I}{w_{i,j}^{1}{{x}_{j}}}+b_{i}^{1} \right),
\label{eq_B3}
\end{equation}
where $f$ denotes the activation function to carry out the nonlinear mapping of the FNN, and the superscript ``$\text{T}$'' denotes the transpose of matrix. After the data transmission of multi hidden layers, the matrixes of weight and bias coefficient connecting the last hidden layer and the output layer are
\begin{equation}
{{\mathbf{W}}^{L+1}} = \left[ \begin{matrix}
   w_{1,1}^{L+1} & w_{1,2}^{L+1} & \cdots  & w_{1,{n_H}}^{L+1}  \\
   w_{2,1}^{L+1} & w_{2,2}^{L+1} & \cdots  & w_{2,{n_H}}^{L+1}  \\
   \vdots  & \vdots  & \ddots  & \vdots   \\
   w_{{n_O},1}^{L+1} & w_{{n_O},2}^{L+1} & \cdots  & w_{{n_O},{n_H}}^{L+1}  \\
\end{matrix} \right], \quad  {{\mathbf{B}}^{L+1}} = \left[ b_{1}^{L},b_{2}^{L+1},\ldots ,b_{n_O}^{L+1} \right],
\label{eq_B4}
\end{equation}
where $w_{ij}^{L+1}\left( i=1,2,\ldots ,{{n}_{O}}; j=1,2,\ldots ,{{n}_{H}} \right)$ denotes the weight coefficient connecting the $i^{\text{th}}$ neuron in the output layer and the $j^{\text{th}}$ neuron in the $L^{\text{th}}$ hidden layer, $b_{i}^{L+1}$ denotes the bias coefficient for the $i^{\text{th}}$ neuron in the output layer and $n_O$ is the number of neurons in the output layer.

The output of the FNN is calculated by
\begin{equation}
{{\mathbf{Y}}^{\text{*}}} = {{\mathbf{W}}^{L+1}}{{\mathbf{H}}^{\text{T}}}+{{\mathbf{B}}^{L+1}}=\left[ y_{1}^{*},y_{2}^{*},\ldots ,y_{{{n}_{O}}}^{*} \right], \quad  y_{i}^{*} = \sum\limits_{j=1}^{{{n}_{H}}}{w_{ij}^{L+1}{{h}_{j}}}+b_{i}^{L+1}.
\label{eq_B5}
\end{equation}

\nocite{*}

\bibliography{PRF}

\begin{thebibliography}{61}%
\makeatletter
\providecommand \@ifxundefined [1]{%
 \@ifx{#1\undefined}
}%
\providecommand \@ifnum [1]{%
 \ifnum #1\expandafter \@firstoftwo
 \else \expandafter \@secondoftwo
 \fi
}%
\providecommand \@ifx [1]{%
 \ifx #1\expandafter \@firstoftwo
 \else \expandafter \@secondoftwo
 \fi
}%
\providecommand \natexlab [1]{#1}%
\providecommand \enquote  [1]{``#1''}%
\providecommand \bibnamefont  [1]{#1}%
\providecommand \bibfnamefont [1]{#1}%
\providecommand \citenamefont [1]{#1}%
\providecommand \href@noop [0]{\@secondoftwo}%
\providecommand \href [0]{\begingroup \@sanitize@url \@href}%
\providecommand \@href[1]{\@@startlink{#1}\@@href}%
\providecommand \@@href[1]{\endgroup#1\@@endlink}%
\providecommand \@sanitize@url [0]{\catcode `\\12\catcode `\$12\catcode
  `\&12\catcode `\#12\catcode `\^12\catcode `\_12\catcode `\%12\relax}%
\providecommand \@@startlink[1]{}%
\providecommand \@@endlink[0]{}%
\providecommand \url  [0]{\begingroup\@sanitize@url \@url }%
\providecommand \@url [1]{\endgroup\@href {#1}{\urlprefix }}%
\providecommand \urlprefix  [0]{URL }%
\providecommand \Eprint [0]{\href }%
\providecommand \doibase [0]{http://dx.doi.org/}%
\providecommand \selectlanguage [0]{\@gobble}%
\providecommand \bibinfo  [0]{\@secondoftwo}%
\providecommand \bibfield  [0]{\@secondoftwo}%
\providecommand \translation [1]{[#1]}%
\providecommand \BibitemOpen [0]{}%
\providecommand \bibitemStop [0]{}%
\providecommand \bibitemNoStop [0]{.\EOS\space}%
\providecommand \EOS [0]{\spacefactor3000\relax}%
\providecommand \BibitemShut  [1]{\csname bibitem#1\endcsname}%
\let\auto@bib@innerbib\@empty
\bibitem [{\citenamefont {Moin}\ and\ \citenamefont
  {Mahesh}(1998)}]{Moin_Mahesh_ARFM_1998}%
  \BibitemOpen
  \bibfield  {author} {\bibinfo {author} {\bibfnamefont {P.}~\bibnamefont
  {Moin}}\ and\ \bibinfo {author} {\bibfnamefont {K.}~\bibnamefont {Mahesh}},\
  }\bibfield  {title} {\enquote {\bibinfo {title} {Direct numerical simulation:
  {A} tool in turbulence research},}\ }\href {\doibase
  10.1146/annurev.fluid.30.1.539} {\bibfield  {journal} {\bibinfo  {journal}
  {Annu. Rev. Fluid Mech.}\ }\textbf {\bibinfo {volume} {30}},\ \bibinfo
  {pages} {539--578} (\bibinfo {year} {1998})}\BibitemShut {NoStop}%
\bibitem [{\citenamefont {Choi}\ and\ \citenamefont
  {Moin}(2012)}]{Choi_Moin_PoF_2012}%
  \BibitemOpen
  \bibfield  {author} {\bibinfo {author} {\bibfnamefont {H.}~\bibnamefont
  {Choi}}\ and\ \bibinfo {author} {\bibfnamefont {P.}~\bibnamefont {Moin}},\
  }\bibfield  {title} {\enquote {\bibinfo {title} {Grid-point requirements for
  large eddy simulation: {C}hapman's estimates revisited},}\ }\href {\doibase
  10.1063/1.3676783} {\bibfield  {journal} {\bibinfo  {journal} {Phys. Fluids}\
  }\textbf {\bibinfo {volume} {24}},\ \bibinfo {pages} {011702} (\bibinfo
  {year} {2012})}\BibitemShut {NoStop}%
\bibitem [{\citenamefont {He}\ \emph {et~al.}(2017)\citenamefont {He},
  \citenamefont {Jin},\ and\ \citenamefont {Yang}}]{He_Jin_Yang_ARFM_2017}%
  \BibitemOpen
  \bibfield  {author} {\bibinfo {author} {\bibfnamefont {G.~W.}\ \bibnamefont
  {He}}, \bibinfo {author} {\bibfnamefont {G.~D.}\ \bibnamefont {Jin}}, \ and\
  \bibinfo {author} {\bibfnamefont {Y.}~\bibnamefont {Yang}},\ }\bibfield
  {title} {\enquote {\bibinfo {title} {Space-time correlations and dynamic
  coupling in turbulent flows},}\ }\href {\doibase
  10.1146/annurev-fluid-010816-060309} {\bibfield  {journal} {\bibinfo
  {journal} {Annu. Rev. Fluid Mech.}\ }\textbf {\bibinfo {volume} {49}},\
  \bibinfo {pages} {51--70} (\bibinfo {year} {2017})}\BibitemShut {NoStop}%
\bibitem [{\citenamefont {Piomelli}\ and\ \citenamefont
  {Balaras}(2002)}]{Piomelli_Balaras_ARFM_2002}%
  \BibitemOpen
  \bibfield  {author} {\bibinfo {author} {\bibfnamefont {U.}~\bibnamefont
  {Piomelli}}\ and\ \bibinfo {author} {\bibfnamefont {E.}~\bibnamefont
  {Balaras}},\ }\bibfield  {title} {\enquote {\bibinfo {title} {Wall-layer
  models for large-eddy simulations},}\ }\href {\doibase
  10.1146/annurev.fluid.34.082901.144919} {\bibfield  {journal} {\bibinfo
  {journal} {Annu. Rev. Fluid Mech.}\ }\textbf {\bibinfo {volume} {34}},\
  \bibinfo {pages} {349--374} (\bibinfo {year} {2002})}\BibitemShut {NoStop}%
\bibitem [{\citenamefont {Cabot}\ and\ \citenamefont
  {Moin}(2000)}]{Cabot_Moin_FTC_2000}%
  \BibitemOpen
  \bibfield  {author} {\bibinfo {author} {\bibfnamefont {W.}~\bibnamefont
  {Cabot}}\ and\ \bibinfo {author} {\bibfnamefont {P.}~\bibnamefont {Moin}},\
  }\bibfield  {title} {\enquote {\bibinfo {title} {Approximate wall boundary
  conditions in the large-eddy simulation of high {R}eynolds number flow},}\
  }\href {\doibase 10.1023/A:1009958917113} {\bibfield  {journal} {\bibinfo
  {journal} {Flow Turbul. Combust.}\ }\textbf {\bibinfo {volume} {63}},\
  \bibinfo {pages} {269--291} (\bibinfo {year} {2000})}\BibitemShut {NoStop}%
\bibitem [{\citenamefont {Davidson}(2009)}]{Davidson_IJHF_2009}%
  \BibitemOpen
  \bibfield  {author} {\bibinfo {author} {\bibfnamefont {L.}~\bibnamefont
  {Davidson}},\ }\bibfield  {title} {\enquote {\bibinfo {title} {Large eddy
  simulations: {H}ow to evaluate resolution},}\ }\href {\doibase
  10.1016/j.ijheatfluidflow.2009.06.006} {\bibfield  {journal} {\bibinfo
  {journal} {Int. J. Heat Fluid Flow}\ }\textbf {\bibinfo {volume} {30}},\
  \bibinfo {pages} {1016--1025} (\bibinfo {year} {2009})}\BibitemShut {NoStop}%
\bibitem [{\citenamefont {Piomelli}(2008)}]{Piomelli_PAS_2008}%
  \BibitemOpen
  \bibfield  {author} {\bibinfo {author} {\bibfnamefont {U.}~\bibnamefont
  {Piomelli}},\ }\bibfield  {title} {\enquote {\bibinfo {title} {Wall-layer
  models for large-eddy simulations},}\ }\href {\doibase
  10.1016/j.paerosci.2008.06.001} {\bibfield  {journal} {\bibinfo  {journal}
  {Prog. Aerosp. Sci.}\ }\textbf {\bibinfo {volume} {44}},\ \bibinfo {pages}
  {437--446} (\bibinfo {year} {2008})}\BibitemShut {NoStop}%
\bibitem [{\citenamefont {Bose}\ and\ \citenamefont
  {Park}(2018)}]{Bose_Park_ARFM_2018}%
  \BibitemOpen
  \bibfield  {author} {\bibinfo {author} {\bibfnamefont {S.~T.}\ \bibnamefont
  {Bose}}\ and\ \bibinfo {author} {\bibfnamefont {G.~I.}\ \bibnamefont
  {Park}},\ }\bibfield  {title} {\enquote {\bibinfo {title} {Wall-modeled
  large-eddy simulation for complex turbulent flows},}\ }\href {\doibase
  10.1146/annurev-fluid-122316-045241} {\bibfield  {journal} {\bibinfo
  {journal} {Annu. Rev. Fluid Mech.}\ }\textbf {\bibinfo {volume} {50}},\
  \bibinfo {pages} {535--561} (\bibinfo {year} {2018})}\BibitemShut {NoStop}%
\bibitem [{\citenamefont {Duraisamy}\ \emph {et~al.}(2019)\citenamefont
  {Duraisamy}, \citenamefont {Iaccarino},\ and\ \citenamefont
  {Xiao}}]{Duraisamy_etal_ARFM_2019}%
  \BibitemOpen
  \bibfield  {author} {\bibinfo {author} {\bibfnamefont {K.}~\bibnamefont
  {Duraisamy}}, \bibinfo {author} {\bibfnamefont {G.}~\bibnamefont
  {Iaccarino}}, \ and\ \bibinfo {author} {\bibfnamefont {H.}~\bibnamefont
  {Xiao}},\ }\bibfield  {title} {\enquote {\bibinfo {title} {Turbulence
  modeling in the age of data},}\ }\href {\doibase
  10.1146/annurev-fluid-010518-040547} {\bibfield  {journal} {\bibinfo
  {journal} {Annu. Rev. Fluid Mech.}\ }\textbf {\bibinfo {volume} {51}},\
  \bibinfo {pages} {357--377} (\bibinfo {year} {2019})}\BibitemShut {NoStop}%
\bibitem [{\citenamefont {Brunton}\ \emph {et~al.}(2020)\citenamefont
  {Brunton}, \citenamefont {Noack},\ and\ \citenamefont
  {Koumoutsakos}}]{Brunton_etal_ARFM_2019}%
  \BibitemOpen
  \bibfield  {author} {\bibinfo {author} {\bibfnamefont {S.~L.}\ \bibnamefont
  {Brunton}}, \bibinfo {author} {\bibfnamefont {B.~R.}\ \bibnamefont {Noack}},
  \ and\ \bibinfo {author} {\bibfnamefont {P.}~\bibnamefont {Koumoutsakos}},\
  }\bibfield  {title} {\enquote {\bibinfo {title} {Machine learning for fluid
  mechanics},}\ }\href {\doibase 10.1146/annurev-fluid-010719-060214}
  {\bibfield  {journal} {\bibinfo  {journal} {Annu. Rev. Fluid Mech.}\ }\textbf
  {\bibinfo {volume} {52}},\ \bibinfo {pages} {477--508} (\bibinfo {year}
  {2020})}\BibitemShut {NoStop}%
\bibitem [{\citenamefont {Pope}(2000)}]{Pope_Cambridge_2000}%
  \BibitemOpen
  \bibfield  {author} {\bibinfo {author} {\bibfnamefont {S.~B.}\ \bibnamefont
  {Pope}},\ }\href {\doibase 10.1017/CBO9780511840531} {\emph {\bibinfo {title}
  {Turbulent flows}}}\ (\bibinfo  {publisher} {Cambridge University Press,
  Cambridge},\ \bibinfo {year} {2000})\BibitemShut {NoStop}%
\bibitem [{\citenamefont {Wang}\ and\ \citenamefont
  {Moin}(2002)}]{Wang_Moin_PoF_2002}%
  \BibitemOpen
  \bibfield  {author} {\bibinfo {author} {\bibfnamefont {M.}~\bibnamefont
  {Wang}}\ and\ \bibinfo {author} {\bibfnamefont {P.}~\bibnamefont {Moin}},\
  }\bibfield  {title} {\enquote {\bibinfo {title} {Dynamic wall modeling for
  large-eddy simulation of complex turbulent flows},}\ }\href {\doibase
  10.1063/1.1476668} {\bibfield  {journal} {\bibinfo  {journal} {Phys. Fluids}\
  }\textbf {\bibinfo {volume} {14}},\ \bibinfo {pages} {2043} (\bibinfo {year}
  {2002})}\BibitemShut {NoStop}%
\bibitem [{\citenamefont {Kawai}\ and\ \citenamefont
  {Larsson}(2013)}]{Kawai_Larsson_PoF_2013}%
  \BibitemOpen
  \bibfield  {author} {\bibinfo {author} {\bibfnamefont {S.}~\bibnamefont
  {Kawai}}\ and\ \bibinfo {author} {\bibfnamefont {J.}~\bibnamefont
  {Larsson}},\ }\bibfield  {title} {\enquote {\bibinfo {title} {Dynamic
  non-equilibrium wall-modeling for large eddy simulation at high {R}eynolds
  numbers},}\ }\href {\doibase 10.1063/1.4775363} {\bibfield  {journal}
  {\bibinfo  {journal} {Phys. Fluids}\ }\textbf {\bibinfo {volume} {25}},\
  \bibinfo {pages} {015105} (\bibinfo {year} {2013})}\BibitemShut {NoStop}%
\bibitem [{\citenamefont {Park}\ and\ \citenamefont
  {Moin}(2014)}]{Park_Moin_PoF_2014}%
  \BibitemOpen
  \bibfield  {author} {\bibinfo {author} {\bibfnamefont {G.~I.}\ \bibnamefont
  {Park}}\ and\ \bibinfo {author} {\bibfnamefont {P.}~\bibnamefont {Moin}},\
  }\bibfield  {title} {\enquote {\bibinfo {title} {An improved dynamic
  non-equilibrium wall-model for large eddy simulation},}\ }\href {\doibase
  10.1063/1.4861069} {\bibfield  {journal} {\bibinfo  {journal} {Phys. Fluids}\
  }\textbf {\bibinfo {volume} {26}},\ \bibinfo {pages} {015108} (\bibinfo
  {year} {2014})}\BibitemShut {NoStop}%
\bibitem [{\citenamefont {Yang}\ \emph
  {et~al.}(2015{\natexlab{a}})\citenamefont {Yang}, \citenamefont {Sadique},
  \citenamefont {Mittal},\ and\ \citenamefont {C.}}]{Yang_etal_PoF_2015}%
  \BibitemOpen
  \bibfield  {author} {\bibinfo {author} {\bibfnamefont {X.~I.~A.}\
  \bibnamefont {Yang}}, \bibinfo {author} {\bibfnamefont {J.}~\bibnamefont
  {Sadique}}, \bibinfo {author} {\bibfnamefont {R.}~\bibnamefont {Mittal}}, \
  and\ \bibinfo {author} {\bibfnamefont {Meneveau}\ \bibnamefont {C.}},\
  }\bibfield  {title} {\enquote {\bibinfo {title} {Integral wall model for
  large eddy simulations of wall-bounded turbulent flows},}\ }\href {\doibase
  10.1063/1.4908072} {\bibfield  {journal} {\bibinfo  {journal} {Phys. Fluids}\
  }\textbf {\bibinfo {volume} {27}},\ \bibinfo {pages} {025112} (\bibinfo
  {year} {2015}{\natexlab{a}})}\BibitemShut {NoStop}%
\bibitem [{\citenamefont {Chung}\ and\ \citenamefont
  {Pullin}(2009)}]{Chung_Pullin_JFM_2009}%
  \BibitemOpen
  \bibfield  {author} {\bibinfo {author} {\bibfnamefont {D.}~\bibnamefont
  {Chung}}\ and\ \bibinfo {author} {\bibfnamefont {D.~I.}\ \bibnamefont
  {Pullin}},\ }\bibfield  {title} {\enquote {\bibinfo {title} {Large-eddy
  simulation and wall modelling of turbulent channel flow},}\ }\href {\doibase
  10.1017/S0022112009006867} {\bibfield  {journal} {\bibinfo  {journal} {J.
  Fluid Mech.}\ }\textbf {\bibinfo {volume} {631}},\ \bibinfo {pages}
  {281--309} (\bibinfo {year} {2009})}\BibitemShut {NoStop}%
\bibitem [{\citenamefont {Cheng}\ \emph {et~al.}(2015)\citenamefont {Cheng},
  \citenamefont {Pullin},\ and\ \citenamefont
  {Samtaney}}]{Cheng_etal_JFM_2015}%
  \BibitemOpen
  \bibfield  {author} {\bibinfo {author} {\bibfnamefont {W.}~\bibnamefont
  {Cheng}}, \bibinfo {author} {\bibfnamefont {D.~I.}\ \bibnamefont {Pullin}}, \
  and\ \bibinfo {author} {\bibfnamefont {R.}~\bibnamefont {Samtaney}},\
  }\bibfield  {title} {\enquote {\bibinfo {title} {Large-eddy simulation of
  separation and reattachment of a flat plate turbulent boundary layer},}\
  }\href {\doibase 10.1017/jfm.2015.604} {\bibfield  {journal} {\bibinfo
  {journal} {J. Fluid Mech.}\ }\textbf {\bibinfo {volume} {785}},\ \bibinfo
  {pages} {78--108} (\bibinfo {year} {2015})}\BibitemShut {NoStop}%
\bibitem [{\citenamefont {Bose}\ and\ \citenamefont
  {Moin}(2014)}]{Bose_Moin_PoF_2014}%
  \BibitemOpen
  \bibfield  {author} {\bibinfo {author} {\bibfnamefont {S.~T.}\ \bibnamefont
  {Bose}}\ and\ \bibinfo {author} {\bibfnamefont {P.}~\bibnamefont {Moin}},\
  }\bibfield  {title} {\enquote {\bibinfo {title} {A dynamic slip boundary
  condition for wall-modeled large-eddy simulation},}\ }\href {\doibase
  10.1063/1.4849535} {\bibfield  {journal} {\bibinfo  {journal} {Phys. Fluids}\
  }\textbf {\bibinfo {volume} {26}},\ \bibinfo {pages} {015104} (\bibinfo
  {year} {2014})}\BibitemShut {NoStop}%
\bibitem [{\citenamefont {Bae}\ \emph {et~al.}(2018)\citenamefont {Bae},
  \citenamefont {Lozano-Dur{\' a}n}, \citenamefont {Bose},\ and\ \citenamefont
  {Moin}}]{Bae_etal_PRF_2018}%
  \BibitemOpen
  \bibfield  {author} {\bibinfo {author} {\bibfnamefont {H.~J.}\ \bibnamefont
  {Bae}}, \bibinfo {author} {\bibfnamefont {A.}~\bibnamefont {Lozano-Dur{\'
  a}n}}, \bibinfo {author} {\bibfnamefont {S.~T.}\ \bibnamefont {Bose}}, \ and\
  \bibinfo {author} {\bibfnamefont {P.}~\bibnamefont {Moin}},\ }\bibfield
  {title} {\enquote {\bibinfo {title} {Turbulence intensities in large-eddy
  simulation of wall-bounded flows},}\ }\href {\doibase
  10.1103/PhysRevFluids.3.014610} {\bibfield  {journal} {\bibinfo  {journal}
  {Phys. Rev. Fluids}\ }\textbf {\bibinfo {volume} {3}},\ \bibinfo {pages}
  {014610} (\bibinfo {year} {2018})}\BibitemShut {NoStop}%
\bibitem [{\citenamefont {Larsson}\ \emph {et~al.}(2016)\citenamefont
  {Larsson}, \citenamefont {Kawai}, \citenamefont {Bodart},\ and\ \citenamefont
  {Bermejo-Moreno}}]{Larsson_etal_MER_2016}%
  \BibitemOpen
  \bibfield  {author} {\bibinfo {author} {\bibfnamefont {J.}~\bibnamefont
  {Larsson}}, \bibinfo {author} {\bibfnamefont {S.}~\bibnamefont {Kawai}},
  \bibinfo {author} {\bibfnamefont {J.}~\bibnamefont {Bodart}}, \ and\ \bibinfo
  {author} {\bibfnamefont {I.}~\bibnamefont {Bermejo-Moreno}},\ }\bibfield
  {title} {\enquote {\bibinfo {title} {Large eddy simulation with modeled
  wall-stress: {R}ecent progress and future directions},}\ }\href {\doibase
  10.1299/mer.15-00418} {\bibfield  {journal} {\bibinfo  {journal} {Mechanical
  Engineering Reviews}\ }\textbf {\bibinfo {volume} {3}},\ \bibinfo {pages}
  {15--00418} (\bibinfo {year} {2016})}\BibitemShut {NoStop}%
\bibitem [{\citenamefont {Frere}\ \emph {et~al.}(2018)\citenamefont {Frere},
  \citenamefont {Hillewaert}, \citenamefont {Chatelain},\ and\ \citenamefont
  {Winckelmans}}]{Frere_etal_FTC_2018}%
  \BibitemOpen
  \bibfield  {author} {\bibinfo {author} {\bibfnamefont {A.}~\bibnamefont
  {Frere}}, \bibinfo {author} {\bibfnamefont {K.}~\bibnamefont {Hillewaert}},
  \bibinfo {author} {\bibfnamefont {L.~P.}\ \bibnamefont {Chatelain}}, \ and\
  \bibinfo {author} {\bibfnamefont {G.}~\bibnamefont {Winckelmans}},\
  }\bibfield  {title} {\enquote {\bibinfo {title} {High {R}eynolds number
  airfoil: {F}rom wall-resolved to wall-modeled {LES}},}\ }\href {\doibase
  10.1007/s10494-018-9972-9} {\bibfield  {journal} {\bibinfo  {journal} {Flow
  Turbul. Combust.}\ }\textbf {\bibinfo {volume} {101}},\ \bibinfo {pages}
  {457--476} (\bibinfo {year} {2018})}\BibitemShut {NoStop}%
\bibitem [{\citenamefont {Bae}\ \emph {et~al.}(2019)\citenamefont {Bae},
  \citenamefont {Lozano-Duran}, \citenamefont {Bose},\ and\ \citenamefont
  {Moin}}]{Bae_etal_JFM_2019}%
  \BibitemOpen
  \bibfield  {author} {\bibinfo {author} {\bibfnamefont {H.~J.}\ \bibnamefont
  {Bae}}, \bibinfo {author} {\bibfnamefont {A.}~\bibnamefont {Lozano-Duran}},
  \bibinfo {author} {\bibfnamefont {S.~T.}\ \bibnamefont {Bose}}, \ and\
  \bibinfo {author} {\bibfnamefont {P.}~\bibnamefont {Moin}},\ }\bibfield
  {title} {\enquote {\bibinfo {title} {Dynamic slip wall model for large-eddy
  simulation},}\ }\href {\doibase 10.1017/jfm.2018.838} {\bibfield  {journal}
  {\bibinfo  {journal} {J. Fluid Mech.}\ }\textbf {\bibinfo {volume} {859}},\
  \bibinfo {pages} {400--432} (\bibinfo {year} {2019})}\BibitemShut {NoStop}%
\bibitem [{\citenamefont {Ling}\ \emph
  {et~al.}(2016{\natexlab{a}})\citenamefont {Ling}, \citenamefont {Kurzawski},\
  and\ \citenamefont {Templeton}}]{Ling_etal_JFM_2016}%
  \BibitemOpen
  \bibfield  {author} {\bibinfo {author} {\bibfnamefont {J.}~\bibnamefont
  {Ling}}, \bibinfo {author} {\bibfnamefont {A.}~\bibnamefont {Kurzawski}}, \
  and\ \bibinfo {author} {\bibfnamefont {J.}~\bibnamefont {Templeton}},\
  }\bibfield  {title} {\enquote {\bibinfo {title} {Reynolds averaged turbulence
  modelling using deep neural networks with embedded invariance},}\ }\href
  {\doibase 10.1017/jfm.2016.615} {\bibfield  {journal} {\bibinfo  {journal}
  {J. Fluid Mech.}\ }\textbf {\bibinfo {volume} {807}},\ \bibinfo {pages}
  {155--166} (\bibinfo {year} {2016}{\natexlab{a}})}\BibitemShut {NoStop}%
\bibitem [{\citenamefont {Zhou}\ \emph {et~al.}(2019)\citenamefont {Zhou},
  \citenamefont {He}, \citenamefont {Wang},\ and\ \citenamefont
  {Jin}}]{Zhou_etal_CAF_2019}%
  \BibitemOpen
  \bibfield  {author} {\bibinfo {author} {\bibfnamefont {Z.~D.}\ \bibnamefont
  {Zhou}}, \bibinfo {author} {\bibfnamefont {G.~W.}\ \bibnamefont {He}},
  \bibinfo {author} {\bibfnamefont {S.~Z.}\ \bibnamefont {Wang}}, \ and\
  \bibinfo {author} {\bibfnamefont {G.~D.}\ \bibnamefont {Jin}},\ }\bibfield
  {title} {\enquote {\bibinfo {title} {Subgrid-scale model for large-eddy
  simulation of isotropic turbulent flows using an artificial neural
  network},}\ }\href {\doibase 10.1016/j.compfluid.2019.104319} {\bibfield
  {journal} {\bibinfo  {journal} {Comput. Fluids}\ }\textbf {\bibinfo {volume}
  {195}},\ \bibinfo {pages} {104319} (\bibinfo {year} {2019})}\BibitemShut
  {NoStop}%
\bibitem [{\citenamefont {King}\ \emph {et~al.}(2018)\citenamefont {King},
  \citenamefont {Hennigh}, \citenamefont {Mohan},\ and\ \citenamefont
  {Chertkov}}]{King_etal_arXiv_2018}%
  \BibitemOpen
  \bibfield  {author} {\bibinfo {author} {\bibfnamefont {R.}~\bibnamefont
  {King}}, \bibinfo {author} {\bibfnamefont {O.}~\bibnamefont {Hennigh}},
  \bibinfo {author} {\bibfnamefont {A.}~\bibnamefont {Mohan}}, \ and\ \bibinfo
  {author} {\bibfnamefont {M.}~\bibnamefont {Chertkov}},\ }\bibfield  {title}
  {\enquote {\bibinfo {title} {From deep to physics-informed learning of
  turbulence: {D}iagnostics},}\ }\href@noop {} {\bibfield  {journal} {\bibinfo
  {journal} {arXiv:1810.07785v2}\ } (\bibinfo {year} {2018})}\BibitemShut
  {NoStop}%
\bibitem [{\citenamefont {Lee}\ and\ \citenamefont
  {You}(2019)}]{Lee_You_JFM_2019}%
  \BibitemOpen
  \bibfield  {author} {\bibinfo {author} {\bibfnamefont {S.}~\bibnamefont
  {Lee}}\ and\ \bibinfo {author} {\bibfnamefont {D.}~\bibnamefont {You}},\
  }\bibfield  {title} {\enquote {\bibinfo {title} {Data-driven prediction of
  unsteady flow over a circular cylinder using deep learning},}\ }\href
  {\doibase 10.1017/jfm.2019.700} {\bibfield  {journal} {\bibinfo  {journal}
  {J. Fluid Mech.}\ }\textbf {\bibinfo {volume} {879}},\ \bibinfo {pages}
  {217--254} (\bibinfo {year} {2019})}\BibitemShut {NoStop}%
\bibitem [{\citenamefont {Fukami}\ \emph {et~al.}(2019)\citenamefont {Fukami},
  \citenamefont {Fukagata},\ and\ \citenamefont
  {Taira}}]{Fukami_Fukagata_Taira_JFM_2019}%
  \BibitemOpen
  \bibfield  {author} {\bibinfo {author} {\bibfnamefont {K.}~\bibnamefont
  {Fukami}}, \bibinfo {author} {\bibfnamefont {K.}~\bibnamefont {Fukagata}}, \
  and\ \bibinfo {author} {\bibfnamefont {K.}~\bibnamefont {Taira}},\ }\bibfield
   {title} {\enquote {\bibinfo {title} {Super-resolution reconstruction of
  turbulent flows with machine learning},}\ }\href {\doibase
  10.1017/jfm.2019.238} {\bibfield  {journal} {\bibinfo  {journal} {J. Fluid
  Mech.}\ }\textbf {\bibinfo {volume} {870}},\ \bibinfo {pages} {106--120}
  (\bibinfo {year} {2019})}\BibitemShut {NoStop}%
\bibitem [{\citenamefont {Liu}\ \emph {et~al.}(2020)\citenamefont {Liu},
  \citenamefont {Tang}, \citenamefont {Huang},\ and\ \citenamefont
  {Lu}}]{Liu_etal_PoF_2020}%
  \BibitemOpen
  \bibfield  {author} {\bibinfo {author} {\bibfnamefont {B.}~\bibnamefont
  {Liu}}, \bibinfo {author} {\bibfnamefont {J.~P.}\ \bibnamefont {Tang}},
  \bibinfo {author} {\bibfnamefont {H.~B.}\ \bibnamefont {Huang}}, \ and\
  \bibinfo {author} {\bibfnamefont {X.~Y.}\ \bibnamefont {Lu}},\ }\bibfield
  {title} {\enquote {\bibinfo {title} {Deep learning methods for
  super-resolution reconstruction of turbulent flows},}\ }\href {\doibase
  10.1063/1.5140772} {\bibfield  {journal} {\bibinfo  {journal} {Phys. Fluids}\
  }\textbf {\bibinfo {volume} {32}},\ \bibinfo {pages} {025105} (\bibinfo
  {year} {2020})}\BibitemShut {NoStop}%
\bibitem [{\citenamefont {Ling}\ \emph
  {et~al.}(2016{\natexlab{b}})\citenamefont {Ling}, \citenamefont {Jones},\
  and\ \citenamefont {Templeton}}]{Ling_etal_JCP_2016}%
  \BibitemOpen
  \bibfield  {author} {\bibinfo {author} {\bibfnamefont {J.}~\bibnamefont
  {Ling}}, \bibinfo {author} {\bibfnamefont {R.}~\bibnamefont {Jones}}, \ and\
  \bibinfo {author} {\bibfnamefont {J.}~\bibnamefont {Templeton}},\ }\bibfield
  {title} {\enquote {\bibinfo {title} {Machine learning strategies for systems
  with invariance properties},}\ }\href {\doibase 10.1016/j.jcp.2016.05.003}
  {\bibfield  {journal} {\bibinfo  {journal} {J. Comput. Phys.}\ }\textbf
  {\bibinfo {volume} {318}},\ \bibinfo {pages} {22--35} (\bibinfo {year}
  {2016}{\natexlab{b}})}\BibitemShut {NoStop}%
\bibitem [{\citenamefont {Wang}\ \emph {et~al.}(2017)\citenamefont {Wang},
  \citenamefont {Wu},\ and\ \citenamefont {Xiao}}]{Wang_Wu_Xiao_PRF_2017}%
  \BibitemOpen
  \bibfield  {author} {\bibinfo {author} {\bibfnamefont {J.-X.}\ \bibnamefont
  {Wang}}, \bibinfo {author} {\bibfnamefont {J.-L.}\ \bibnamefont {Wu}}, \ and\
  \bibinfo {author} {\bibfnamefont {H.}~\bibnamefont {Xiao}},\ }\bibfield
  {title} {\enquote {\bibinfo {title} {Physics-informed machine learning
  approach for reconstructing {R}eynolds stress modeling discrepancies based on
  {DNS} data},}\ }\href {\doibase 10.1103/PhysRevFluids.2.034603} {\bibfield
  {journal} {\bibinfo  {journal} {Phys. Rev. Fluids}\ }\textbf {\bibinfo
  {volume} {2}},\ \bibinfo {pages} {034603} (\bibinfo {year}
  {2017})}\BibitemShut {NoStop}%
\bibitem [{\citenamefont {Wu}\ \emph {et~al.}(2018)\citenamefont {Wu},
  \citenamefont {Xiao},\ and\ \citenamefont
  {Paterson}}]{Wu_Xiao_Paterson_2018}%
  \BibitemOpen
  \bibfield  {author} {\bibinfo {author} {\bibfnamefont {J.-L.}\ \bibnamefont
  {Wu}}, \bibinfo {author} {\bibfnamefont {H.}~\bibnamefont {Xiao}}, \ and\
  \bibinfo {author} {\bibfnamefont {E.}~\bibnamefont {Paterson}},\ }\bibfield
  {title} {\enquote {\bibinfo {title} {Physics-informed machine learning
  approach for augmenting turbulence models: {A} comprehensive framework},}\
  }\href {\doibase 10.1103/PhysRevFluids.3.074602} {\bibfield  {journal}
  {\bibinfo  {journal} {Phys. Rev. Fluids}\ }\textbf {\bibinfo {volume} {3}},\
  \bibinfo {pages} {074602} (\bibinfo {year} {2018})}\BibitemShut {NoStop}%
\bibitem [{\citenamefont {Gamahara}\ and\ \citenamefont
  {Hattori}(2017)}]{Gamahara_Hattori_PRF_2017}%
  \BibitemOpen
  \bibfield  {author} {\bibinfo {author} {\bibfnamefont {M.}~\bibnamefont
  {Gamahara}}\ and\ \bibinfo {author} {\bibfnamefont {Y.}~\bibnamefont
  {Hattori}},\ }\bibfield  {title} {\enquote {\bibinfo {title} {Searching for
  turbulence models by artificial neural network},}\ }\href {\doibase
  10.1103/PhysRevFluids.2.054604} {\bibfield  {journal} {\bibinfo  {journal}
  {Phys. Rev. Fluids}\ }\textbf {\bibinfo {volume} {2}},\ \bibinfo {pages}
  {054604} (\bibinfo {year} {2017})}\BibitemShut {NoStop}%
\bibitem [{\citenamefont {Maulik}\ \emph {et~al.}(2018)\citenamefont {Maulik},
  \citenamefont {San}, \citenamefont {Rasheed},\ and\ \citenamefont
  {Vedula}}]{Maulik_etal_PoF_2018}%
  \BibitemOpen
  \bibfield  {author} {\bibinfo {author} {\bibfnamefont {R.}~\bibnamefont
  {Maulik}}, \bibinfo {author} {\bibfnamefont {O.}~\bibnamefont {San}},
  \bibinfo {author} {\bibfnamefont {A.}~\bibnamefont {Rasheed}}, \ and\
  \bibinfo {author} {\bibfnamefont {P.}~\bibnamefont {Vedula}},\ }\bibfield
  {title} {\enquote {\bibinfo {title} {Data-driven deconvolution for large eddy
  simulations of {K}raichnan turbulence},}\ }\href {\doibase 10.1063/1.5079582}
  {\bibfield  {journal} {\bibinfo  {journal} {Phys. Fluids}\ }\textbf {\bibinfo
  {volume} {30}},\ \bibinfo {pages} {125109} (\bibinfo {year}
  {2018})}\BibitemShut {NoStop}%
\bibitem [{\citenamefont {Vollant}\ \emph {et~al.}(2017)\citenamefont
  {Vollant}, \citenamefont {Balarac},\ and\ \citenamefont
  {Corre}}]{Vollant_etal_JOT_2017}%
  \BibitemOpen
  \bibfield  {author} {\bibinfo {author} {\bibfnamefont {A.}~\bibnamefont
  {Vollant}}, \bibinfo {author} {\bibfnamefont {G.}~\bibnamefont {Balarac}}, \
  and\ \bibinfo {author} {\bibfnamefont {C.}~\bibnamefont {Corre}},\ }\bibfield
   {title} {\enquote {\bibinfo {title} {Subgrid-scale scalar flux modelling
  based on optimal estimation theory and machine-learning procedures},}\ }\href
  {\doibase 10.1080/14685248.2017.1334907} {\bibfield  {journal} {\bibinfo
  {journal} {J. Turbul.}\ }\textbf {\bibinfo {volume} {18}},\ \bibinfo {pages}
  {854--878} (\bibinfo {year} {2017})}\BibitemShut {NoStop}%
\bibitem [{\citenamefont {Milano}\ and\ \citenamefont
  {Koumoutsakos}(2002)}]{Milano_Koumoutsakos_JCP_2002}%
  \BibitemOpen
  \bibfield  {author} {\bibinfo {author} {\bibfnamefont {M.}~\bibnamefont
  {Milano}}\ and\ \bibinfo {author} {\bibfnamefont {P.}~\bibnamefont
  {Koumoutsakos}},\ }\bibfield  {title} {\enquote {\bibinfo {title} {Neural
  network modeling for near wall turbulent flow},}\ }\href {\doibase
  10.1006/jcph.2002.7146} {\bibfield  {journal} {\bibinfo  {journal} {J.
  Comput. Phys.}\ }\textbf {\bibinfo {volume} {182}},\ \bibinfo {pages} {1--26}
  (\bibinfo {year} {2002})}\BibitemShut {NoStop}%
\bibitem [{\citenamefont {Yang}\ \emph {et~al.}(2019)\citenamefont {Yang},
  \citenamefont {Zafar}, \citenamefont {Wang},\ and\ \citenamefont
  {Xiao}}]{Yang_etal_PRF_2019}%
  \BibitemOpen
  \bibfield  {author} {\bibinfo {author} {\bibfnamefont {X.~I.~A.}\
  \bibnamefont {Yang}}, \bibinfo {author} {\bibfnamefont {S.}~\bibnamefont
  {Zafar}}, \bibinfo {author} {\bibfnamefont {J.-X.}\ \bibnamefont {Wang}}, \
  and\ \bibinfo {author} {\bibfnamefont {H.}~\bibnamefont {Xiao}},\ }\bibfield
  {title} {\enquote {\bibinfo {title} {Predictive large-eddy-simulation wall
  modeling via physics-informed neural networks},}\ }\href {\doibase
  10.1103/PhysRevFluids.4.034602} {\bibfield  {journal} {\bibinfo  {journal}
  {Phys. Rev. Fluids}\ }\textbf {\bibinfo {volume} {4}},\ \bibinfo {pages}
  {034602} (\bibinfo {year} {2019})}\BibitemShut {NoStop}%
\bibitem [{\citenamefont {Huang}\ \emph {et~al.}(2019)\citenamefont {Huang},
  \citenamefont {Yang},\ and\ \citenamefont {Kunz}}]{Huang_etal_PoF_2019}%
  \BibitemOpen
  \bibfield  {author} {\bibinfo {author} {\bibfnamefont {X.~L.~D.}\
  \bibnamefont {Huang}}, \bibinfo {author} {\bibfnamefont {X.~I.~A.}\
  \bibnamefont {Yang}}, \ and\ \bibinfo {author} {\bibfnamefont {R.~F.}\
  \bibnamefont {Kunz}},\ }\bibfield  {title} {\enquote {\bibinfo {title}
  {Wall-modeled large-eddy simulations of spanwise rotating turbulent
  channels-{C}omparing a physics-based approach and a data-based approach},}\
  }\href {\doibase 10.1063/1.5129178} {\bibfield  {journal} {\bibinfo
  {journal} {Phys. Fluids}\ }\textbf {\bibinfo {volume} {31}},\ \bibinfo
  {pages} {125105} (\bibinfo {year} {2019})}\BibitemShut {NoStop}%
\bibitem [{\citenamefont {Fr{\"o}hlich}\ \emph {et~al.}(2005)\citenamefont
  {Fr{\"o}hlich}, \citenamefont {Mellen}, \citenamefont {Rodi}, \citenamefont
  {Temmerman},\ and\ \citenamefont {Leschziner}}]{Frohlich_etal_JFM_2005}%
  \BibitemOpen
  \bibfield  {author} {\bibinfo {author} {\bibfnamefont {J.}~\bibnamefont
  {Fr{\"o}hlich}}, \bibinfo {author} {\bibfnamefont {C.~P.}\ \bibnamefont
  {Mellen}}, \bibinfo {author} {\bibfnamefont {W.}~\bibnamefont {Rodi}},
  \bibinfo {author} {\bibfnamefont {L.}~\bibnamefont {Temmerman}}, \ and\
  \bibinfo {author} {\bibfnamefont {M.~A.}\ \bibnamefont {Leschziner}},\
  }\bibfield  {title} {\enquote {\bibinfo {title} {Highly resolved large-eddy
  simulation of separated flow in a channel with streamwise periodic
  constrictions},}\ }\href {\doibase 10.1017/S0022112004002812} {\bibfield
  {journal} {\bibinfo  {journal} {J. Fluid Mech.}\ }\textbf {\bibinfo {volume}
  {526}},\ \bibinfo {pages} {19--66} (\bibinfo {year} {2005})}\BibitemShut
  {NoStop}%
\bibitem [{\citenamefont {Temmerman}\ \emph {et~al.}(2003)\citenamefont
  {Temmerman}, \citenamefont {Leschziner}, \citenamefont {Mellen},\ and\
  \citenamefont {Fr{\"o}hlich}}]{Temmerman_etal_IJHF_2003}%
  \BibitemOpen
  \bibfield  {author} {\bibinfo {author} {\bibfnamefont {L.}~\bibnamefont
  {Temmerman}}, \bibinfo {author} {\bibfnamefont {M.~A.}\ \bibnamefont
  {Leschziner}}, \bibinfo {author} {\bibfnamefont {C.~P.}\ \bibnamefont
  {Mellen}}, \ and\ \bibinfo {author} {\bibfnamefont {J.}~\bibnamefont
  {Fr{\"o}hlich}},\ }\bibfield  {title} {\enquote {\bibinfo {title}
  {Investigation of wall-function approximations and subgrid-scale models in
  large eddy simulation of separated flow in a channel with streamwise periodic
  constrictions},}\ }\href {\doibase 10.1016/S0142-727X(02)00222-9} {\bibfield
  {journal} {\bibinfo  {journal} {Int. J. Heat Fluid Flow}\ }\textbf {\bibinfo
  {volume} {24}},\ \bibinfo {pages} {157--180} (\bibinfo {year}
  {2003})}\BibitemShut {NoStop}%
\bibitem [{\citenamefont {Manhart}\ \emph {et~al.}(2008)\citenamefont
  {Manhart}, \citenamefont {Peller},\ and\ \citenamefont
  {Brun}}]{Manhart_etal_TCFD_2008}%
  \BibitemOpen
  \bibfield  {author} {\bibinfo {author} {\bibfnamefont {M.}~\bibnamefont
  {Manhart}}, \bibinfo {author} {\bibfnamefont {N.}~\bibnamefont {Peller}}, \
  and\ \bibinfo {author} {\bibfnamefont {C.}~\bibnamefont {Brun}},\ }\bibfield
  {title} {\enquote {\bibinfo {title} {Near-wall scaling for turbulent boundary
  layers with adverse pressure gradient},}\ }\href {\doibase
  10.1007/s00162-007-0055-0} {\bibfield  {journal} {\bibinfo  {journal} {Theor.
  Comput. Fluid Dyn.}\ }\textbf {\bibinfo {volume} {22}},\ \bibinfo {pages}
  {243--260} (\bibinfo {year} {2008})}\BibitemShut {NoStop}%
\bibitem [{\citenamefont {Duprat}\ \emph {et~al.}(2011)\citenamefont {Duprat},
  \citenamefont {Balarac}, \citenamefont {M{\' e}tais}, \citenamefont
  {Congedo},\ and\ \citenamefont {O.}}]{Duprat_etal_PoF_2011}%
  \BibitemOpen
  \bibfield  {author} {\bibinfo {author} {\bibfnamefont {C.}~\bibnamefont
  {Duprat}}, \bibinfo {author} {\bibfnamefont {G.}~\bibnamefont {Balarac}},
  \bibinfo {author} {\bibfnamefont {O.}~\bibnamefont {M{\' e}tais}}, \bibinfo
  {author} {\bibfnamefont {P.~M.}\ \bibnamefont {Congedo}}, \ and\ \bibinfo
  {author} {\bibfnamefont {Brugi{\` e}re}\ \bibnamefont {O.}},\ }\bibfield
  {title} {\enquote {\bibinfo {title} {A wall-layer model for large-eddy
  simulations of turbulent flows {with/out} pressure gradient},}\ }\href
  {\doibase 10.1063/1.3529358} {\bibfield  {journal} {\bibinfo  {journal}
  {Phys. Fluids}\ }\textbf {\bibinfo {volume} {23}},\ \bibinfo {pages} {015101}
  (\bibinfo {year} {2011})}\BibitemShut {NoStop}%
\bibitem [{\citenamefont {Breuer}\ \emph {et~al.}(2009)\citenamefont {Breuer},
  \citenamefont {Peller}, \citenamefont {Rapp},\ and\ \citenamefont
  {Manhart}}]{Breuer_etal_CAF_2009}%
  \BibitemOpen
  \bibfield  {author} {\bibinfo {author} {\bibfnamefont {M.}~\bibnamefont
  {Breuer}}, \bibinfo {author} {\bibfnamefont {N.}~\bibnamefont {Peller}},
  \bibinfo {author} {\bibfnamefont {Ch.}\ \bibnamefont {Rapp}}, \ and\ \bibinfo
  {author} {\bibfnamefont {M.}~\bibnamefont {Manhart}},\ }\bibfield  {title}
  {\enquote {\bibinfo {title} {Flow over periodic hills - {N}umerical and
  experimental study in a wide range of {R}eynolds numbers},}\ }\href {\doibase
  10.1016/j.compfluid.2008.05.002} {\bibfield  {journal} {\bibinfo  {journal}
  {Comput. Fluids}\ }\textbf {\bibinfo {volume} {38}},\ \bibinfo {pages}
  {433--457} (\bibinfo {year} {2009})}\BibitemShut {NoStop}%
\bibitem [{\citenamefont {Rapp}\ and\ \citenamefont
  {Manhart}(2011)}]{Rapp_Manhart_EF_2011}%
  \BibitemOpen
  \bibfield  {author} {\bibinfo {author} {\bibfnamefont {Ch.}\ \bibnamefont
  {Rapp}}\ and\ \bibinfo {author} {\bibfnamefont {M.}~\bibnamefont {Manhart}},\
  }\bibfield  {title} {\enquote {\bibinfo {title} {Flow over periodic hills: an
  experimental study},}\ }\href {\doibase 10.1007/s00348-011-1045-y} {\bibfield
   {journal} {\bibinfo  {journal} {Exp. Fluids}\ }\textbf {\bibinfo {volume}
  {51}},\ \bibinfo {pages} {247--269} (\bibinfo {year} {2011})}\BibitemShut
  {NoStop}%
\bibitem [{\citenamefont {Krank}\ \emph {et~al.}(2018)\citenamefont {Krank},
  \citenamefont {Kronbichler},\ and\ \citenamefont
  {Wall}}]{Krank_etal_FTC_2018}%
  \BibitemOpen
  \bibfield  {author} {\bibinfo {author} {\bibfnamefont {B.}~\bibnamefont
  {Krank}}, \bibinfo {author} {\bibfnamefont {M.}~\bibnamefont {Kronbichler}},
  \ and\ \bibinfo {author} {\bibfnamefont {W.~A.}\ \bibnamefont {Wall}},\
  }\bibfield  {title} {\enquote {\bibinfo {title} {Direct numerical simulation
  of flow over periodic hills up to ${R}e_h$=10,595},}\ }\href {\doibase
  10.1007/s10494-018-9941-3} {\bibfield  {journal} {\bibinfo  {journal} {Flow
  Turbul. Combust.}\ }\textbf {\bibinfo {volume} {101(2)}},\ \bibinfo {pages}
  {521--551} (\bibinfo {year} {2018})}\BibitemShut {NoStop}%
\bibitem [{\citenamefont {Xiao}\ \emph {et~al.}(2020)\citenamefont {Xiao},
  \citenamefont {Wu}, \citenamefont {Laizet},\ and\ \citenamefont
  {Duan}}]{Xiao_etal_CAF_2020}%
  \BibitemOpen
  \bibfield  {author} {\bibinfo {author} {\bibfnamefont {H.}~\bibnamefont
  {Xiao}}, \bibinfo {author} {\bibfnamefont {J.-L.}\ \bibnamefont {Wu}},
  \bibinfo {author} {\bibfnamefont {S.}~\bibnamefont {Laizet}}, \ and\ \bibinfo
  {author} {\bibfnamefont {L.}~\bibnamefont {Duan}},\ }\bibfield  {title}
  {\enquote {\bibinfo {title} {Flows over periodic hills of parameterized
  geometries: {A} dataset for data-driven turbulence modeling from direct
  simulations},}\ }\href {\doibase 10.1016/j.compfluid.2020.104431} {\bibfield
  {journal} {\bibinfo  {journal} {Comput. Fluids}\ }\textbf {\bibinfo {volume}
  {200}},\ \bibinfo {pages} {104431} (\bibinfo {year} {2020})}\BibitemShut
  {NoStop}%
\bibitem [{\citenamefont {Yang}\ \emph
  {et~al.}(2015{\natexlab{b}})\citenamefont {Yang}, \citenamefont
  {Sotiropoulos}, \citenamefont {Conzemius}, \citenamefont {Wachtler},\ and\
  \citenamefont {Strong}}]{Yang_etal_WE_2015}%
  \BibitemOpen
  \bibfield  {author} {\bibinfo {author} {\bibfnamefont {X.~L.}\ \bibnamefont
  {Yang}}, \bibinfo {author} {\bibfnamefont {F.}~\bibnamefont {Sotiropoulos}},
  \bibinfo {author} {\bibfnamefont {R.~J.}\ \bibnamefont {Conzemius}}, \bibinfo
  {author} {\bibfnamefont {J.~N.}\ \bibnamefont {Wachtler}}, \ and\ \bibinfo
  {author} {\bibfnamefont {M.~B.}\ \bibnamefont {Strong}},\ }\bibfield  {title}
  {\enquote {\bibinfo {title} {Large-eddy simulation of turbulent flow past
  wind turbines/farms: the {V}irtual {W}ind {S}imulator ({VWiS})},}\ }\href
  {\doibase 10.1002/we.1802} {\bibfield  {journal} {\bibinfo  {journal} {Wind
  Energy}\ }\textbf {\bibinfo {volume} {18(12)}},\ \bibinfo {pages}
  {2025--2045} (\bibinfo {year} {2015}{\natexlab{b}})}\BibitemShut {NoStop}%
\bibitem [{\citenamefont {Yang}\ and\ \citenamefont
  {Sotiropoulos}(2018)}]{Yang_Sotiropoulos_WE_2018}%
  \BibitemOpen
  \bibfield  {author} {\bibinfo {author} {\bibfnamefont {X.~L.}\ \bibnamefont
  {Yang}}\ and\ \bibinfo {author} {\bibfnamefont {F.}~\bibnamefont
  {Sotiropoulos}},\ }\bibfield  {title} {\enquote {\bibinfo {title} {A new
  class of actuator surface models for wind turbines},}\ }\href {\doibase
  10.1002/we.2162} {\bibfield  {journal} {\bibinfo  {journal} {Wind Energy}\
  }\textbf {\bibinfo {volume} {21(5)}},\ \bibinfo {pages} {285--302} (\bibinfo
  {year} {2018})}\BibitemShut {NoStop}%
\bibitem [{\citenamefont {Kang}\ and\ \citenamefont
  {Sotiropoulos}(2012)}]{Kang_Sotiropoulos_AWR_2012}%
  \BibitemOpen
  \bibfield  {author} {\bibinfo {author} {\bibfnamefont {S.}~\bibnamefont
  {Kang}}\ and\ \bibinfo {author} {\bibfnamefont {F.}~\bibnamefont
  {Sotiropoulos}},\ }\bibfield  {title} {\enquote {\bibinfo {title} {Numerical
  modeling of {3D} turbulent free surface flow in natural waterways},}\ }\href
  {\doibase 10.1016/j.advwatres.2012.01.012} {\bibfield  {journal} {\bibinfo
  {journal} {Adv. Water Resour.}\ }\textbf {\bibinfo {volume} {40}},\ \bibinfo
  {pages} {23--36} (\bibinfo {year} {2012})}\BibitemShut {NoStop}%
\bibitem [{\citenamefont {Kang}\ \emph {et~al.}(2014)\citenamefont {Kang},
  \citenamefont {Yang},\ and\ \citenamefont
  {Sotiropoulos}}]{Kang_etal_JFM_2014}%
  \BibitemOpen
  \bibfield  {author} {\bibinfo {author} {\bibfnamefont {S.}~\bibnamefont
  {Kang}}, \bibinfo {author} {\bibfnamefont {X.~L.}\ \bibnamefont {Yang}}, \
  and\ \bibinfo {author} {\bibfnamefont {F.}~\bibnamefont {Sotiropoulos}},\
  }\bibfield  {title} {\enquote {\bibinfo {title} {On the onset of wake
  meandering for an axial flow turbine in a turbulent open channel flow},}\
  }\href {\doibase 10.1017/jfm.2014.82} {\bibfield  {journal} {\bibinfo
  {journal} {J. Fluid Mech.}\ }\textbf {\bibinfo {volume} {744}},\ \bibinfo
  {pages} {376--403} (\bibinfo {year} {2014})}\BibitemShut {NoStop}%
\bibitem [{\citenamefont {Khosronejad}\ and\ \citenamefont
  {Sotiropoulos}(2014)}]{Khosronejad_Sotiropoulos_JFM_2014}%
  \BibitemOpen
  \bibfield  {author} {\bibinfo {author} {\bibfnamefont {A.}~\bibnamefont
  {Khosronejad}}\ and\ \bibinfo {author} {\bibfnamefont {F.}~\bibnamefont
  {Sotiropoulos}},\ }\bibfield  {title} {\enquote {\bibinfo {title} {Numerical
  simulation of sand waves in a turbulent open channel flow},}\ }\href
  {\doibase 10.1017/jfm.2014.335} {\bibfield  {journal} {\bibinfo  {journal}
  {J. Fluid Mech.}\ }\textbf {\bibinfo {volume} {753}},\ \bibinfo {pages}
  {150--216} (\bibinfo {year} {2014})}\BibitemShut {NoStop}%
\bibitem [{\citenamefont {Yang}\ \emph
  {et~al.}(2017{\natexlab{a}})\citenamefont {Yang}, \citenamefont
  {Khosronejad},\ and\ \citenamefont {Sotiropoulos}}]{Yang_etal_RE_2017}%
  \BibitemOpen
  \bibfield  {author} {\bibinfo {author} {\bibfnamefont {X.~L.}\ \bibnamefont
  {Yang}}, \bibinfo {author} {\bibfnamefont {A.}~\bibnamefont {Khosronejad}}, \
  and\ \bibinfo {author} {\bibfnamefont {F.}~\bibnamefont {Sotiropoulos}},\
  }\bibfield  {title} {\enquote {\bibinfo {title} {Large-eddy simulation of a
  hydrokinetic turbine mounted on an erodible bed},}\ }\href {\doibase
  10.1016/j.renene.2017.07.007} {\bibfield  {journal} {\bibinfo  {journal}
  {Renewable Energy}\ }\textbf {\bibinfo {volume} {113}},\ \bibinfo {pages}
  {1419--1433} (\bibinfo {year} {2017}{\natexlab{a}})}\BibitemShut {NoStop}%
\bibitem [{\citenamefont {Yang}\ and\ \citenamefont
  {Sotiropoulos}(2019{\natexlab{a}})}]{Yang_Sotiropoulos_BE_2019}%
  \BibitemOpen
  \bibfield  {author} {\bibinfo {author} {\bibfnamefont {X.~L.}\ \bibnamefont
  {Yang}}\ and\ \bibinfo {author} {\bibfnamefont {F.}~\bibnamefont
  {Sotiropoulos}},\ }\bibfield  {title} {\enquote {\bibinfo {title} {On the
  dispersion of contaminants released far upwind of a cubical building for
  different turbulent inflows},}\ }\href {\doibase
  10.1016/j.buildenv.2019.02.003} {\bibfield  {journal} {\bibinfo  {journal}
  {Build. Environ.}\ }\textbf {\bibinfo {volume} {154}},\ \bibinfo {pages}
  {324--335} (\bibinfo {year} {2019}{\natexlab{a}})}\BibitemShut {NoStop}%
\bibitem [{\citenamefont {Yang}\ and\ \citenamefont
  {Sotiropoulos}(2019{\natexlab{b}})}]{Yang_Sotiropoulos_PRF_2019}%
  \BibitemOpen
  \bibfield  {author} {\bibinfo {author} {\bibfnamefont {X.~L.}\ \bibnamefont
  {Yang}}\ and\ \bibinfo {author} {\bibfnamefont {F.}~\bibnamefont
  {Sotiropoulos}},\ }\bibfield  {title} {\enquote {\bibinfo {title} {Wake
  characteristics of a utility-scale wind turbine under coherent inflow
  structures and different operating conditions},}\ }\href {\doibase
  10.1103/PhysRevFluids.4.024604} {\bibfield  {journal} {\bibinfo  {journal}
  {Phys. Rev. Fluids}\ }\textbf {\bibinfo {volume} {4(2)}},\ \bibinfo {pages}
  {024604} (\bibinfo {year} {2019}{\natexlab{b}})}\BibitemShut {NoStop}%
\bibitem [{\citenamefont {Foti}\ \emph {et~al.}(2019)\citenamefont {Foti},
  \citenamefont {Yang}, \citenamefont {Shen},\ and\ \citenamefont
  {Sotiropoulos}}]{Foti_etal_JFM_2019}%
  \BibitemOpen
  \bibfield  {author} {\bibinfo {author} {\bibfnamefont {D.}~\bibnamefont
  {Foti}}, \bibinfo {author} {\bibfnamefont {X.~L.}\ \bibnamefont {Yang}},
  \bibinfo {author} {\bibfnamefont {L.}~\bibnamefont {Shen}}, \ and\ \bibinfo
  {author} {\bibfnamefont {F.}~\bibnamefont {Sotiropoulos}},\ }\bibfield
  {title} {\enquote {\bibinfo {title} {Effect of wind turbine nacelle on
  turbine wake dynamics in large wind farms},}\ }\href {\doibase
  10.1017/jfm.2019.206} {\bibfield  {journal} {\bibinfo  {journal} {J. Fluid
  Mech.}\ }\textbf {\bibinfo {volume} {869}},\ \bibinfo {pages} {1--26}
  (\bibinfo {year} {2019})}\BibitemShut {NoStop}%
\bibitem [{\citenamefont {Khosronejad}\ \emph {et~al.}(2020)\citenamefont
  {Khosronejad}, \citenamefont {Flora},\ and\ \citenamefont
  {Kang}}]{Khosronejad_etal_JHE_2020}%
  \BibitemOpen
  \bibfield  {author} {\bibinfo {author} {\bibfnamefont {A.}~\bibnamefont
  {Khosronejad}}, \bibinfo {author} {\bibfnamefont {K.}~\bibnamefont {Flora}},
  \ and\ \bibinfo {author} {\bibfnamefont {S.}~\bibnamefont {Kang}},\
  }\bibfield  {title} {\enquote {\bibinfo {title} {Effect of inlet turbulent
  boundary conditions on scour predictions of coupled {LES} and morphodynamics
  in a field-scale river: Bankfull flow conditions},}\ }\href {\doibase
  10.1061/(ASCE)HY.1943-7900.0001719} {\bibfield  {journal} {\bibinfo
  {journal} {J. Hydraul. Eng.}\ }\textbf {\bibinfo {volume} {146(4)}},\
  \bibinfo {pages} {04020020} (\bibinfo {year} {2020})}\BibitemShut {NoStop}%
\bibitem [{\citenamefont {Germano}\ \emph {et~al.}(1991)\citenamefont
  {Germano}, \citenamefont {Piomelli}, \citenamefont {Moin},\ and\
  \citenamefont {Cabot}}]{Germano_etal_PoF_1991}%
  \BibitemOpen
  \bibfield  {author} {\bibinfo {author} {\bibfnamefont {M.}~\bibnamefont
  {Germano}}, \bibinfo {author} {\bibfnamefont {U.}~\bibnamefont {Piomelli}},
  \bibinfo {author} {\bibfnamefont {P.}~\bibnamefont {Moin}}, \ and\ \bibinfo
  {author} {\bibfnamefont {W.~H.}\ \bibnamefont {Cabot}},\ }\bibfield  {title}
  {\enquote {\bibinfo {title} {A dynamic subgrid-scale eddy viscosity model},}\
  }\href {\doibase 10.1063/1.857955} {\bibfield  {journal} {\bibinfo  {journal}
  {Phys. Fluids A}\ }\textbf {\bibinfo {volume} {3}},\ \bibinfo {pages} {1760}
  (\bibinfo {year} {1991})}\BibitemShut {NoStop}%
\bibitem [{\citenamefont {Ge}\ and\ \citenamefont
  {Sotiropoulos}(2007)}]{Ge_Sotiropoulos_JCP_2007}%
  \BibitemOpen
  \bibfield  {author} {\bibinfo {author} {\bibfnamefont {L.}~\bibnamefont
  {Ge}}\ and\ \bibinfo {author} {\bibfnamefont {F.}~\bibnamefont
  {Sotiropoulos}},\ }\bibfield  {title} {\enquote {\bibinfo {title} {A
  numerical method for solving the {3D} unsteady incompressible
  {N}avier-{S}tokes equations in curvilinear domains with complex immersed
  boundaries},}\ }\href {\doibase 10.1016/j.jcp.2007.02.017} {\bibfield
  {journal} {\bibinfo  {journal} {J. Comput. Phys.}\ }\textbf {\bibinfo
  {volume} {225(2)}},\ \bibinfo {pages} {1782--1809} (\bibinfo {year}
  {2007})}\BibitemShut {NoStop}%
\bibitem [{\citenamefont {Kang}\ \emph {et~al.}(2011)\citenamefont {Kang},
  \citenamefont {Lightbody}, \citenamefont {Hill},\ and\ \citenamefont
  {Sotiropoulos}}]{Kang_etal_AWR_2011}%
  \BibitemOpen
  \bibfield  {author} {\bibinfo {author} {\bibfnamefont {S.}~\bibnamefont
  {Kang}}, \bibinfo {author} {\bibfnamefont {A.}~\bibnamefont {Lightbody}},
  \bibinfo {author} {\bibfnamefont {C.}~\bibnamefont {Hill}}, \ and\ \bibinfo
  {author} {\bibfnamefont {F.}~\bibnamefont {Sotiropoulos}},\ }\bibfield
  {title} {\enquote {\bibinfo {title} {High-resolution numerical simulation of
  turbulence in natural waterways},}\ }\href {\doibase
  10.1016/j.advwatres.2010.09.018} {\bibfield  {journal} {\bibinfo  {journal}
  {Adv. Water Resour.}\ }\textbf {\bibinfo {volume} {34(1)}},\ \bibinfo {pages}
  {98--113} (\bibinfo {year} {2011})}\BibitemShut {NoStop}%
\bibitem [{\citenamefont {Yang}\ \emph
  {et~al.}(2017{\natexlab{b}})\citenamefont {Yang}, \citenamefont {Bose},\ and\
  \citenamefont {Moin}}]{Yang_etal_CTR_2017}%
  \BibitemOpen
  \bibfield  {author} {\bibinfo {author} {\bibfnamefont {X.}~\bibnamefont
  {Yang}}, \bibinfo {author} {\bibfnamefont {S.}~\bibnamefont {Bose}}, \ and\
  \bibinfo {author} {\bibfnamefont {P.}~\bibnamefont {Moin}},\ }\bibfield
  {title} {\enquote {\bibinfo {title} {A physics-based interpretation of the
  slip-wall {LES} model},}\ }\href@noop {} {\bibfield  {journal} {\bibinfo
  {journal} {Center for Turbulence Research}\ }\textbf {\bibinfo {volume}
  {Annual Research Briefs (Stanford University, Stanford)}},\ \bibinfo {pages}
  {65--74} (\bibinfo {year} {2017}{\natexlab{b}})}\BibitemShut {NoStop}%
\bibitem [{\citenamefont {Goodfellow}\ \emph {et~al.}(2016)\citenamefont
  {Goodfellow}, \citenamefont {Bengio},\ and\ \citenamefont
  {Courville}}]{Goodfellow_etal_2016}%
  \BibitemOpen
  \bibfield  {author} {\bibinfo {author} {\bibfnamefont {I.}~\bibnamefont
  {Goodfellow}}, \bibinfo {author} {\bibfnamefont {Y.}~\bibnamefont {Bengio}},
  \ and\ \bibinfo {author} {\bibfnamefont {A.}~\bibnamefont {Courville}},\
  }\href {\doibase www.deeplearningbook.org} {\emph {\bibinfo {title} {Deep
  Learning}}}\ (\bibinfo  {publisher} {MIT Press, Cambridge, MA},\ \bibinfo
  {year} {2016})\BibitemShut {NoStop}%
\bibitem [{\citenamefont {Kingma}\ and\ \citenamefont
  {Ba}(2014)}]{Kingma_Ba_arXiv_2014}%
  \BibitemOpen
  \bibfield  {author} {\bibinfo {author} {\bibfnamefont {D.~P.}\ \bibnamefont
  {Kingma}}\ and\ \bibinfo {author} {\bibfnamefont {J.}~\bibnamefont {Ba}},\
  }\bibfield  {title} {\enquote {\bibinfo {title} {Adam: A method for
  stochastic optimization},}\ }\href@noop {} {\bibfield  {journal} {\bibinfo
  {journal} {arXiv:1412.6980v9}\ } (\bibinfo {year} {2014})}\BibitemShut
  {NoStop}%
\end{thebibliography}%

\end{document}